\documentclass[11pt,a4paper]{article}
\usepackage{cite}

\usepackage{amsmath, amsthm,mathtools,setspace, hyperref, url,amssymb,upgreek,textgreek,slashed}

 \usepackage[mathscr]{eucal}
\usepackage{ifpdf}
\ifpdf
  \usepackage[pdftex]{graphicx}
  \usepackage{epstopdf}
\else
  \usepackage[dvips]{graphicx}
\fi
\parskip=\baselineskip

\begin{document}

\begin{titlepage}
\begin{flushright}

\end{flushright}

\vskip 1.5in
\begin{center}
{\bf\Large{Gauge Theory and the Analytic Form}\vskip.5cm 
\bf\Large{Of The Geometric Langlands Program}}

\vskip
0.5cm { Davide Gaiotto} \vskip 0.05in {\small{ \textit{Perimeter Institute}
\vskip -.4cm
{\textit{Waterloo, Ontario, Canada N2L 2Y5}}
}}\vskip
0.5cm { Edward Witten} \vskip 0.05in {\small{ \textit{Institute for Advanced Study}\vskip -.4cm
{\textit{Einstein Drive, Princeton, NJ 08540 USA}}}
}
\end{center}
\vskip 0.5in
\baselineskip 16pt
\begin{abstract} We present a gauge-theoretic interpretation of the ``analytic'' version of the geometric Langlands program, 
in which Hitchin Hamiltonians and Hecke operators are viewed as concrete operators acting on a Hilbert space of quantum states.
The gauge theory ingredients required to understand this construction -- such as electric-magnetic duality between Wilson and 't Hooft line
operators in four-dimensional gauge theory -- are the same ones that enter in understanding via gauge theory the more familiar formulation
of geometric Langlands, but now these ingredients are organized and applied in a novel fashion.   
   \end{abstract}
\date{May, 2020}
\end{titlepage}
\def\SO{{\mathrm{SO}}}
\def\G{{\text{\sf G}}}
\def\frak{\mathfrak}
\def\la{\langle}
\def\ra{\rangle}
\def\Spinc{{\mathrm{Spin}}_c}
\def\g{{\mathfrak g}}
\def\cl{{\mathrm{cl}}}
\def\m{{\sf m}}
\def\veps{\varepsilon}
\def\Re{{\mathrm{Re}}}
\def\Im{{\mathrm{Im}}}
\def\SU{{\mathrm{SU}}}
\def\SL{{\mathrm{SL}}}
\def\U{{\mathrm U}}
\def\M{{\mathcal M}}
\def\d{{\mathrm d}}
\def\g{{\mathfrak g}}
\def\su{{\mathfrak {su}}}
\def\CS{{\mathrm{CS}}}
\def\Z{{\Bbb Z}}
\def\cB{{\mathcal B}}
\def\zZ{{\mathcal Z}}
\def\DD{{\mathscr D}}
\def\R{{\Bbb R}}
\def\sF{{\sf F}}
\def\cS{{\mathcal S}}
\def\sB{{\sf B}}
\def\sA{{\sf A}}
\def\sD{{\mathcal D}}
\def\dD{{\mathrm D}}
\def\F{{ \mathscr F}}
\def\cF{{\mathcal F}}
\def\J{{\mathcal J}}
\def\Bbb{\mathbb}
\def\Tr{{\rm Tr}}
\def\ad{{\mathrm{ad}}}
\def\j{{\sf j}}
\def\16{{\bf 16}}
\def\1{{(1)}}
\def\bCP{{\Bbb{CP}}}
\def\2{{(2)}}
\def\3{{\bf 3}}
\def\4{{\bf 4}}
\def\free{{\mathrm{free}}}
\def\sg{{\mathrm g}}
\def\J{{\mathcal J}}
\def\i{{\mathrm i}}
\def\h{\widehat}
\def\b{\overline}
\def\u{u}
\def\D{D}
\def\Rf{{\eurm{R}}}
\def\sp{{\sigma}}
\def\E{{\mathcal E}}
\def\O{{\mathcal O}}
\def\PF{{\mathit{P}\negthinspace\mathit{F}}}
\def\tr{{\mathrm{tr}}}
\def\be{\begin{equation}}
\def\ee{\end{equation}}
 \def\Sp{{\mathrm{Sp}}}
  \def\PSp{{\mathrm{PSp}}}
 \def\Spin{{\mathrm{Spin}}}
 \def\SL{{\mathrm{SL}}}
 \def\SU{{\mathrm{SU}}}
 \def\SO{{\mathrm{SO}}}
 \def\PGL{{\mathrm{PGL}}}
 \def\ll{\langle\langle}
 \def\frL{{{\mathfrak L}}}
 \def\RR{{\mathcal R}}
\def\rr{\rangle\rangle}
\def\la{\langle}
\def\CP{{C\negthinspace P}}
\def\sCP{{\sf{CP}}}
\def\I{{\mathcal I}}
\def\ra{\rangle}
\def\T{{\mathcal T}}
\def\V{{\mathcal V}}
\def\bar{\overline}
\def\spinc{{\mathrm{spin}_c}}
\def\dim{{\mathrm{dim}}}
\def\v{v}
\def\Pic{{\mathrm{Pic}}}

\def\RP{{\Bbb{RP}}}

\def\tilde{\widetilde}
\def\t{\widetilde}
\def\R{{\Bbb{R}}}
\def\N{{\mathcal N}}
\def\B{{\mathcal B}}
\def\H{{\mathcal H}}
\def\hat{\widehat}
\def\Pf{{\mathrm{Pf}}}
\def\bM{{\overline\M}}
\def\PSL{{\mathrm{PSL}}}
\def\PSU{{\mathrm{PSU}}}
\def\Im{{\mathrm{Im}}}
\def\Gr{{\mathrm{Gr}}}

\font\tencmmib=cmmib10 \skewchar\tencmmib='177
\font\sevencmmib=cmmib7 \skewchar\sevencmmib='177
\font\fivecmmib=cmmib5 \skewchar\fivecmmib='177
\newfam\cmmibfam
\textfont\cmmibfam=\tencmmib \scriptfont\cmmibfam=\sevencmmib
\scriptscriptfont\cmmibfam=\fivecmmib
\def\cmmib#1{{\fam\cmmibfam\relax#1}}
\numberwithin{equation}{section}
\def\lmark{{\mathrm L}}
\def\neg{\negthinspace}
\def\d{\mathrm d}
\def\C{{\Bbb C}}
\def\K{{\sf K}}
\def\cc{{\mathrm{cc}}}
\def\bB{{\bar {\mathscr B}}}
\def\HH{{\mathbb H}}
\def\P{{\mathcal P}}
\def\Q{{\mathcal Q}}
\def\NS{{\sf{NS}}}
\def\Ra{{\sf{R}}}
\def\sV{{\sf V}}
\def\CP{{\mathrm{CP}}}
\def\Hom{{\mathrm{Hom}}}
\def\Z{{\Bbb Z}}
\def\op{{\mathrm{op}}}
\def\bop{{\overline{\mathrm{op}}}}
\def\bA{\bar{\mathscr A}}
\def\A{{\mathscr A}}
\def\cA{{\mathcal A}}
\def\B{{\mathscr B}}
\def\S{{\mathcal S}}
\def\bar{\overline}
\def\sc{{\mathrm{sc}}}
\def\Max{{\mathrm{Max}}}
\def\CS{{\mathrm{CS}}}
\def\ga{\gamma}
\def\bg{\bar\ga}
\def\S{{\mathcal S}}
\def\W{{\mathcal W}}
\def\M{{\mathcal M}}
\def\bM{{\overline \M}}
\def\L{{\mathcal L}}
\def\sM{{\sf M}}
\def\gst{\mathrm{g}_{\mathrm{st}}}
\def\gstt{\widetilde{\mathrm{g}}_{\mathrm{st}}}
\def\hbbar{\pmb{\hbar}}
\def\G{{\mathcal G}}
\def\Sym{{\mathrm{Sym}}}
\def\UU{{\mathcal U}}
\def\Bun{{\mathcal M}(G,C)}
\def\be{\begin{equation}}
\def\ee{\end{equation}}

\tableofcontents

\section{Introduction}\label{intro}

Geometric Langlands duality as originally formulated by Beilinson and Drinfeld \cite{BD} is a relationship
between categories associated to moduli spaces of fields on a Riemann surface $C$.
 Many ingredients that enter in formulating and analyzing this
duality  are familiar in quantum field theory.   In particular, two-dimensional conformal field theory plays a prominent role, as reviewed in \cite{F}.
The relation of geometric Langlands duality to quantum field theory can be understood more fully by formulating the subject
in terms of a twisted version of $\N=4$ super Yang-Mills theory \cite{KW}.  From that point of view, geometric Langlands duality is
deduced from electric-magnetic duality between the $\N=4$ theory for a compact gauge group $G$
and the same theory based on the Langlands or GNO dual group $G^\vee$ (the complexifications of these groups will be called $G_\C$ and $G^\vee_\C$).  
Twisting of $\N=4$ super Yang-Mills
theory produces a  four-dimensional topological field theory, which naturally \cite{BaDo,L1,Fr} assigns a number (the partition function) to a four-manifold,
a vector space (the space of physical states) to a three-manifold, and a category of branes or boundary conditions to a two-manifold.\footnote{\label{partial}
This idealized description
does not take into account the fact that the complex Lie groups $G_\C$ and $G^\vee_\C$ and the associated moduli spaces are not compact.   
Because of this noncompactness, one likely gets only  a partial topological field theory. 
Branes and spaces of physical states can be defined, but 
it is not clear that the integrals that formally would define partition functions associated
to four-manifolds can really be defined satisfactorily. (A somewhat similar situation arises in Donaldson theory of four-manifolds, though the
details are quite different: one does not get a complete topological field theory, as the partition function cannot be suitably defined for all four-manifolds.)  
The noncompactness also means that to define branes and spaces of physical states, 
one has to specify the allowed asymptotic behavior of a brane or a wavefunction.  Each (reasonable) choice leads to a different version of the duality. 
Possibilities include the Betti and de Rham versions  \cite{BN}.}   
The usual geometric Langlands duality is a duality between the categories associated to two-manifolds. 

Recently  an analytic version of geometric Langlands duality has been discovered by Etingof, Frenkel, and Kazhdan \cite{EFK,EFK2,EFK3}, stimulated in part by
a number of mathematical \cite{L} and physical \cite{T} developments.   Rather than categories and functors acting on categories, one considers
a Hilbert space of quantum states and self-adjoint operators such as quantum Hitchin Hamiltonians acting on this Hilbert space.    Very roughly, the usual formulation of geometric Langlands duality involves deformation quantization of the algebra of holomorphic functions on
 the moduli space $\M_H(G,C)$ of $G$ Higgs bundles on a Riemann surface $C$, while the analytic version
of the theory involves ordinary quantization of the same moduli space, viewed now as a real symplectic manifold.   As usual, quantization means that
 a suitable class of smooth functions -- in general neither holomorphic
nor antiholomorphic -- become operators on a Hilbert space. See also \cite{Teschner:2010je,Balasubramanian:2017gxc,Nekrasov:2010ka,Nekrasov:2009rc,Nekrasov:2011bc,Nekrasov:2014yra,Jeong:2018qpc,Bonelli:2011na} for prior work on the gauge theory interpretation of the spectrum of quantum Hitchin Hamiltonians. 

The goal of the present article is to place the analytic version of geometric Langlands duality in the gauge theory framework.   The first step is simply to
understand what one should do  in that framework to study  the quantization of $\M_H$ viewed as a real symplectic manifold, as opposed
to its deformation quantization when viewed as a complex symplectic manifold.    The basic idea here is that the problem of quantization of a real symplectic manifold $M$
is part of the $A$-model of a suitable complexification $Y$ of $M$ (if such a $Y$ exists) \cite{GuW}.   We have explored this construction in more detail elsewhere 
as background to the present article \cite{GW}.   For the application to geometric Langlands, 
we want to quantize the Higgs bundle moduli space $\M_H$ viewed as a real symplectic
manifold with one of its real symplectic structures.   A suitable complexification of $\M_H$ is simply the product of two copies of $\M_H$ with opposite complex
structures.   With this as the starting point for understanding the quantization of $\M_H$, we will show that the analytic version of geometric Langlands
can be understood by assembling in a novel fashion the same  gauge theory 
ingredients that have been used previously for  understanding the more traditional version of geometric Langlands.

The organization of this article is as follows.   In Section \ref{basic}, we explain the basic setup for formulating the analytic version of geometric Langlands
duality in terms of $\N=4$ super Yang-Mills theory.    In Section \ref{hitchval}, we explain the predictions of electric-magnetic duality for the joint eigenvalues of the
Hitchin Hamiltonians, and in Section \ref{wtw}, we analyze Hecke or 't Hooft operators and the dual Wilson operators.  In Section \ref{wkb}, we show that the
joint eigenfunctions of the Hitchin Hamiltonians satisfy a quantum-deformed WKB condition.   In Section \ref{realb}, we discuss the quantization of real forms
of the Higgs bundle moduli space.    
Starting in Section \ref{avatar}, 
we use chiral algebras that arise at junctions between supersymmetric boundary conditions in order to build an interesting class of wavefunctions,
whose spectral decomposition is controlled by the geometry of $S$-dual boundary conditions. These wavefunctions are typically not eigenfunctions, 
but we expect them to play a useful role 
in the context of the analytic version of the geometric Langlands program. 
In Section \ref{sec:chiral}, we study Hecke operators by relating them to spectral flow automorphisms of Kac-Moody algebras.   The relationship between Hecke operators
and spectral flow automorphisms of chiral algebras, which is important in this analysis,
has been observed previously from a different but related perspective \cite{Teschner:2010je}.
 In Section \ref{sec:symp}, we explore in detail an example with remarkable properties, leading to a wavefunction which intertwines three copies of the Hitchin Hamiltonians and Hecke operators for $SL_2$.

Some further issues are treated in appendices. In Appendix \ref{coho}, we analyze the cohomology of local operators inserted along the canonical coisotropic
brane.   The result shows that Hitchin's classical integrable system can be quantized.    In Appendix \ref{bmodel}, we explain that electric-magnetic duality together with positivity of the Hilbert space
inner product in quantization of $\M_H$ imply that the intersections of the varieties $L_\op$ and $L_\bop$ that parametrize holomorphic and antiholomorphic
opers must be isolated and transverse (as conjectured in \cite{EFK} and proved in some cases).    We further point out that electric-magnetic duality implies
a natural normalization for the joint eigenfunctions of the Hitchin Hamiltonians and argue that the Hilbert space norm of a normalized wavefunction is given
by, roughly speaking, the torsion of the associated oper bundle.  In Appendix \ref{somex}, we explore some examples of differential equations satisfied by line
operators.   In Appendix \ref{localmodel}, we construct a local model to analyze the singular behavior of the eigenfunctions of the Hitchin Hamiltonians
along the divisor of not very stable bundles.  

Throughout this article, $C$ is a Riemann surface of genus $g>1$.    All considerations can be naturally extended to the case of bundles with parabolic
structure, but for simplicity we will omit this generalization (which has been developed in \cite{EFK,EFK2,EFK3}).  With sufficiently many parabolic
points, there is a quite similar theory for genus $g=0,1$.   The cases $g=0,1$ without parabolic structure (or with too few parabolic points for $g=0$)
require a different treatment because any low energy description requires gauge fields, not just $\sigma$-model fields.

\section{Basic Setup}\label{basic}

This section is devoted to an explanation of the basic framework in which we will study the analytic version of geometric Langlands,
and a review of part of the background.

\subsection{Quantization Via Branes}\label{background}

Here we briefly summarize some things that have been described more fully elsewhere \cite{GuW,GW}.   First we explain deformation quantization via
branes and then we describe quantization.

Let $Y$ be a complex symplectic manifold, with complex structure $I$ and
holomorphic symplectic form $\Omega$.   We view $Y$ as a real symplectic manifold with the real symplectic
form $\omega_Y=\Im\,\Omega$.   We assume that $Y$ is such that a quantum $\sigma$-model with target $Y$ exists (as an ultraviolet-complete
quantum field theory), and we consider the $A$-model obtained
by twisting this $\sigma$-model in a standard way.   It is not known in general for what class of $Y$'s the $\sigma$-model does exist, but a sufficient condition is believed
to be that the complex symplectic structure of $Y$ can be extended to a complete hyper-Kahler structure.   The examples important for the present article are
the cases that $Y$ is the Higgs bundle moduli space $\M_H$, which indeed admits a complete hyper-Kahler metric \cite{H}, or a product $\M_H\times \M_H$ (with
opposite complex structures on the two factors and complex conjugate symplectic structures), which of course also has a complete hyper-Kahler metric.

In general, the $A$-model of a symplectic manifold $Y$, in addition to the usual Lagrangian branes whose support is middle-dimensional in $Y$,
can have coisotropic branes, supported on a coisotropic  submanifold of $Y$ that is above the middle dimension \cite{KO}.    The simplest case and the only
case that we will need in the present article is a rank 1 coisotropic $A$-brane whose support is all of $Y$.   Let $\sB$ be the $B$-field of the $\sigma$-model,
and consider a brane with support $Y$ whose Chan-Paton or $\CP$ bundle is a  line bundle $\L\to Y$, with a unitary connection $\sA$ of curvature $\sF=\d\sA$.  
The Kapustin-Orlov condition
for this data to define an $A$-brane is  that $\I=\omega_Y^{-1}(\sF+\sB)$ should be an integrable complex structure on $Y$.   Two solutions
of this condition immediately present themselves.    One choice is $\sF+\sB=\Re\,\Omega$, $\I=I$.   The brane constructed this way is what we will call
the canonical coisotropic $A$-brane, $\B_\cc$.  A second choice is $\sF+\sB=-\Re\,\Omega$, $\I=-I$.    This leads to what we will call the conjugate
canonical coisotropic $A$-brane, $\bB_\cc$.    To treat the two cases symmetrically, in the present article it is convenient to take\footnote{This choice was made
in \cite{GuW}.   However, when only one of $\B_\cc$, $\bB_\cc$ is relevant, it can be simpler to take $\sF=0$, $\sB=\Re\,\Omega$.}
 $\sB=0$, $\sF=\pm \Re\,\Omega$.   This choice is only possible if there exists a complex line bundle $\L\to Y$ with curvature $\Re\,\Omega$.   In our
 application to the Higgs bundle moduli space, $\Re\,\Omega$ is cohomologically trivial, so $\L$ exists and can  be assumed to be topologically trivial. 
 
Now consider the algebra $\A=\Hom(\B_\cc,\B_\cc)$ (which in physical terms is the space of $(\B_\cc,\B_\cc)$ open strings, with an associative
multiplication that comes by joining of open strings).    Suppose that $\Omega=\Omega_0/\hbar$, where we keep $\Omega_0$ fixed and vary $\hbar$.
For $\hbar\to 0$, $\A$ reduces to the commutative algebra $\A_0$ of holomorphic functions on $Y$ in complex structure $I$.   In order $\hbar$,
the multiplication law in $\A$ differs from the commutative multiplication law in $\A_0$ by the Poisson bracket $\{f,g\}=(\Omega^{-1})^{ij}\partial_i f\partial_j g$.
So $\A$ can be viewed as a deformation quantization of $\A_0$.   $\bA=\Hom(\bB_\cc,\bB_\cc)$ is  related in the same way to the commutative algebra $\bA_0$
of holomorphic functions on $Y$ in complex structure $-I$, or equivalently antiholomorphic functions in complex structure $I$, and can be viewed as a deformation
quantization of $\bA_0$.   

Generically, deformation quantization is a formal procedure that has to be defined over a ring of formal power series in $\hbar$.   When a quantum $\sigma$-model
of $Y$ exists, it is expected that $\hbar$ can be set to a complex value (such as 1), rather than being treated as a formal power series variable.  
In the present example, one can give a more direct explanation of this.   The Higgs bundle moduli space $\M_H(G,C)$ has a $\C^*$ symmetry, rescaling
the Higgs field by $\varphi\to \lambda\varphi$, $\lambda\in \C^*$.     This operation rescales $\Omega$ in the same way.  Of course, this is possible only
because $\Omega$ is cohomologically trivial.  The ring $\A_0$ is generated by functions that scale with a definite degree 
(namely the Hitchin Hamiltonians) and the scaling symmetry implies
that all the relations in the deformed ring $\A$ are polynomials in the deformation parameter $\hbar$. Hence it makes sense to set $\hbar=1$.
Similar remarks will apply  when we consider quantization rather than deformation quantization, given that the important branes considered  are $\C^*$-invariant:
the scaling symmetry will imply that certain semiclassical formulas are actually exact.

Deformation quantization is particularly interesting if $Y$ is an affine variety, with lots of holomorphic functions.  As already noted, in our main examples, 
 $Y$ will be  $\M_H$ or  $\M_H\times \M_H$.   $\M_H$ is far from being
an affine variety.  The ring $\A_0$ in the case $Y=\M_H$ is simply the ring of functions on the base of the Hitchin fibration; in other words, the global holomorphic functions
on $\M_H$ are simply the functions of Hitchin's Poisson-commuting Hamiltonians.    Hitchin's Poisson-commuting Hamiltonians can be quantized
to commuting differential operators, acting on sections of the line bundle $K^{1/2}$, where $K$ is the canonical bundle of $\M_H$.
  This was shown by Hitchin \cite{H2}
for $G=\SU(2)$ and by Beilinson and Drinfeld \cite{BD} in general.  In particular, the ring $\A$ is commutative.  
The fact that the Poisson-commuting classical Hamiltonians can be quantized to
commuting differential operators can also be seen in the gauge theory language, as we will discuss in Section \ref{hh} and Appendix \ref{coho}.  For $Y=\M_H\times\M_H$,
deformation quantization just gives the product of this answer with its complex conjugate.

In the present article, we are really interested in quantization rather than deformation quantization.   How to modify the story just described 
to encompass quantization has been explained in \cite{GuW,GW}.   Suppose that $M$ is a real symplectic manifold, with symplectic form $\omega_M$,
that we wish to quantize.  (In our main application, $M$ will be $\M_H$, viewed as a real symplectic manifold, and $Y$ will be $\M_H\times \M_H$.)     If $M$ has a complexification $Y$ that obeys certain conditions, then quantization of $M$ is part of the $A$-model of $Y$.
$Y$ should be a complex symplectic manifold with a holomorphic symplectic form $\Omega$ whose restriction to $M$ is $\omega_M$.   Moreover,
$Y$ should have an antiholomorphic involution\footnote{An involution is an automorphism whose square is 1.}  $\tau$ with $M$ as a component of its fixed point set.   These conditions imply that $\tau^*\Omega=\bar\Omega$.   For the example  $Y=\M_H\times \M_H$, the antiholomorphic involution is the map that exchanges the two factors.
Finally, the quantum $\sigma$-model of $Y$ must exist.   Under these conditions, a choice of a prequantum line bundle  ${\frL}\to M$, in the 
sense of geometric quantization,\footnote{A prequantum line bundle over a symplectic manifold $M$ with symplectic form $\omega$ is a complex line
bundle $\frL\to M$ with a unitary connection of curvature $\omega$.}
 determines an $A$-brane $\B$ with support $M$.    (The details of this depend on the choice that was made in satisfying 
the condition $\sF+\sB=\Re\,\Omega$ to
define the brane $\B_\cc$. We will be specific in Section \ref{cm}.)
The $A$-model answer for the Hilbert space obtained by
quantizing $M$ with symplectic structure $\omega_M$ and prequantum line bundle $\frL$ is $\H=\Hom(\B,\B_\cc)$. 
The definition of the hermitian inner product on $\H$ is described in Section \ref{cm}.

 $\H$ is always a module for $\A=\Hom(\B_\cc,\B_\cc)$. 
This can be described by saying that those functions on $M$ that can be analytically continued to holomorphic functions on $Y$ are quantized
to give operators on $\H$.    In addition, under mild conditions, a correspondence\footnote{A correspondence between $M$ and itself is simply
a Lagrangian submanifold of $M_1\times M_2$, where $M_1$ and $M_2$ are two copies of $M$, with respective symplectic structures $\omega_M$ and 
$-\omega_M$.    A  holomorphic correspondence between $Y$ and itself is defined similarly.}
between $M$ and itself that can be analytically continued to a holomorphic correspondence between
$Y$ and itself can be quantized in a natural way to give an operator on $\H$.   
 In our main example with $M=\M_H$, the ring $\A$
will be relatively small, consisting of polynomials in the holomorphic and antiholomorphic Hitchin Hamiltonians.   By way of compensation, there is an ample
supply of correspondences -- the Hecke correspondences -- that will provide additional operators on $\H$.  Quantizing these correspondences is simple
because of the $\C^*$ scaling symmetry that was invoked in the discussion of deformation quantization.   

We have described this in the context of two-dimensional $\sigma$-models, as is appropriate for quantization of a fairly general real symplectic manifold $M$.
However, in the particular case that $M$ is the Higgs bundle moduli space $\M_H(G,C)$ for gauge group $G$ on a Riemann surface  $C$, a four-dimensional
picture is available and gives much more complete understanding.   For this, the starting point is $\N=4$ super Yang-Mills theory, in four dimensions,
with gauge group $G$.  One restricts to four-manifolds of the form $\Sigma\times C$, where $\Sigma$ is an arbitrary two-manifold
but $C$ is kept fixed.   At low energies, the four-dimensional gauge theory reduces for many purposes, assuming that $G$ has trivial center,
to a supersymmetric $\sigma$-model on $\Sigma$ with target  $\M_H(G,C)$
\cite{BJSV,JWS}.   A certain twisting of the $\N=4$ theory in four dimensions produces a topological field 
theory\footnote{This is really a partial topological field theory, as remarked in footnote \ref{partial}.}  that, when we specialize
to four-manifolds of the form $\Sigma\times C$, reduces to an $A$-model on $\Sigma$ with target $\M_H(G,C)$.   The brane $\B_\cc$ of the $\sigma$-model
originates in the four-dimensional gauge theory as a boundary condition that is a simple deformation of Neumann boundary 
conditions for the gauge field, extended to the whole supermultiplet in a half-BPS fashion; see Section 12.4 of \cite{KW}. 

The four-dimensional lift of the $\B_\cc$ boundary condition  has an interesting and important feature.    Though it is a brane in an $A$-model that
has full four-dimensional topological symmetry, the definition of the brane $\B_\cc$ depends on a choice of complex structure of $C$.
The deformed Neumann boundary condition is a {\it holomorphic-topological} local boundary condition for the four-dimensional 
topological field theory.\footnote{Concretely, the boundary condition breaks some of the bulk supersymmetries 
of the physical theory. As a consequence, some translation generators which are $Q$-exact in the bulk cease to be $Q$-exact in 
the presence of the boundary and the boundary condition is not topological in the four-dimensional sense.  Another explanation of the dependence of $\B_\cc$
on a choice of complex structure is simply that $\B_\cc$ 
is defined with $\sF+\sB=\omega_J$, whose definition depends on the complex structure of $C$.   See the beginning of Section \ref{cm} for a statement of which
structures on $\M_H$ do or do not depend on a choice of complex structure. } 
Boundary local operators  supported
at a point $p\in C$ can depend holomorphically  on $p$, even though they depend topologically on the one remaining boundary 
direction. This  will be important whenever we 
discuss the four-dimensional lift of our constructions.  Although the $A$-model has full four-dimensional topological invariance, in the presence of the brane
$\B_\cc$, only two-dimensional topological invariance is available.   For example, in proving the commutativity of the Hitchin Hamiltonians and the Hecke
operators, we will use only two-dimensional topological invariance.

If $G$ has a  nontrivial center $\zZ(G)$, then the assertion about a reduction to a $\sigma$-model must be slightly modified.   A more precise statement is that upon
compactification on $C$, the
four-dimensional gauge theory reduces, for many purposes, to the product of a $\sigma$-model with target $\M_H(G,C)$ and a gauge theory with the finite
gauge group $\zZ(G)$ (acting trivially on $\M_H(G,C)$).   This finite gauge group will play no role until we want to compute the eigenvalues of Hitchin Hamiltonians
and Hecke operators, so in much of the following it will not be mentioned.

A purely two-dimensional description via a $\sigma$-model with target $\M_H(G,C)$ (possibly extended by the finite gauge group) is useful for many purposes. But information is lost in the reduction to
two dimensions, and the four-dimensional picture is needed for a complete account of the duality.\footnote{To be more precise, a complete formulation of the duality
is possible in four dimensions.  A fuller explanation of the duality  comes from a certain starting point in six 
dimensions \cite{WittenLecture}.}
In standard mathematical treatments, an analogous statement is that geometric Langlands duality must be  formulated
in terms of the ``stack'' of $G_\C$-bundles or of $G_\C^\vee$ local systems, rather than in terms of a finite-dimensional moduli space.  
There is a simple relation between the two statements.
It was shown by Atiyah and Bott \cite{AB} that the space of all $G$-valued connections on a smooth $G$-bundle  $E\to C$, with the action
of the group of  complexified gauge transformations, provides a model of the stack of holomorphic $G_\C$-bundles.  
That is because the $(0,1)$ part of any connection gives $E$ (or more precisely its complexification) a complex structure, making it a holomorphic
$G_\C$ bundle over $C$.
A gauge field on a 
$G$-bundle   over $ \Sigma\times C$ 
determines, in particular, a family, parametrized by $\Sigma$, of gauge fields on $C$.    So any such connection  determines a  map from $\Sigma$ to the stack of
$G_\C$ bundles over $C$.   Thus $\N=4$ super Yang-Mills theory on $\Sigma\times C$ can be understood as a supersymmetric $\sigma$-model on $\Sigma$
with the target being the stack of $G_\C$ bundles over $C$.  (In this formulation, the theory is a two-dimensional supersymmetric
gauge theory on $\Sigma$ coupled to matter fields.
 The matter 
fields are gauge fields on $C$ and their superpartners, and the gauge group is the group $\h G$ of maps of $C$ to the finite-dimensional group $G$. A  gauge transformation in the two-dimensional theory is a map from $\Sigma$ to $\h G$; in the four-dimensional description, this is interpreted
 as a map of the four-manifold $\Sigma\times C$ to $G$.
To keep these statements simple, we have assumed that all bundles are trivialized.)
 The mathematical statement that the correct formulation involves stacks corresponds to the
quantum field theory statement that the correct formulation is in four dimensions.

\subsection{Quantizing A Complex Manifold}\label{cm}

We view $\M_H(G,C)$ as a complex manifold in the complex structure, called $I$ by Hitchin \cite{H}, in which it parametrizes Higgs bundles over $C$.
$I$ is one of a triple of complex structures $I,J,K$ that, along with the corresponding Kahler forms $\omega_I,\omega_J,\omega_K$, 
make a hyper-Kahler structure on $C$.   In complex structure $J$, $\M_H(G,C)$ parametrizes flat bundles over $C$ with structure group $G_\C$.
Complex structure $J$ and the corresponding holomorphic symplectic form $\Omega_J=\omega_K+\i\omega_I$ are topological in the sense that they
depend on $C$ only as an oriented two-manifold, while the other structures $I$, $K$, and $\omega_J$ depend on a choice of complex structure on $C$.

The complex symplectic form of $\M_H(G,C)$, in complex structure $I$, is $\Omega_I=\omega_J+\i\omega_K$.    We want to view $\M_H$ as a real symplectic
manifold with the real symplectic structure $\omega=\omega_J=\Re\,\Omega_I$, and quantize it.

As we have  explained in Section \ref{background}, the first step in quantizing any real symplectic manifold $M$ via branes is to pick a suitable complexification
of it.    Here we are in a special situation, since $M$ is actually a complex manifold.   
In such a case, there
 is a standard way to proceed, described in Section 5 of \cite{GW}.
   
 In spelling out the details, rather than specializing to $\M_H$, we will consider the general problem of quantizing a complex symplectic manifold viewed as a real
 symplectic manifold.    In this particular context, we will write $Y$ (rather than $M$) for the manifold that is being quantized.   The reason for this  is
 that $Y$ is our generic notation for a complex symplectic manifold, and in the approach to quantization that we will describe, the complex symplectic
 structure plays an important role.     Our notation is particularly natural in the
  ``unfolded'' version of the construction (see fig. \ref{folding} below).   Hopefully, this usage will cause no confusion.

For a complexification of $Y$, we take
$\h Y=Y_1\times Y_2$, where $Y_1$ and $Y_2$ are two copies of $Y$, with opposite complex structures $I$ and $-I$. The complex
structure of $\h Y$ is thus a direct sum $\I=I\oplus (-I)$.     The involution  $\tau:\h Y\to\h Y$
that exchanges the two factors is antiholomorphic, and its fixed point set is a copy of $Y$, embedded as the diagonal in $\h Y=Y_1\times Y_2$.
We endow $\h Y$ with the complex symplectic form $\h\Omega=\frac{1}{2}\Omega\boxplus \frac{1}{2}\bar\Omega$; in other words, the symplectic
form of $\h Y=Y_1\times Y_2$ is $\frac{1}{2}\Omega$ on the first factor and $\frac{1}{2}\bar\Omega$ on the second factor.  This definition ensures
that the restriction of $\h\Omega$ to the diagonal is $\Re\,\Omega$.   Suppose that the complex symplectic structure of $Y$ can be extended to a complete
hyper-Kahler metric.   Then the complex symplectic structure of $\h Y$ can likewise be extended to a complete hyper-Kahler metric,
 namely  a product hyper-Kahler metric on $\h Y=Y_1\times Y_2$.
So all conditions are satisfied, and  the $A$-model of $\h Y$ with real symplectic form 
$\Im\,\h\Omega=\frac{1}{2}\Im\,\Omega\boxplus (-\frac{1}{2}\Im\,\Omega)$
is suitable for quantizing $Y$ with the symplectic structure $\omega_Y=\Re\,\Omega$.  

To define coisotropic branes, we have to satisfy
 the Kapustin-Orlov condition that $\omega^{-1} (\sF+\sB)$ should be an integrable complex structure.  In doing so, we  will take $\sB=0$, as this will make
it possible to treat the two factors of $\h Y$ symmetrically.
   Since the complex structures
of $Y_1$ and $Y_2$ are respectively $I$ and $-I$ and the $A$-model symplectic structures are respectively $\frac{1}{2}\Im\,\Omega$ and $-\frac{1}{2}\Im\,\Omega$,
we can define canonical coisotropic branes  $\B_{\cc,1}$ and $\B_{\cc,2}$ on $Y_1$ and $Y_2$ by taking in each case a $\CP$ bundle  $\L$ with curvature
 $\sF=\frac{1}{2}\omega_J$.   (In our application to the Higgs bundle moduli space, $\Re\,\Omega$ is exact and $b_1(Y)=0$, so a topologically trivial line bundle over $Y$
 with curvature $\frac{1}{2}\omega_J$ exists and is unique up to isomorphism.)  
 One then defines on $\h Y$ the product brane $\h\B_\cc=\h\B_{\cc,1}\times
 \h\B_{\cc,2}$, with $\CP$ bundle $\h\L=\L\boxtimes\L$.
         We also define a Lagrangian brane $\B$ supported on $Y$ with trivial $\CP$ bundle.   Following the general logic, the Hilbert space for quantization of $Y$ with the real symplectic structure $\Re\,\Omega$ is $\H=\Hom(\B,\h\B_\cc)$.   The prequantum line bundle in this situation is $\frL=\h\L|_Y\cong \L^2$.   Since $\L$ has curvature
         $\frac{1}{2}\Re\,\Omega$, $\L^2$ has curvature $\Re\,\Omega$, so it is an appropriate prequantum line bundle for quantization of $Y$ with symplectic
         structure $\Re\,\Omega$.

 \begin{figure}
 \begin{center}
   \includegraphics[width=4in]{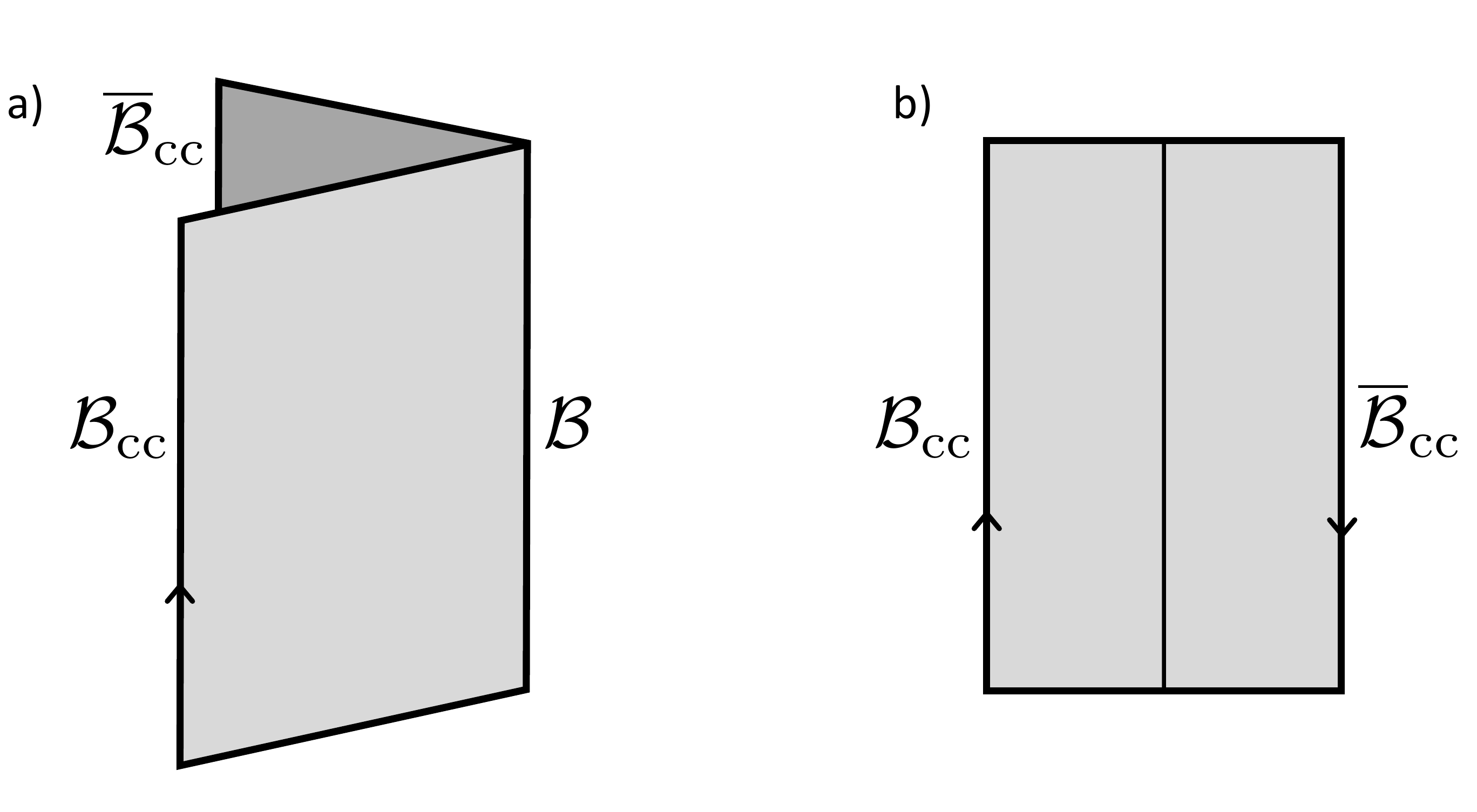}
 \end{center}
\caption{\small (a) In the folded construction, we have two copies of a $\sigma$-model on a strip.   The two copies are decoupled except on the
right boundary, where they are glued together by a brane $\B$ that is supported on the diagonal in $\h Y=Y_1\times Y_2$. $\B$ has trivial $\CP$
bundle so this gluing of the two copies is its only effect.  (b)   After unfolding,
we have a single copy of the $\sigma$-model on a strip of twice the width.   No trace of $\B$ remains.   \label{folding}}
\end{figure}

   However, 
   this definition has a useful variant.   To compute $\Hom(\B,\h\B_\cc)$, we study the $\sigma$ model with target $\h Y$ on
 a strip $\Sigma$, with boundary conditions set by $\h\B_\cc$ of the left boundary of the strip and by $\B$ on the right boundary, as in fig. \ref{folding}(a).
 The two factors of $\h Y=Y_1\times Y_2$ are decoupled in the bulk of the $\sigma$-model, since the metric on $\h Y$ is a product; they are also
 decoupled on the left boundary, since the brane $\h\B_\cc=\B_{\cc,1}\times \B_{\cc,2}$ is likewise a product.   So away from the right boundary of the strip, we can
 think of $\Sigma$ as having two sheets, one of which is mapped to $Y_1$ and one to $Y_2$, as in the figure.   The two sheets are coupled only
 on the right boundary, where, as $\B$ is supported on the diagonal in $Y_1\times Y_2$, the two sheets are ``glued together'' and map to the same point in $Y$.
 This suggests that we should ``unfold'' the picture (fig. \ref{folding}(b)).   After this unfolding, we simply have a single sheet of twice the width that is
 mapped to a single copy of $Y$.   Unfolding reverses the orientation of one of the two sheets of the folded picture, and this orientation reversal
changes the sign of the $A$-model
 symplectic form.   In the folded picture, the $A$-model symplectic form was $\frac{1}{2}\omega_K$ on $Y_1$ and $-\frac{1}{2}\omega_K$ on $Y_2$;
 hence after unfolding, the $A$-model symplectic form is $\frac{1}{2}\omega_K$ everywhere.   In other words,  the unfolded picture involves the ordinary
 $A$-model of a single copy of $Y$ with symplectic form $\frac{1}{2}\omega_K$.  Before unfolding, the branes $\B_{\cc,1}$ and $\B_{\cc,2}$ both
 have $\CP$ bundles with curvature $\frac{1}{2} \omega_J$.    Reversing the orientation of $\Sigma$ replaces the $\CP$ bundle of a brane with its dual
 (or its inverse, in the rank 1 case), and so reverses the sign of the $\CP$ curvature.   Hence in the unfolded picture, the boundaries are labeled
 by branes $\B_\cc$ and $\bB_\cc$ whose respective $\CP$ bundles are lines bundles $\L$ and $\L^{-1}$ with curvatures
  $\frac{1}{2}\omega_J$ and $-\frac{1}{2}\omega_J$.   These are
 the conjugate canonical coisotropic branes that were introduced in Section \ref{background}.  The prequantum line bundle
is still $\frL=\L^2$. 
   $\A=\Hom(\B_\cc,\B_\cc)$ is a deformation
 quantization of the commutative algebra $\A_0$ of holomorphic functions on $Y$, and $\b\A=\Hom(\bB_\cc,\bB_\cc)$ is similarly a deformation
 quantization of the algebra $\bA_0$ of antiholomorphic functions on $Y$.    What in the folded picture was $\H=\Hom(\B,\h\B_\cc)$ becomes in the
 unfolded picture $\H=\Hom(\bar\B_\cc,\B_\cc)$.

 With either description of $\H$, we need to define a hermitian product on $\H$.     For definiteness
 we use the folded language.\footnote{For more detail on the following, see Section 2.7 of \cite{GW}.}    Topological
 field theory would give us in general  a nondegenerate bilinear (not hermitian) pairing $(~,~)$ between $\H=\Hom(\B,\h\B_\cc)$ and its dual space
 $\H'=\Hom(\h\B_\cc,\B)$.     To get a hermitian pairing on $\H$, we need to compose this bilinear pairing with an antilinear map from $\H$ to $\H'$.
 Such a map in the underlying physical theory is provided by the $\sf{CPT}$ symmetry $\Theta$.   But $\Theta$ is not an $A$-model symmetry;
 it maps the $A$-model to a conjugate $A$-model with the opposite sign of the symplectic form.   The  involution $\tau$ that exchanges the two factors
 of $\h Y$
 also exchanges the $A$-model with its conjugate, since it is antisymplectic 
 (it reverses the sign of the $A$-model symplectic form), so $\Theta_\tau=\Theta\tau$ is an antilinear symmetry
 of the $A$-model.   Finally, because the branes $\B$ and $\B_\cc$ are $\Theta_\tau$-invariant, $\Theta_\tau$ maps $\H$ to $\H'$ and
 we can define a nondegenerate hermitian pairing on $\H$ by 
 \be\label{hermdef}\la \psi,\psi'\ra= (\Theta_\tau\psi,\psi').\ee   
For  general $\Theta_\tau$-invariant branes,  
such a pairing is not positive-definite.   For the specific case of quantizing a cotangent bundle, which is our main example on the $A$-model
 side, one expects positivity.   The $B$-model analog of this construction uses an antiholomorphic (not antisymplectic) involution $\tau$.  
 Positivity of the pairing in this case is subtle and is discussed in Appendix \ref{bmodel}.    
 
  In the folded picture, $\tau$ and therefore also
 $\Theta_\tau$ exchanges the two factors of $\h Y=Y_1\times Y_2$; in the unfolded picture, they exchange the two ends of the strip.   Exchanging the
 two ends of the strip reverses the orientation of the strip and therefore would change the sign of $\sB$.   Hence in a description with $\sB\not=0$,
 the definition of the inner product is less natural (one would need to accompany $\Theta_\tau$ with a $B$-field gauge transformation).  That is why we took
 $\sB=0$ in solving the Kapustin-Orlov conditions for rank 1 coisotropic branes.

 \subsection{Quantizing The Higgs Bundle Moduli Space}\label{aphiggs}
 
 For our application to quantization of $ \M_H(G,C)$ for some gauge group $G$ and 
 Riemann surface $C$, we really want to study the four-dimensional
 version of this construction.   This means that we study the $\N=4$ super Yang-Mills theory, with gauge group $G$, on $\Sigma\times C$, where
 $\Sigma$ is the strip of fig \ref{folding}(b).   The boundary conditions on the left and right of the strip are set by the gauge theory versions of $\B_\cc$ and
 $\bB_\cc$.   A detailed description of $\B_\cc$ in four-dimensional gauge theory language was given in Section 12.4 of \cite{KW}.    We can describe
 a Higgs bundle on $C$ by a pair $(A,\phi)$, where $A$ is gauge field, that is, a connection on a $G$-bundle $E\to G$,
  and $\phi$ is a 1-form valued in the adjoint bundle $\ad(E)$.  In this description, $\bB_\cc$ is obtained from $\B_\cc$ by $(A,\phi)\to (A,-\phi)$ (suitably extended
  to the rest of the four-dimensional supermultiplet).     This is a familiar involution of the Higgs bundle moduli space that acts holomorphically in complex
  structure $I$ and antiholomorphically in complex structures $J$ and $K$.

Although we have used machinery of gauge theory and branes to construct a Hilbert space $\H$ associated to quantization of $\M_H(G,C)$,   the actual
output of this construction is completely unsurprising.  A dense open set in $\M_H(G,C)$ is a cotangent bundle $T^*\M(G,C)$, where $\M$ is the moduli
space of semistable holomorphic $G$-bundles over $C$.   Geometric quantization -- or simply elementary quantum mechanics -- suggests that the
Hilbert space that we should associate to quantization of $T^*\M(G,C)$ should be the space of $\lmark^2$ half-densities on $\M(G,C)$.    

The reason to speak of $\lmark^2$ half-densities rather than $\lmark^2$ functions is that on a general smooth manifold $N$ without some choice of 
a measure,\footnote{The space $\M(G,C)$ actually does have a natural measure, namely the one associated to its real symplectic structure when viewed
as a moduli space of flat bundles over $C$ with compact structure group $G$. This is also the measure induced by its embedding in the hyper-Kahler
manifold $\M_H(G,C)$.  However, this measure is not part of the $A$-model and does not naturally appear
in $A$-model constructions such as the definition of the Hilbert space $\H$.} there is no way to integrate a function
so there is no natural Hilbert space of $\lmark^2$ functions.   A density on a  manifold $N$ is a section of a trivial real line bundle  $\K$
and can be written locally in any
coordinate system as $|\d x^1\d x^2\cdots \d x^w| \,f(x^1,x^2,\cdots, x^w)$, where  $f(x^1,\cdots, x^w)$ is a function and  
$|\d x^1\d x^2\cdots \d x^w|$ is a measure, not a differential form.  $\K$ has a square root $\K^{1/2}$, also trivial, whose sections are locally
described in a given coordinate system by functions $g(x^1,x^2,\cdots, x^w) $ that transform under a change of coordinates in such a way
that $|\d x^1\d x^2\cdots \d x^w| \,g(x^1,x^2,\cdots, x^w)^2$ is invariant.   It is convenient to formally write 
\be\label{formw} h=|\d x^1\d x^2\cdots \d x^w|^{1/2} g(x^1,x^2,\cdots, x^w)\ee
 for a section of $\K^{1/2}$.   Complex-valued half-densities, which are described by the same formula where locally $g$ is a complex-valued function,
form a Hilbert space in an obvious way: \be\label{ob} ||h||^2=\int |\d x^1\d x^2\cdots\d x^w|\, |g(x^1,x^2\cdots x^w)|^2.\ee
   Now, motivated by the application to the
complex manifold $\M(G,C)$, let us describe the bundle of densities or half-densities on a complex manifold $N$.   If $N$ has complex
dimension $n=w/2$ and local holomorphic coordinates $z^1,z^2,\cdots, z^n$, then $|\d z^1 \d z^2 \cdots \d z^n \d \bar z^1\d\bar z^2\cdots \d\bar z^n|$
is a  density on $N$, in other words a section of $\K\to N$.   On the other hand, $\d z^1 \d z^2 \cdots \d z^n $ is a section of the holomorphic canonical bundle $K\to N$,
and $ \d \bar z^1\d\bar z^2\cdots \d\bar z^n$ is a section of the complex conjugate line bundle $\bar K\to N$ (which can also be viewed as the canonical
line bundle of $N$ if $N$ is endowed with the opposite complex structure).   So $\K$ can be identified with $K\otimes \bar K$; more precisely
$K\otimes \bar K\cong \K\otimes _\R\C$, that is, $K\otimes \bar K$ is the complexification of $\K$, the bundle of complex-valued densities.   Similarly $K$ always has a square root at least locally, and for any choice of local square
root of $K$, we have $\K^{1/2}\cong K^{1/2}\otimes \bar K{}^{1/2}$; more precisely $K^{1/2}\otimes \bar K^{1/2}\cong \K^{1/2}\otimes_\R \C$, that is,
$K^{1/2}\otimes \bar K^{1/2}$ is the bundle of complex-valued half-densities.  As long as $\bar K{}^{1/2}$ is the complex conjugate of $K^{1/2}$, this relation holds
for any choice of $K^{1/2}$.

In Section 3 of \cite{GW}, criteria were described under which brane quantization of $M=T^*N$, with its standard symplectic structure as a cotangent
bundle,  leads to a Hilbert space of $\lmark^2$ half-densities on $N$.
Beyond requiring that $M=T^*N$ has a complexification $Y$ that is suitable for brane quantization, the necessary condition 
is that $Y$ should be the cotangent bundle of a complexification $W$ of $N$ (and $Y$ should have the natural complex symplectic structure
of a cotangent bundle).    This condition is automatically satisfied if $M$ and $N$ are already complex manifolds and $Y$ is defined
as the product of two copies of $M$ with opposite complex structures.

The Higgs bundle moduli space $\M_H(G,C)$ contains $T^*\M(G,C)$ as a dense open set, but is not actually isomorphic to $T^*\M(G,C)$.   One would not
expect a measure zero discrepancy to be important in the definition of a Hilbert space of $\lmark^2$ wavefunctions.   The construction in \cite{GW} maps
the Hilbert space $\H$ obtained in quantizing $\M_H(G,C)$  to a space of half-densities on $\M(G,C)$ without requiring that $T^*\M(G,C)$ is literally all of $\M_H(G,C)$.

The Hilbert space $\H$ of half-densities on $\M(G,C)$ 
was already introduced in \cite{EFK} without any reference to branes and $\sigma$-models or gauge theories.   The  interpretation via branes
makes it possible to apply  electric-magnetic duality and other methods of  gauge theory.  We will see an example next in discussing the Hitchin 
Hamiltonians.

\subsection{Hitchin Hamiltonians}\label{hh}

As we have seen, in brane quantization of a complex manifold (such as $\M_H(G,C)$) that we quantize as a real symplectic manifold, the Hilbert space has an unfolded description as
$\H=\Hom(\bar\B_\cc,\B_\cc)$.   $\H$ admits a left action of\footnote{In the folded picture, we have instead a left action of both algebras
$\Hom(\B_{\cc,1},\B_{\cc,1})$ and $\Hom(\B_{\cc,2},\B_{\cc,2})$. A left action of an algebra is the same as a right action of the opposite algebra. (The notion
of the opposite algebra
is explained in Appendix C of \cite{GW}.)   Unfolding reverses
the orientation of one sheet in fig. \ref{folding} and hence replaces one of the two algebras with its opposite.   Of course, which algebra
acts on the left and which on the right depends on some choices.   None of this will be important in the present article as $\A$ and $\b\A$ will be commutative and
hence isomorphic to their opposites.}
$\A=\Hom(\B_\cc,\B_\cc)$ and a right action of $\bar\A=\Hom(\bar\B_\cc,\bar\B_\cc)$.
$\A$ and $\bar\A$ are quantum deformations of the commutative rings $\A_0$ and $\bar\A_0$ of holomorphic functions on $Y$.  In a general case,  these
deformations can be noncommutative, but 
in the particular case of $\M_H(G,C)$, it turns out that $\A$ and $\bar\A$ are commutative and hence there is no distinction between a left action and a right action.

Concretely, the ring $\A_0$ of holomorphic functions on $\M_H(G,C)$ in complex structure $I$
is simply the ring of functions on the base of the Hitchin fibration \cite{H,NH}.
Consider a solution
$(A,\phi)$ of Hitchin's equations, where $A$ is a connection on a $G$-bundle $E\to C$ and $\phi$ is a $1$-form valued in $\ad(E)$.  Let $\varphi$ be the holomorphic
Higgs field, that is,  the $(1,0)$ part of $\phi$.  Hitchin's equations give $\bar\partial_A\varphi=0$.
So if $\P$ be  an invariant polynomial on the Lie algebra ${\g}_\C$ of
$G_\C$, homogeneous of some degree $s$, then $\P(\varphi)$ is a holomorphic section of $K_C^s$ (with $K_C$ the canonical bundle of $C$; we will
also set $T_C=K_C^{-1}$).  Given
any $(0,1)$-form $\alpha$ on $C$ with values in $T_C^{s-1}$, we can define
\be\label{juno} H_{\P,\alpha}=\int_C \alpha \,\P(\varphi). \ee
This is a holomorphic function on $\M_H$ and depends only on the cohomology class of $\alpha$ in $H^1(C,T_C^{s-1})$.
The dimension of $H^1(C,T_C^{s-1})$ is $(s+1)(g-1)$, and this is the number of linearly independent functions
$H_{\P,\alpha}$  for a given $\P$.   For $G=\SU(2)$, the ring $\A_0$ of holomorphic functions on $\M_H(G,C)$ is generated by the $H_{\P,\alpha}$
where $\P(\varphi)=\Tr\,\varphi^2$.   More generally, a simple Lie group $G$ of rank $r$ has $r$ independent Casimir operators, corresponding to $r$ 
homogeneous polynomials $\P_j$, $j=1,\cdots,r$ of various degrees,  and $\A_0$ is generated by the $H_{\P_j,\alpha_j}$.   For example, if $G=\SU(N)$, we can 
take the generating polynomials to  be $\Tr\,\varphi^{k}$, $k=2,3,\cdots,N$.   

The functions $H_{\P_j,\alpha_j}$ are Poisson-commuting, since the holomorphic symplectic structure of $\M_H(G,C)$ in complex
structure $I$ is such that $\varphi_z$ and $A_{\bar z}$ are conjugate variables, and in particular
 any functions constructed from $\varphi$ only (and not $A$) are Poisson-commuting.      These Poisson-commuting functions 
  are the Hamiltonians of Hitchin's classical integrable system.

The quantum deformation from $\A_0$ to $\A$ is unobstructed in the sense that every element of $\A_0$ can be quantum-deformed   to an element of $\A$.  
This statement means, concretely, that if  $\P$ is an invariant polynomial on $\g$ homogeneous
of some 
degree $s$, and $H_{\P,\alpha}$ is a corresponding Hitchin Hamiltonian, then there is a differential operator $D_{\P,\alpha}$, acting on sections of $K^{1/2}\to \M(G,C)$,
whose leading symbol is equal to $H_{\P,\alpha}$.   The passage from $H_{\P,\alpha}$ to $D_{\P,\alpha}$ is not entirely canonical, since specifying the desired leading
symbol of $D_{\P,\alpha}$ leaves one free to add to $D_{\P,\alpha}$ a globally-defined holomorphic differential operator of degree less than $s$.   
For $G_\C=\SL(2,\C)$, one
can take $\P$ to be of degree 2, and then the only globally-defined holomorphic differential operators of lower degree are the operators of degree 0 -- the complex
constants.  For groups of higher rank, in general there are more possibilities.
Mathematically,  the fact that the deformation is unobstructed
 follows from the fact that, for any simple $G$,  $H^1(\M_H,\O)=0$, by virtue of which there is no potential obstruction in the deformation.
A gauge theory version of this argument is given in Appendix \ref{coho}, elaborating on a previous discussion in \cite{KW}.   

From the point of view of the $\sigma$-model, or the underlying gauge theory, the deformation from $H_{\P,\alpha}$ to $D_{\P,\alpha}$ arises from an
expansion in powers of $\hbar$.   This expansion terminates after finitely many steps, since we define $\A_0$ to consist of functions whose restriction to
a fiber of the cotangent bundle is a polynomial.   A specific definition of the $\sigma$-model or the gauge theory gives a specific recipe for passing from $H_{\P,\alpha}$
to $D_{\P,\alpha}$, but this depends on data (such as a Riemannian metric on $C$, not just a complex structure) that is not part of the $A$-model.

 \begin{figure}
 \begin{center}
   \includegraphics[width=4in]{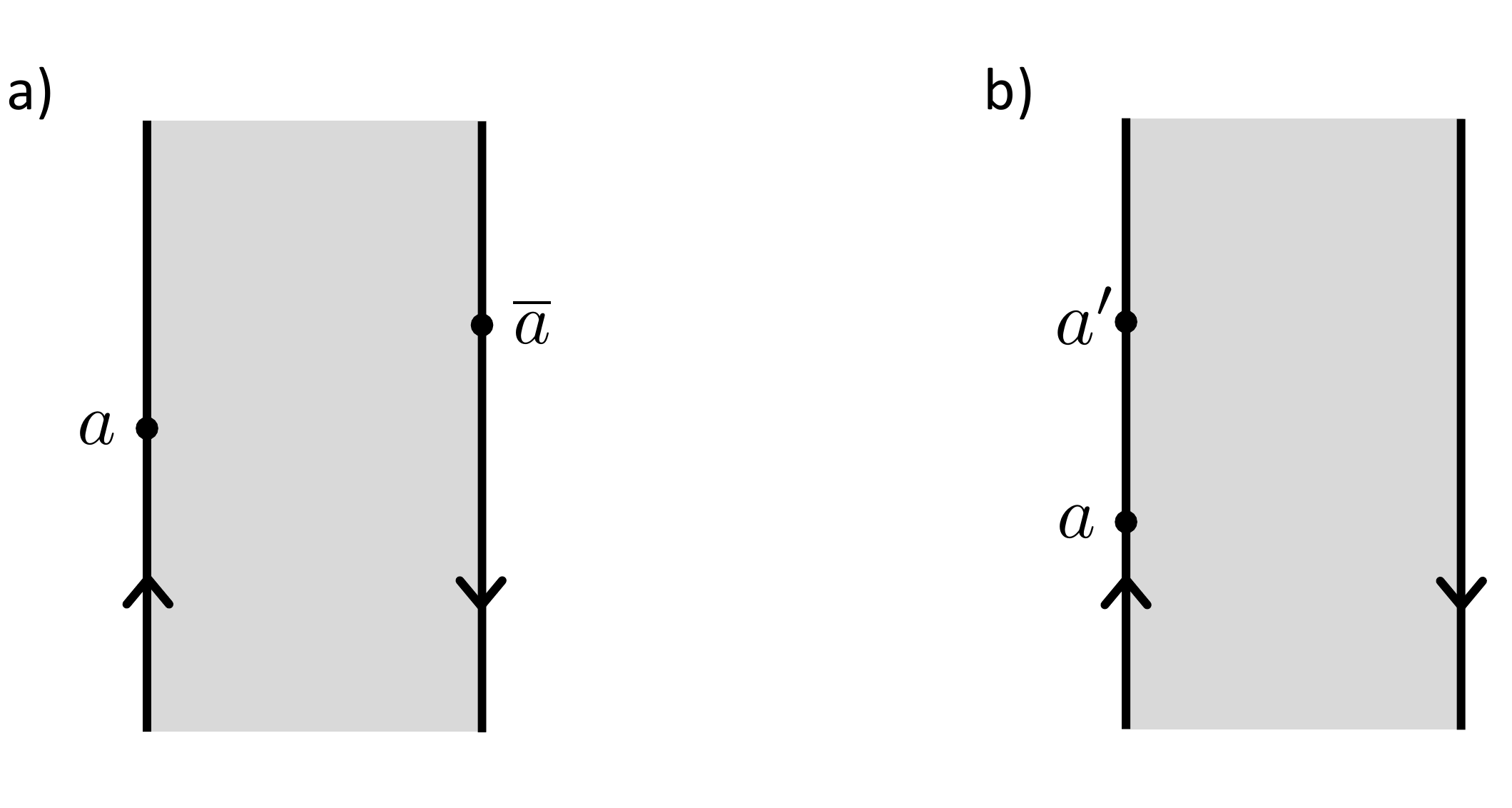}
 \end{center}
\caption{\small  (a) As a general statement in two-dimensional topological field theory, $\A=\Hom(\B_\cc,\B_\cc)$ commutes with $\b\A=\Hom(\b\B_\cc,\b\B_\cc)$
in acting on $\H=\Hom(\b\B_\cc,\B_\cc)$, because elements $a\in\A$ and $\b a\in \b\A$ are inserted on opposite boundaries.   Diffeomorphism
inariance does not allow any natural notion of which is inserted ``first.''  (b) In general, in two-dimensional topological field theory, $\A$ (and similarly
$\b\A$) can be noncommutative, because elements $a,a'\in\A$ are inserted on the same boundary with a well-defined order, relative
to the boundary orientation.   However, in the present context there are two additional dimensions, comprising the Riemann surface $C$, not drawn
in the two-dimensional picture.  One can assume that $a$ and $a'$ have disjoint support in $C$. Hence they can be moved up and down past
each other without singularity and must commute. \label{sliding}}
\end{figure} 
 
Next we would like to explain why $\A$ is commutative, like $\A_0$.   This was originally proved for $\SU(2)$ by Hitchin \cite{H2} and for general simple $G$ by
Beilinson and Drinfeld \cite{BD}.   We will give a four-dimensional explanation similar to that in \cite{KW}.   Of course the same considerations apply to $\bar\A$.
As a warmup, we first explain why $\A$  commutes with $\b\A$.  This is clear  from the fact (fig. \ref{sliding}(a)) that an element 
$a\in \A$ is inserted on the left boundary of the strip, while an element $\bar a\in \b\A$ is inserted on the right boundary. In two-dimensional 
topological field theory, we are free to slide these insertions up and down along the boundary independently.  There is no meaningful relative time-ordering
between the two boundary insertions and they must commute.   By contrast, consider the insertion of a pair of elements $a,a'\in\A$ (fig \ref{sliding}(b)).  Here, as
a general statement in two-dimensional topological field theory, there is a meaningful time-ordering between $a$ and $a'$.
If we try to slide one of them past the other in time, there may be a discontinuity when they cross, and therefore in general we may have $aa'\not=a'a$.
However, in the present problem, we are really not in two dimensions but in four dimensions; there are two
extra dimensions, making up the Riemann surface $C$, that are not shown in the figure.   The definition of $H_{\P,\alpha}$ in eqn. (\ref{juno}) depended only
on the cohomology class of $\alpha$ in $H^1(C,T_C^{s-1})$.   We can choose a representative with support in an arbitrarily selected small open ball in $C$.
Therefore, when we consider a pair of elements $a,a'\in \A$, we can assume that they are represented by operators that
 have disjoint support in $C$.   Hence we can slide the two operators
up and down past each other in the two-dimensional picture of fig. \ref{sliding}(b) without any singularity.  Accordingly, they commute.

We can elaborate slightly on the four-dimensional origin of the quantum Hitchin hamiltonians. The integrands $\P(\varphi)$ in the classical Hitchin Hamiltonians depend holomorphically on $C$. As we review in Section \ref{sec:chiral}, the quantum Hitchin Hamiltonians $D_{\P,\alpha}$ can also be written as \be\label{qjuno} D_{\P,\alpha}=\int_C \alpha \,{\cal D}_{\P} \ee
in terms of certain differential operators ${\cal D}_{\P}$ which act on the bundle locally at a point $p\in C$ and depend holomorphically on $p$.
We identify ${\cal D}_{\P}(p)$ as the action of a four-dimensional boundary local operator $O_{\P}(p)$. The holomorphic-topological nature of the boundary condition insures that such boundary local operators commute in the topological direction and have non-singular operator product expansion (OPE) with each other. 

As was explained in Section \ref{aphiggs}, the quantum Hilbert space  $\H=\Hom(\b\B_\cc,\B_\cc)$ is the space of 
$\lmark^2$ half-densities on $\M(G,C)$,  or equivalently the space of $\lmark^2$ sections of $K^{1/2}\otimes \bar K{}^{1/2}$.
We recall that this  happens because brane quantization of $\M_H(G,C)$ is equivalent to quantizing it as a cotangent bundle $T^*\M(G,C)$.   In quantizing a cotangent bundle $T^*W$, a function whose restriction to a fiber of the cotangent bundle is a polynomial of degree $s$
becomes a differential operator of degree $s$ acting on half-densities on $W$.   If, as in the case of interest here, $W$ is a complex manifold, then more
specifically holomorphic functions on $T^*W$ become holomorphic differential operators on $W$.  From a holomorphic point of view, one usually says
that holomorphic functions on $T^*W$ (with polynomial dependence on the fibers) become holomorphic differential operators acting on sections of $K^{1/2}$.
(The role of $K^{1/2}$ is explained from a $\sigma$-model point of view in \cite{GW}, Section 3.2 and Appendix C.)  
In the case of the Hitchin Hamiltonians, the fact that they can be quantum deformed to differential operators acting on sections of $K^{1/2}$, and not on
sections of any other holomorphic line bundle, is part of the standard story \cite{H2,BD}.
 However, the antiholomorphic line bundle $\bar K{}^{1/2}$ is invisible
to a holomorphic differential operator, since its transition functions are antiholomorphic and commute with holomorphic differential operators.   So the holomorphic
differential operators that are obtained by deformation quantization of holomorphic functions on $T^*W$ can naturally act on $K^{1/2}\otimes \bar K{}^{1/2}$,
or equivalently on the bundle $\K^{1/2}$ of half-densities.  Similarly, under deformation quantization, antiholomorphic functions on $T^*W$ become
antiholomorphic differential operators that can act on $\K^{1/2}$.

So $\A$ and $\bA$ become, respectively, algebras of holomorphic and antiholomorphic differential operators acting on half-densities on $\M(G,C)$.
From this point of view, the statement that $\A$ and $\bA$ commute just reflects the fact that holomorphic differential operators commute with antiholomorphic ones.

\subsection{The Duals Of The Coisotropic Branes}\label{dualco}

In order to be able to apply duality to this problem, we need one more ingredient.   We need 
 to understand the duals of the $A$-branes $\B_\cc$ and $\bB_\cc$ in the $B$-model of $\M_H(G^\vee,C)$.    To be specific, here we mean
the  $B$-model in the
complex structure that is called $J$ in \cite{H}, in which $\M_H(G^\vee,C)$ parametrizes flat bundles over $C$ with structure group $G^\vee_\C$.   We denote
the  connection on the flat bundle as $\cA=A+\i\phi$, where $(A,\phi)$ are the unitary connection and Higgs field that appear in Hitchin's equations.
A general $B$-brane is a coherent
sheaf, or a complex of coherent sheaves, on $\M_H(G^\vee,C)$.   
However, the $A$-branes $\B_\cc$ and $\bB_\cc$ have additional properties that imply that the dual $B$-branes must be rather special.
To explain this, we recall that the Higgs bundle moduli spaces are hyper-Kahler manifolds,
with complex structures $I,J,K$ that obey the usual quaternion relations, and a triple of corresponding Kahler forms $\omega_I,\omega_J,\omega_K$ and
complex symplectic forms $\Omega_I=\omega_J+\i \omega_K$, etc.   Geometric Langlands duality in general maps the $A$-model of $\M_H(G,C)$
with symplectic structure $\omega_K$ to the $B$-model of $\M_H(G^\vee,C)$ in complex structure $J$.   When we speak of the $A$-model or the $B$-model
without further detail, these are the models we mean.
A generic brane in either of these models is merely an $A$-brane or $B$-brane
of the appropriate type.   However, many of the branes that are most important in geometric Langlands have additional properties.   For example, a brane
supported on a point in $\M_H(G^\vee,C)$ is a brane of type $(B,B,B)$, that is, it is a $B$-brane in each of complex structures $I,J$, and $K$ (or any linear
combination).    The dual of a brane of type $(B,B,B)$ is a brane of type $(B,A,A)$; in the case of the Higgs bundle moduli space,
 the dual of a rank 1 brane supported at a point is a brane
supported on a fiber of the Hitchin fibration, with a rank 1 flat $\CP$ bundle.   These branes are the Hecke eigensheaves which are central objects of
study in the geometric Langlands program; they
 will be discussed in Section \ref{wkb}.   In the case at hand,
$\B_\cc$ and $\bB_\cc$ are branes of type $(A,B,A)$; they are $A$-branes of types $I$ and $K$, by virtue of the Kapustin-Orlov conditions for
coisotropic branes, and they are $B$-branes of type $J$, because the curvature $\pm\frac{1}{2}\omega_J$ of their $\CP$ bundles is of type $(1,1)$
in complex structure $J$, so that those  bundles are holomorphic  in complex structure $J$.   In general, the dual of a brane of type
$(A,B,A)$ is a brane of type $(A,B,A)$, so the duals of $\B_\cc$ and $\bB_\cc$ will be branes of that type.   The simplest kind of brane of type $(A,B,A)$
is given by the structure sheaf of a complex Lagrangian submanifold in complex structure $J$.   In more physical language, these are branes supported
on a complex Lagrangian submanifold with a trivial $\CP$ bundle. And indeed, the duals of $\B_\cc$ and $\bB_\cc$ are  of this type. These duals 
were first identified (in a different formulation) by Beilinson and Drinfeld \cite{BD}, with the help of conformal field theory at critical level $k=-h^\vee$.  
A gauge-theory explanation involves duality between  the D3-NS5  and  D3-D5 systems of string theory \cite{GWknots}.  

The complex Lagrangian submanifold supporting the dual of $\B_\cc$ parametrizes flat $G^\vee_C$ bundles which satisfy a ``holomorphic oper'' 
condition. We will denote it as $L_\op$. Similarly, the dual of $\bB_\cc$ is supported on a complex Lagrangian submanifold $L_\bop$  that parametrizes
flat $G^\vee_C$ bundles which satisfy an ``antiholomorphic oper'' condition. 

The holomorphic oper condition can be stated rather economically as a global constraint on the holomorphic type of the bundle, i.e. on the $(0,1)$ part of $\cA$,
as we will do here, or in a more local way, as we will do in Section \ref{localoper}.   Both formulations are standard mathematically.
In a four-dimensional gauge theory, $S$-duality maps a deformed Neumann boundary condition to a deformed ``Nahm pole'' boundary condition, 
which  imposes directly the local constraints \cite{GWknots}. 

One general way to define a complex Lagrangian submanifold of $\M_H(G^\vee,C)$ is to consider all flat 
$G^\vee_C$ bundles $E^\vee\to C$ with some fixed holomorphic type.  Specifying the holomorphic type of a bundle is 
equivalent to specifying the $(0,1)$ part of $\cA$.   Here and in several
analogous cases considered momentarily, we  place no constraint
on the $(1,0)$ part of the connection except that the full connection should be flat.  While preserving the flatness of the connection on $E^\vee$,
we are free to add to the $(1,0)$ part of the connection an arbitrary $\bar\partial_\cA$-closed form representing an
element of $H^0(C,K_C\otimes \ad(E^\vee))$.    The dimension of this space is half the dimension of $\M_H(G^\vee,C)$, so flat $G^\vee_\C$
bundles 
of a specified holomorphic type are a middle-dimensional submanifold $L$ of $\M_H(G^\vee,C)$.   $L$ is  a complex Lagrangian submanifold, because
the holomorphic symplectic structure $\Omega_J$ of $\M_H(G^\vee,C)$ in complex structure $J$ is a pairing between the $(1,0)$ and $(0,1)$ parts of $\cA$
and vanishes if the $(0,1)$ part is specified.  Once one picks a base point in $L$, $L$ is isomorphic to the vector space $H^0(C,K_C\otimes \ad(E^\vee))$.

We can define a second family of complex Lagrangian submanifolds, in the same complex structure on $\M_H(G^\vee,C)$, by specifying the antiholomorphic
structure of a flat $G^\vee_\C$ bundle.   This amounts to specifying the $(1,0)$ part of $\cA$, and letting the $(0,1)$ part vary.    It leads to a complex
Lagrangian submanifold for the same reasons as before.   It may come as a slight surprise that fixing either the $(1,0)$ or the $(0,1)$ part of $\cA$ is
a holomorphic condition in complex structure $J$.   Indeed, complex structure $J$ on the Higgs bundle moduli space is not sensitive to the complex
structure of the two-manifold $C$, and treats the $(1,0)$ and $(0,1)$ parts of $\cA$ in a completely symmetric way.    

The submanifolds $L_\op$ and $L_\bop$ can be defined by specifying a particular choice of the holomorphic or antiholomorphic structure of a flat $E^\vee$ bundle.
First we explain the definition for the case $G_\C=\SL(2,\C)$. For this group, an oper  is a flat bundle  $E^\vee\to C$ that, as a holomorphic bundle, 
is  a nontrivial extension of $K_C^{-1/2}$ by $K_C^{1/2}$:
\be\label{nonex}0\to K_C^{1/2}\to E^\vee\to K_C^{-1/2}\to 0. \ee   
There is a unique such nontrivial extension, up to isomorphism.    The family of  flat bundles of this holomorphic type
 is therefore a complex Lagrangian submanifold
that we will call $L_\op$.   In making this definition, we have made a choice of $K_C^{1/2}$, or equivalently a choice of spin structure on $C$.
Indeed, for $\SL(2,\C)$, the definition of an oper depends on such  a choice of spin structure (though we do not indicate this in the notation for $L_\op$).
We return to this point in Section \ref{toposubt}.

It is possible to give a simple description of $L_\op$ once one picks a base point, that is, a particular  $\SL(2,\C)$ bundle $E^\vee$
of oper type with flat connection  $\cA_0$.   In deforming $E^\vee$ as an oper, we may as well keep the $(0,1)$ part of $\cA_0$ fixed, since we are required to 
keep it fixed up to a complex gauge transformation.  But we can modify the $(1,0)$ part of $\cA_0$.  To do this while preserving the flatness of $\cA_0$,
we  add to $\cA_0$ a $\bar\partial_{\cA_0}$-closed $(1,0)$-form, that is, an element of $H^0(C,K_C\otimes \ad(E^\vee))$.  Using the exact sequence
(\ref{nonex}), one can show that $H^0(C,K_C\otimes \ad(E^\vee))$ is isomorphic to the space of quadratic differentials on $C$.  This is the base of the Hitchin
fibration for $\SL(2,\C)$, so $L_\op$ is isomorphic
 to the base of the Hitchin fibration.   This isomorphism is not entirely canonical as it depends on the choice of a base point in $L_\op$.   A similar
 reasoning applies for other groups.
 
 Similarly, an antiholomorphic oper for $G_\C=\SL(2,\C)$, or anti-oper for short, is a flat bundle $E^\vee$ that, antiholomorphically, is a nontrivial extension of 
$\bar K_C^{-1/2}$ by $\bar K_C^{1/2}$:
\be\label{onex} 0\to\bar K_C^{1/2}\to E^\vee \to \bar K_C^{-1/2}\to 0. \ee
The family of such flat bundles is another complex Lagrangian submanifold, which we will call $L_\bop$.   It is noncanonically isomorphic to the base
of the Hitchin fibration, with the opposite complex structure.

In general, if $G_\C^\vee$ is a simple complex Lie group, there is a notion of a ``principal embedding'' of Lie algebras ${\su}(2)\to {\g^\vee}$.  For example, if $G^\vee=\SL(n,\C)$,
the principal embedding is such that the $n$-dimensional representation of $\g^\vee$ transforms irreducibly under $\su(2)$; the corresponding principal
subgroup is a copy of $\SL(2,\C)$ or $\SO(3,\C)$ depending on whether $n$ is even or odd.   For brevity we will sometimes ignore this subtlety and
 refer simply to a principal
$\SL(2,\C)$ subgroup, though the global form of the group is sometimes $\SO(3,\C)$.

A  $G_\C^\vee$
oper is a flat $G_\C^\vee$ bundle $E_\C^\vee$ that, as a holomorphic bundle, is equivalent to a principal embedding of an $\SL(2,\C)$ oper bundle,
that is, a principal embedding of a rank two bundle that is  a nontrivial extension
of the form in eqn. (\ref{nonex}). For $G^\vee_\C=\SL(n,\C)$, this means that an oper bundle is, holomorphically, 
the $(n-1)^{th}$ symmetric tensor power of such a nontrivial extension, and therefore
has a subbundle isomorphic to $K_C^{(n-1)/2}$:
\be\label{nex}0\to K_C^{(n-1)/2}\to E^\vee\to \cdots \ee
(and a filtration by powers of $K_C$).
  Again,
the $(1,0)$ part of the connection on $E_\C^\vee$ is not restricted except by requiring the full connection to be flat.   Similarly an antiholomorphic $G_\C^\vee$ oper
is a flat $G_\C^\vee$ bundle that, as an antiholomorphic bundle, is equivalent to a principal embedding of an antiholomorphic $\SL(2,\C)$ oper bundle.   

At the $\sigma$-model level, the  duals of $\B_\cc$ and $\bB_\cc$ are the structure sheaves of $L_\op$ and $L_\bop$; that is, they are 
rank 1 branes $\B_\op$ and $\B_\bop$   supported on $L_\op$ and $L_\bop$ with trivial $\CP$ bundles.

\subsection{The Local Constraints}\label{localoper}

To describe a more local characterization of an oper, we consider first the  case $G^\vee_\C=\SL(2,\C)$. The extension structure of $E^\vee$ implies the existence of  a
global holomorphic section $s$ of $E^\vee \otimes K_C^{-1/2}$. Denote as $D$ the $(1,0)$ part of the connection. 
The $\SL(2,\C)$-invariant combination $s \wedge D s$ is a global holomorphic function on $C$. This function must be non-zero: 
if it vanished, we could write $D s= a s$ and $a$ would define a holomorphic flat connection on $K_C^{-1/2}$, which does not exist (for $C$ of genus greater than 1 or
in lower genus in the presence of parabolic structure).  

Without loss of generality, we can normalize $s$ so that $s \wedge D s=1$. This fixes $s$ up to sign.
The local version of the holomorphic oper condition for $G^\vee_\C=\SL(2,\C)$ is precisely the condition that $E^\vee \otimes K_C^{-1/2}$ admits a
globally defined holomorphic section such that $s \wedge D s=1$.

Taking a derivative, we have $s \wedge D^2 s =0$ and thus $s$ satisfies a second order differential equation
\begin{equation}\label{stresst}
D^2 s + t s=0
\end{equation}
for some ``classical stress tensor'' $t$ on $C$. 
Under a change of local coordinate on $C$, $t$ transforms as a stress tensor, not as a quadratic differential.    Not coincidentally, eqn. (\ref{stresst})
can be viewed as a classical limit of the 
 Belavin-Polyakov-Zamolodchikov (BPZ) differential equations for the correlator of a degenerate field in two-dimensional
conformal field theory. The classical stress tensor can be used to define a set of generators of the algebra of holomorphic 
functions on the oper manifold, consisting of functions of the form
\be\label{opjuno2} f_{t,\alpha}=\int_C \alpha \,t ,\ee
with $\alpha$ being a $(0,1)$-form with values in $T_C$.

The case of $G^\vee_\C=\SL(n,\C)$ can be analyzed  similarly.   In this case, the oper structure of $E^\vee$ (eqn. (\ref{nex})) implies the existence of a
 global holomorphic section  $s$ of $E^\vee \otimes K_C^{(1-n)/2}$.   Then  $s \wedge D s \wedge \cdots D^{n-1} s$ is a global holomorphic function on $C$ which cannot vanish, for a similar reason to what we explained for $n=2$.   We can normalize $s$ so that $s \wedge D s \wedge \cdots D^{n-1} s=1$; this uniquely fixes $s$,
 up to the possibility of multiplying by an $n^{th}$ root of 1.   Since  $0=D(s \wedge D s \wedge \cdots D^{n-1} s)=s\wedge Ds \wedge \cdots \wedge  D^{n-2}s\wedge D^ns$, 
 we learn that $s$ satisfies a degree $n$ differential equation 
\begin{equation}
D^n s + t_2 D^{n-2} s+  \cdots + t_n s =0.
\end{equation}
We can define a set of generators of the algebra of  holomorphic functions on the oper manifold, consisting of functions of the form
\be\label{opjunon} f_{t_k,\alpha}=\int_C \alpha \,t_k \ee
with $\alpha$ being a $(0,1)$-form with values in $T_C^{k-1}$, $k=2,\cdots,n$. 

For general $G^\vee_\C$, there is no distinguished representation as convenient as the $n$-dimensional representation of $\SL(n,\C)$.  However,
given an oper bundle $E^\vee$,
we can consider associated bundles $E^\vee_R$ in any irreducible representation $R$ of $G^\vee_\C$.  
By the definition of an oper, the structure group of $E^\vee_R$ as a holomorphic bundle reduces to a rank 1 subgroup  $H_\C\subset G^\vee_\C$;
this subgroup is a copy of either $\SL(2,\C)$ or $\SO(3,\C)$, depending on $G^\vee_\C$ and $R$.  Let $R_n$ be the $n$-dimensional irreducible representation of $H_\C$ ($n$ is any positive integer
or any odd positive integer for $\SL(2,\C)$ or $\SO(3,\C)$, respectively).   As a representation of $H_\C$, we have $R\cong \oplus_{n=0}^\infty Q_n\otimes R_n$, where $Q_n$ are some vector
spaces, almost all of which vanish.  Actually, if $N$ is the largest integer for which $Q_n$ is nonzero, then $Q_N$ is 1-dimensional and we can replace $Q_N\otimes
R_N$ with $R_N$.   So $R\cong R_N\oplus_{n=0}^{N-1}Q_n\otimes R_n$.   In this decomposition, a highest weight vector of $R_N$ with respect to a Borel
subgroup $B_H$ of $H$ is a highest weight vector of $G^\vee$ with respect to the Borel subgroup $B_{G^\vee}$ of $G^\vee$ that contains $B_H$. 
The associated bundle $E^\vee_R$ 
has a similar decomposition as holomorphic bundle 
\be\label{regasso} E^\vee_{R}= E^\vee_{R,N}\oplus \left(\oplus_{n=1}^{N-1} Q_n\otimes E^\vee_{R_n}\right),  \ee
where $E^\vee_{R_n}$ is the holomorphic bundle associated to an $H_\C$ oper in the $n$-dimensional representation. 
For each $n$ we get from eqn. (\ref{nex}) a canonical image $s_{R,n}$ of the vector space $Q_n$ into the space of global holomorphic sections 
of $E^\vee_{R} \otimes K_C^{(1-n)/2}$, or equivalently a holomorphic map 
\be\label{holmap} s_{R,n}: Q_n \otimes K_C^{(n-1)/2} \to E^\vee_{R}. \ee

Of particular importance here is the ``highest weight'' object $s_{R,N}$, which we will just denote as $s_R$:
\be\label{linmap}s_R:K_C^{(N-1)/2}\to E^\vee_R. \ee
Here $N$ is defined by saying that a highest weight vector of the representation $R$ (for some Borel subgroup $B$) is in an $N$-dimensional
representation of a principal $\SL(2,\C)$ subgroup (which has a Borel subgroup contained in $B$).   For $G^\vee=\SL(n,\C)$ and $R$ the $n$-dimensional
representation or its dual, $N=n$.

A partial analogue to the $s \wedge D s=1$ condition is the condition that the collection $D^m s_{R,n}$ for $m< n\leq N$ should 
span $E^\vee_{R}$ at each point of $C$. The derivatives $D^n s_{R,n}$ can then be expanded out in terms of the $D^m s_{R,n}$ with $m<n$, giving rise to an intricate collection of differential equations. (See Appendix \ref{somex} for some examples.)   The coefficients of the differential equations 
are holomorphic functions on the oper manifold and can be expressed as polynomials in derivatives of a collection of observables $T_{\cal P}$ 
which have the same labels as the integrands for the quantum Hitchin Hamiltonians for $G^\vee_\C$. Tensor products of the form $D^m s_{R,n} \otimes D^{m'} s_{R',n'}$ 
can also be expanded in the basis of $D^{m''} s_{R'',n''}$ for all $R''$ that appear in the decomposition of the tensor product $R \otimes R'$. 
Coefficients  in this expansion
 are also holomorphic functions on the oper manifold and can be expressed as polynomials in derivatives of a collection of observables $T_{\cal P}$. 

The collection of observables  $T_{\cal P}$ is sometimes called the classical $W$-algebra for $G^\vee_\C$. In that language, the differential equations satisfied by the 
$s_{R,n}$ are a classical analogue of the BPZ equations,
and the tensor product expansion is analogous to the operator product expansion (OPE) of degenerate fields.  

In four-dimensional gauge theory, the deformed Neumann boundary condition is $S$-dual to the deformed Nahm pole boundary condition,
which is also holomorphic-topological. This boundary condition involves a certain prescribed singularity for the gauge theory fields at the boundary. 
Effectively, the singular boundary conditions of the physical theory impose an oper boundary condition in the topologically twisted theory. 
The gauge-invariant information contained in the subleading behaviour of the fields is captured by boundary local operators which match the $T_{\cal P}$ 
observables and are $S$-dual to the corresponding local operators at the deformed Neumann boundary condition which are employed in 
the definition of the quantum Hitchin Hamiltonians.\footnote{Local operators at the oper
boundary conditions  also include holomorphic
forms on the oper manifold.   These arise from  fermionic partners of the
$T_{\cal P}$; their duals are the operators $\cal P'$ described in Appendix \ref{coho}.} The $s_{R,n}$ also appear naturally in gauge theory, as we will illustrate in Section \ref{wilop}.

Finally, in order to gain further intuition on the various relations satisfied by the $s_{R,n}$, we 
note that the oper manifold has a simpler ``classical'' cousin given by Hitchin's section of the Hitchin fibration.   
For $\SL(2,\C)$, the Hitchin section parametrizes Higgs bundles $(E,\varphi)$ such that $E$ is a direct sum $K_C^{1/2} \oplus K_C^{-1/2}$.
In other words, the Hitchin section is what we get if we work in complex structure $I$ (rather than $J$) and ask for the extension in eqn. (\ref{nonex}) to be split
(as opposed to a non-split extension, leading to an oper bundle).  For any $G^\vee_\C$, the Hitchin section parametrizes pairs $(E,\varphi)$ such that $E$ 
is induced by the principal embedding of $K_C^{1/2} \oplus K_C^{-1/2}$.
  For the Hitchin section, one can deduce local conditions analogous to what we have explained
for opers, but using the 
 Higgs field $\varphi$ instead of the holomorphic derivative $D$. 
For example, the $\SL(2,\C)$ Hitchin section of Higgs bundle moduli space  is characterized locally by the existence of a holomorphic section $s$  of $E$
which satisfies $s \wedge \varphi s=1$ along with  $\varphi^2 s = \frac12 \Tr\, \varphi^2 s$  (the latter equation holds simply because $\varphi^2=\frac12 \Tr\, \varphi^2$
for $\SL(2,\C)$). 
Comparing to the oper case, $D$ is replaced by $\varphi$ and the stress tensor $t$ is replaced by the quadratic differential
$\frac12\Tr\,\varphi^2$.
In general, the associated bundle $E^\vee_R$ 
in a representation $R$ of $G^\vee$ will have sections $s_{R,n}$, $n\leq N$, such that $\varphi^m s_{R,n}$ for $m<n$ 
span $E^\vee_{R}$. Hence $\varphi^n s_{R,n}$ can be expanded in terms of $\varphi^m s_{R,n}$ for $m<n$; likewise
for two representations $R$, $R'$,  $\varphi^m s_{R,n} \otimes \varphi^{m'} s_{R',n'}$ 
can be expanded out in terms of $\varphi^{m''} s_{R'',n''}$ for all $R''$ in $R \otimes R'$, with coefficients built from the 
gauge-invariant polynomials $\P(\varphi)$. The oper relations are a deformation of these.   With an extension of this analysis, one can recover the assertion of section 
\ref{dualco} that $L_\op$ is noncanonically isomorphic to the base of the Hitchin fibration.

\subsection{Some Topological Subtleties}\label{toposubt}

The definition of a $G_\C$ oper depends on the choice of a spin structure on $C$ if the image of the principal embedding is an $\SL(2,\C)$ subgroup of $G_\C^\vee$,
but not if it is an $\SO(3,\C)$ subgroup.   (For example, for $G_\C^\vee=\SL(n,\C)$, the notion of an oper depends on a choice of spin structure precisely if
$n$ is even.)    This dependence on spin structure for some groups is in tension with the claim that $\B_\op$ is the dual of $\B_\cc$, since the definition of $\B_\cc$ 
did not seem to depend on
a choice of spin structure.   The resolution of this point was essentially described in Section 8 of \cite{FG}.  We will explain the details for groups of
rank 1.    In the context of the twisted version of $\N=4$ super
Yang-Mills theory that is relevant to geometric Langlands, the electric-magnetic dual of $\SO(3)$ gauge theory is not standard $\SU(2)$ gauge theory,
but what is sometimes called $\Spin\cdot\SU(2)$ gauge theory.   For any $d>0$, the group $\Spin(d)\cdot\SU(2)$ is a double cover of $\SO(d)\times \SO(3)$
that restricts to a nontrivial double cover of either factor.   In particular, $\Spin(d)\cdot\SU(2)$ has projections to $\SO(d)$ and to $\SO(3)$:
\be\label{dobl} \begin{matrix}& & \Spin(d)\cdot\SU(2) && \cr
                                                 &\swarrow &&\searrow &\cr
                                                  \SO(d) &&&& \SO(3). \end{matrix}\ee
 By a $\Spin(d)\cdot \SU(2)$ structure on a $d$-manifold $M$, we mean a principal bundle over $ M$ with that structure group such that the projection to the first
 factor $\SO(d)$    gives the frame bundle of $M$ (the principal bundle associated to the tangent bundle $TM$ of $M$).  Likewise a $\Spin(d)\cdot\SU(2)$ connection
on a Riemannian manifold $M$ is  a connection with that structure group that when restricted to the first factor is the Levi-Civita connection of the tangent bundle of $M$. 
   Assuming that $M$ is spin, a down-to-earth
 description of a $\Spin(d)\cdot\SU(2)$ structure on $M$ is as follows: once a spin structure is picked on $M$, a $\Spin(d)\cdot   \SU(2)$ bundle is equivalent
 to an $\SU(2)$ bundle     $E^\vee\to M$; if the spin structure of $M$ is twisted by a line bundle $\ell$ such that $\ell^2$ is trivial, then $E^\vee$ is replaced by $E^\vee\otimes \ell$.    With this
 characterization, it is evident that although the variety $L_\op$ of opers is not canonically defined in $\SU(2)$ gauge theory, it is canonically defined in 
 $\Spin(4)\cdot \SU(2)$ gauge theory.   Indeed, bearing in mind that $\ell\cong\ell^{-1}$, if $E^\vee$ appears in the exact sequence defining an oper
 with some choice of $K^{1/2}$, then $E^\vee\otimes \ell$ appears in a similar exact sequence with $K^{1/2}$ replaced by $K^{1/2}\otimes \ell$.
 
What is the dual of $\SU(2)$ gauge theory, as opposed to $\Spin(4)\cdot \SU(2)$ gauge theory?
 The answer \cite{FG} is that the dual is $\SO(3)$ gauge theory, but with an extra factor $\Delta=(-1)^{\int_M w_2(M) w_2(E)}$ included in the definition of the path
 integral.   Here $w_2(M)$ and $w_2(E)$ are respectively the second Stieffel-Whitney classes of $TM$ and of an $\SO(3)$ bundle $E\to M$.
 If the $B$-model description is by $\SU(2)$ gauge theory, and therefore the definition of $L_\op$ requires a spin structure on $C$, then the $A$-model
 description is by $\SO(3)$ gauge theory with the additional factor $\Delta$, and this must ensure that the definition of $\B_\cc$ similarly requires a choice of
 spin structure on $C$.   That happens as follows.
 If $M=\Sigma\times C$ where $\Sigma$ and $C$ are oriented two-manifolds without boundary, then $w_2(M)=0$ and $\Delta=1$, so the factor of $\Delta$ in
 the path integral
 has no consequence.   But suppose $M$ has a boundary $\partial M$ with $\B_\cc$ boundary conditions.   To define the topological invariant $\int_M w_2(M) w_2(E)$,
 one needs a trivialization of the class $w_2(M) w_2(E)$ along $\partial M$.   $\B_\cc$ boundary conditions are a version of free boundary conditions 
 for the gauge field, so with $\B_\cc$ boundary conditions, there is no restriction on $w_2(E)$ along $\partial M$.  But we can trivialize $w_2(M)w_2(E)$ 
 along $\partial M$ by trivializing $w_2(M)$, that is, by picking a spin structure along $\partial M$.    In our application, $\partial M$ is the product of a Riemann
 surface $C$ with a contractible one-manifold (the boundary of the strip) and what is needed is a spin structure on $C$.    Thus the $A$-model
 dual of $\SU(2)$ gauge theory  is an $\SO(3)$ gauge theory in which, despite appearances, the definition of $\B_\cc$ (or similarly $\b\B_\cc$) requires a choice of spin structure on $C$.

 These remarks
 have analogs for all groups  such that the definition of an oper requires a choice of spin structure.   
They do not have analogs  in the usual physics of $\N=4$ super Yang-Mills theory because they only come into play after topological twisting.   Untwisted
 $\N=4$ super Yang-Mills theory has fermion fields whose definition requires a spin structure on $M$.  When a spin structure is present, the difference between
 $\Spin(4)\cdot \SU(2)$ gauge theory and $\SU(2)$ gauge theory disappears. Likewise, the choice of a spin structure  trivializes $\Delta$.  
 
 \subsection{Topological Aspects of the Oper Boundary Condition In Gauge Theory}\label{opergauge}
 
 When the center $\zZ(G^\vee)$ of $G^\vee$ is nontrivial, the description of the dual of $\B_\cc$ as the brane $\B_\op$ 
(and the analogous
statement for  $\b\B_\cc$) needs a slight refinement.   $\B_\op$ is the dual of $\B_\cc$ in the $\sigma$-model of $\M_H(G^\vee,C)$, but we should recall
 that the low energy description  also contains a $\zZ(G^\vee)$ gauge field.   Along a boundary labeled by $\B_\op$, the $\zZ(G^\vee)$ gauge field is trivialized.
 As explained momentarily, this
  condition ensures that when quantized on $I\times C$, where $I$ is an interval with $\B_\op$ and $\B_\bop$ boundary conditions, the theory  supports
 a discrete electric charge along $I$.
 Following the logic of Section 7 of \cite{KW}, this condition is dual to the fact that on the $A$-model side, a $G$-bundle over $C$ has a characteristic
 class $\zeta \in H^2(C,\pi_1(G))$ and is classified by $\int_C\zeta$ (for $G=\SO(3)$, $\zeta$ is the second Stieffel-Whitney class $w_2(E)$).    
 
 The Nahm pole boundary condition, which is a gauge theory version of $\B_\op$ \cite{GWknots}, reduces the gauge group along the boundary
 from $G^\vee$ to its center 
 $\zZ(G^\vee)$.   The  trivialization of the center along the boundary is an additional condition.
 
 For $G^\vee=\SL(n,\C)$, the oper condition, or the Nahm pole boundary condition, 
 ensures the existence of the object $s$ that was introduced in Section \ref{localoper}, and
 was normalized there, up to an $n^{th}$ root of 1, by the condition condition $s\wedge Ds \wedge \cdots \wedge D^{n-1}s=1$.
An $n^{th}$ root of 1 is an element of $\zZ(G^\vee)$, so a  convenient way to express the fact that the $\zZ(G^\vee)$ gauge invariance
 is trivialized along the boundary is to say that the boundary is equipped with a particular choice of normalized section  $s$. 
 
 When we quantize the theory on a strip $\R\times I$ (times the Riemann surface $C$), with oper and anti-oper boundary conditions, we have 
 such trivializations $s_\ell$ and $s_r$ at the left and right boundaries of the strip.   We are free to make a global gauge transformation by an element $b$
 of  $\zZ(G^\vee)$.   This acts on the pair of trivializations by $(s_\ell,s_r)\to (b s_\ell, b s_r)$, so pairs differing in that way should be considered equivalent.
 However, $\zZ(G^\vee)$ acts on the 
equivalence classes of pairs $(s_\ell,s_r)$ by  
 $(s_\ell,s_r)\to (b s_\ell,s_r)$, $b\in \zZ(G^\vee)$, and this leads to an action of $\zZ(G^\vee)$ on the physical Hilbert space.   This action of $\zZ(G^\vee)$ on $\H$
 in the $B$-model description is dual to the fact that, on the $A$-model side, $\H$ is graded by $\int_C \zeta$.

 For any $G^\vee$,  the trivialization of the $\zZ(G^\vee)$ gauge field on the boundary can be expressed in terms of the objects $s_{R,n}$ that were introduced
 in Section \ref{localoper}, but this is less simple than for $\SL(n,\C)$.

  \section{The Eigenvalues Of The Hitchin Hamiltonians}\label{hitchval} 

\subsection{The Case That The Center Is Trivial}\label{centertriv}

Now we can start to deduce interesting consequences of electric-magnetic duality.

Once one identifies the $B$-model dual of $\B_\cc$ as $\B_\op$, 
as we have done in Section \ref{dualco}, one immediately has a dual description of $\A=\Hom(\B_\cc,\B_\cc)$:
it is  $\Hom(\B_\op,\B_\op)$.    Since $\B_\op$ is a rank 1 Lagrangian brane supported on $L_\op$, $\Hom(\B_\op,\B_\op)$ is just the sum of the $\bar\partial_\cA$ cohomology groups $H^i(L_\op, \O)$.    
In Section \ref{dualco}, we learned that $L_\op$ is noncanonically
 isomorphic to a vector space.   Hence the cohomology $H^i(L_\op,\O)$ vanishes for $i>0$, and $\Hom(L_\op,L_\op)$
is simply the (undeformed!) commutative algebra of holomorphic functions on $L_\op$.    Thus duality with the $B$-model
 gives another explanation that $\A$ must be commutative.
Moreover it shows that the ``spectrum'' of the algebra $\A$, in the abstract sense of the space of its 1-dimensional complex representations, is the 
``variety''\footnote{For our purposes, ``variety'' is just a synonym for ``complex manifold.''} $L_\op$ that parametrizes opers,   as originally shown in \cite{BD}. In the language of previous sections, this is a canonical identification between differential operators $\D_\P$ and functions $T_\P$.  
   Precisely the same argument shows that $\b\A=\Hom(\B_\op,\B_\op)$ is the algebra of holomorphic functions on the variety $L_\bop$
of antiholomorphic opers.  

The variety of opers is noncanonically isomorphic to the base of the Hitchin fibration, as explained in Section \ref{dualco}.
  So the fact that $\A$ is the algebra of holomorphic functions on $L_\op$ is a sort of quantum deformation of the
fact that $\A_0$ is the algebra of holomorphic functions on the base of the Hitchin fibration.   A similar statement holds for $\b\A$, of course.

However, we want to understand the spectrum of $\A\times \bA$ not in the abstract sense already indicated but as concrete operators on
$\H=\Hom(\bB_\cc,\B_\cc)$.   The dual theory gives a dual description by $\H=\Hom(\B_\bop,\B_\op)$.   If the center of $G^\vee$ is trivial,
this can be analyzed just in a $\sigma$-model (rather than a $\sigma$-model with $\zZ(G^\vee)$ gauge fields).   Let us consider this case first.   
Matters are simple because
 the branes involved are  rank 1 Lagrangian branes, supported on the complex Lagrangian manifolds $L_\op$ and $L_\bop$.   In analyzing the problem,
 we will assume that $L_\op$ and $L_\bop$ have only transverse intersections at isolated points.   This is known to 
be true for $\SL(2,\C)$ and in general is one of the conjectures of Etingof, Frenkel, and Kazhdan \cite{EFK,EFK2}.     For the intersections to be isolated
and transverse is actually a prediction of the duality; it is needed in order for the hermitian form on $\H$ to be positive-definite, as expected from the $A$-model
construction in which $\H$ is a Hilbert space of $\lmark^2$ half-densities.  Unfortunately, to explain this requires a fairly detailed discussion of $B$-model quantum 
mechanics, which has been relegated to Appendix \ref{bmodel}.   (In this appendix, we learn that there is actually a further, unproved necessary condition for positivity.)

Let $\Upsilon=L_\op\cap L_\bop$.  Assuming that the intersection points are isolated and transverse,  
$\H=\Hom(\B_\bop,\B_\op)$ simply has a basis with one
basis vector $\psi_u$ for every $u\in \Upsilon$. That is a general statement about intersections of Lagrangian branes in the $B$-model.   Concretely,
since $L_\op$ is the subvariety of $\M_H(G,C)$ that parametrizes flat  bundles that
are holomorphic opers, and $L_\bop$ is the subvariety that parametrizes flat  bundles that are antiholomorphic opers, it follows that
an intersection point represents a flat $G_\C^\vee$ bundle  that is an oper both holomorphically and antiholomorphically.    

We recall that the definition of a hermitian form on $\Hom(\B_\bop,\B_\op)$ makes use of an antiholomorphic involution $\tau$ that acts by $(A,\phi)\to
(A,-\phi)$.    Hence $\tau$ transforms a complex flat connection $\cA=A+\i\phi$ to the complex conjugate flat connection $\b\cA=A-\i\phi$.   Recall that $A$ is
a gauge field in a theory in which the gauge group is the compact form $G^\vee$.  Mathematically, the involution of $G^\vee_\C$ that leaves fixed $G^\vee$ is the product
of complex conjugation with the
Chevalley involution, so $\tau$ acts on $\cA$ as that product  (up to an inner automorphism, which here means a $G^\vee$-valued gauge transformation).  

In the folded construction of the state space $\H$, $\tau$ acts antiholomorphically on $\h Y=Y_1\times Y_2$ by exchanging the two factors.  That means
that in the unfolded construction, $\tau$ exchanges the two ends of the strip of fig. \ref{folding}(b).   It is not difficult to see explicitly why this happens.
If $\cA$ is a complex flat connection that is a holomorphic oper, then $\b\cA$ is a complex flat connection 
that is an antiholomorphic oper, and similarly, if $\cA$ is an antiholomorphic oper, then $\b\cA$ is a holomorphic one.      Thus
$\tau$ exchanges $L_\op$ with $L_\bop$ and likewise exchanges\footnote{The statement that $\tau$ exchanges $\B_\op$ with $\B_\bop$ holds in the underlying physical
$\sigma$-model.  Since $\tau$ acts antiholomorphically on $\M_H(G^\vee,C)$ in the relevant complex structure, it exchanges the $B$-model with a conjugate
$B$-model  and is not a $B$-model symmetry ($\B_\op$ and $\B_\bop$ are valid branes in both the $B$-model and its conjugate).   
The $B$-model symmetry that exchanges 
$\B_\op$ with $\B_\bop$ and is used in defining the hermitian structure (eqn. (\ref{hermdef})) is  $\Theta_\tau=\Theta\tau$, where $\Theta=\sf{CPT}$.}
$\B_\op$ with $\B_\bop$.

Suppose that a point $u\in L_\op\cap L_\bop$ corresponds to a complex flat bundle $E^\vee_\C$  that is an oper both holomorphically and antiholomorphically.   Then
its complex conjugate $\b E^\vee_\C$  is also an oper both holomorphically and antiholomorphically.   If $\b E^\vee_\C$  is not gauge-equivalent to $E^\vee_\C$
as a flat bundle, then $\b E^\vee_\C$ corresponds to a point $\b u\in L_\op\cap L_\bop$ that is distinct from $u$.    If so, $u$ and $\b u$ will correspond to distinct
basis vectors $\psi_u$ and $\psi_{\b u}$ of $\H$, and moreover these will be exchanged by $\Theta_\tau$.  The natural $B$-model pairing is diagonal
in the basis of intersection points:  the basis vectors can be normalized so that for $u,u'\in \Upsilon$, $(\psi_u,\psi_{u'})=\delta_{uu'}$.   Therefore, if $\Theta_\tau$
exchanges two distinct basis vectors $\psi_u$ and $\psi_{\b u}$, then $\psi_u$ and $\psi_{\b u}$ are both null vectors for the hermitian inner product that was defined
in eqn. (\ref{hermdef}).    The duality predicts that this hermitian inner product should be positive-definite, since on the $A$-model side, $\H$ is obtained by quantizing
a cotangent bundle and is a Hilbert space of half-densities.   So we expect that a flat bundle that is an oper both holomorphically and antiholomorphically is
actually real.   This was conjectured in \cite{EFK,EFK2} and was proved by an explicit (but surprisingly non-trivial) computation for $G^\vee=\U(1)$; the result
is also known for $G^\vee=\SU(2)$ \cite{Fal,Go}.

Finally, we can use the duality to predict the spectrum of the holomorphic and antiholomorphic Hitchin Hamiltonians as operators on $\H$.  Let $H_{\P,\alpha}$
be a quantized Hitchin Hamiltonian, that is, an element of $\A=\Hom(\B_\cc,\B_\cc)$.   The duality identifies $\A$ with $\Hom(\B_\op,\B_\op)$ and therefore
identifies $H_{\P,\alpha}$ with a holomorphic function $f_{\P,\alpha}$ on $L_\op$.   Acting on a basis vector $\psi_u$ that corresponds to a point $u\in L_\op\cap
L_\bop$, $H_{\P,\alpha}$ simply acts by multiplication by the corresponding value $f_{\P,\alpha}(u)$.   Similarly, if $H_{\b\P,\b\alpha}\in \b\A=\Hom(\b\B_\cc,\b\B_\cc)$
is an antiholomorphic quantized Hitchin Hamiltonian, then it corresponds under the duality to a  holomorphic function $f_{\bar\P,\b\alpha}$
on $L_\bop$, and it acts on $\psi_u$ as multiplication by $f_{\b\P,\b\alpha}(u)$.    This completes the description of the eigenvalues of the quantized
Hitchin Hamiltonians.   

\subsection{Including the Center}\label{inccenter}

It is not difficult to modify this description to take into account the center of $G_\C^\vee$.   
Consider as usual the $B$-model on $M=\Sigma\times C$.   It localizes on flat bundles over $\Sigma\times C$. 
In our application, $\Sigma=\R\times I$ is contractible, so a flat bundle on $M$ is the pullback of a flat bundle on $C$.
In the case of oper and anti-oper boundary conditions at the two ends of the strip, the flat bundle on $C$ is an oper both
holomorphically and antiholomorphically; thus it is a real oper.

An oper bundle, real or not, is irreducible and its automorphism group consists only of the center $\zZ(G^\vee)$ of the gauge group.
However, as explained in Section \ref{opergauge}, the boundary conditions also give trivializations $s_\ell$ and $s_r$ of the $\zZ(G^\vee)$
gauge symmetry on the two boundaries, modulo gauge transformations that act by $(s_\ell, s_r)\to (b s_\ell,b s_r)$, $b\in \zZ(G^\vee)$. 
   For a given real oper corresponding to a point $u\in \Upsilon$, let $\T_u$ be the set of pairs $s_\ell,s_r$ modulo the
action of $\zZ(G^\vee)$.    
 The $B$-model localizes on the isolated set of points $u,\varepsilon$ with $u\in \Upsilon$, $\varepsilon\in \T_u$.
So the Hilbert space $\H$ in the general case with a nontrivial center has a basis $\psi_{u,\varepsilon}$ for such $u,\varepsilon$.    

One can think of $\varepsilon\in\T_u$ as a sort of global holonomy between the left and right boundaries of the strip.
This refinement involving the torsor $\T_u$ is not important in the dual description of
the algebras $\A$ and $\bA$ via holomorphic functions on $L_\op$ or $L_\bop$, since each algebra acts on only one side of the strip.
   It is similarly not important in the determination of the eigenvalues of the Hitchin
Hamiltonians, which only depends on the interpretation of $\A$ and $\bA$ in terms of functions on $L_\op$ and $L_\bop$, and is not sensitive
to global holonomy across the strip.    It does affect the multiplicity of the eigenvalues, 
since eigenvectors $\psi_{u,\varepsilon}$ with the same $u$
and different $\varepsilon$ have the same eigenvalues of the Hitchin Hamiltonians.   And it will be relevant
in describing the eigenvalues of the 't Hooft or Hecke operators, to which we turn next.

\section{Hecke, 't Hooft, and Wilson Operators}\label{wtw}

\subsection{Line Operators}

In the usual formulation of geometric Langlands \cite{BD}, the main objects of study include the  Hecke functors acting on the category of $A$-branes and
the ``eigenbranes'' of these Hecke functors.

In the gauge theory picture \cite{KW}, the Hecke functors are interpreted in terms of 't Hooft line operators.   Electric-magnetic duality maps 't Hooft
line operators to Wilson line operators, leading to some of the usual statements about geometric Langlands duality.

\begin{figure}
 \begin{center}
   \includegraphics[width=6.2in]{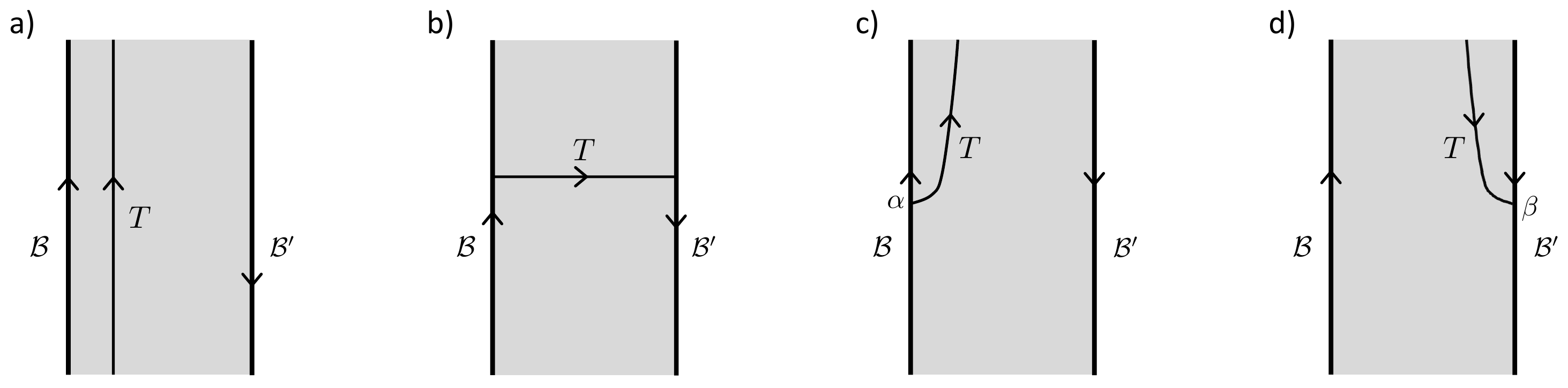}
 \end{center}
\caption{\small (a) A line operator $T$ parallel to the left boundary of the strip, and oriented compatibly.   Moving $T$ to the left, it maps
the boundary condition labeled $\B$ to a composite boundary condition $T\B$.   This is a line operator viewed as a functor on the category
of branes or boundary conditions, as in the usual formulation of geometric Langlands.  (b)  In the analytic approach to geometric
Langlands, the same line operator $T$, running horizontally across the strip, and with some additional data at the endpoints, becomes
an operator acting on physical states.    (c) and (d) The purpose of these drawings is to elucidate the additional data
that is needed at the left and right endpoints in (b).    At the left endpoint, we have an element $\alpha\in \Hom(\B,T\B )$, and at the right endpoint,
an element $\beta\in\Hom(T\B',\B')$.   \label{example1}}
\end{figure}

In general, in two-dimensional topological
field theory, line operators give functors acting on the category of  boundary conditions because a line operator $T$ that runs parallel to a boundary
labeled by a brane $\B$ can be moved  to the boundary, making a composite boundary condition $T\B$ (fig. \ref{example1}(a)).   
Here we assume that the two-manifold and the line operator (or more precisely the one-manifold on which it is supported)
are oriented and that the orientation of the line operator agrees with the orientation of the boundary on which it acts. 
 The same
figure also makes clear the notion of the adjoint of a line operator.   The adjoint $T'$ of a line operator is the same line operator with opposite orientation.
In fig. \ref{example1}(a), we could move the line operator $T$ to the right of the figure.   As its orientation is opposite to the orientation of the right
boundary, this gives an action of the dual line operator $T'$ on the brane $\B'$ that defines the boundary condition on the right boundary.
  So we get $\Hom(\B',T\B)=\Hom(T'\B',\B)$ for any $\B,\B'$.   (Some line operators have the property that $T$ is isomorphic to $T'$; their support
  can be an unoriented 1-manifold.)   
  
  These statements hold in any two-dimensional topological field theory.   Our actual application involves a four-dimensional theory with two additional dimensions
  that comprise a Riemann surface $C$.   Although it is possible to consider an 't Hooft operator (or a dual Wilson operator)
  whose support is an arbitrary curve $\gamma$ in the four-manifold
  $\Sigma\times C$, we will only consider the special case that $\gamma=\ell\times p$, where $p$ is a point in $C$ and $\ell$ is a curve in $\Sigma$.  So
  our line operators will be defined in part by the choice of $p$.
  In addition, in the application to geometric Langlands, an 't Hooft operator is labeled by a  finite-dimensional irreducible
  representation $R$ of $G^\vee$ (or equivalently of $G^\vee_\C$).   When we want to indicate
  this data, we denote the 't Hooft operator as $T_{R,p}$.
Similarly the  dual Wilson operator depending on the representation $R$ and the point $p$ will be denoted as $W_{R,p}$.

 \begin{figure}
 \begin{center}
   \includegraphics[width=3.4in]{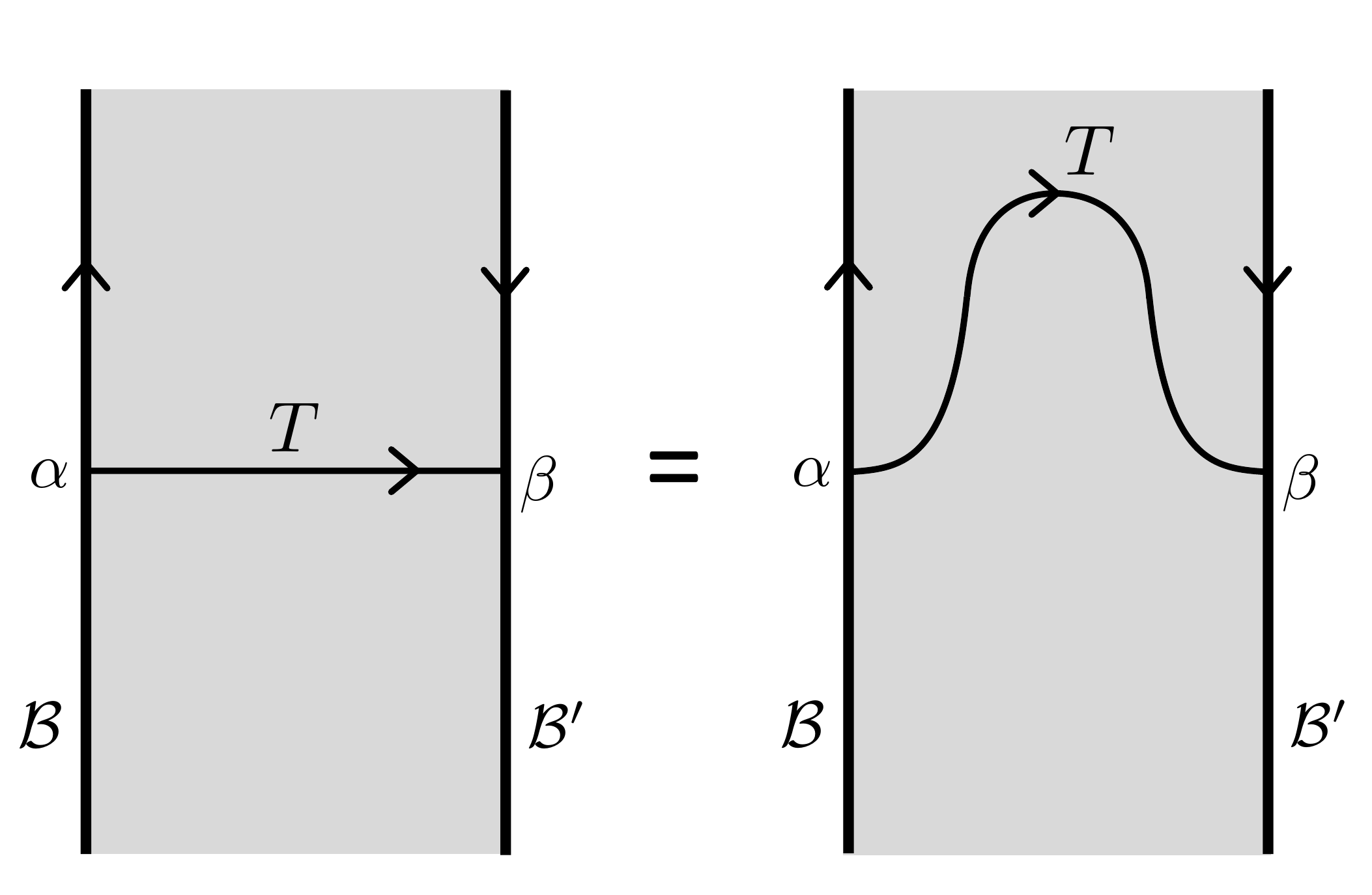}
 \end{center}
\caption{\small  This picture illustrates an algebraic manipulation described in the text.    Reading the drawing on the right from bottom to top,
one first encounters the operations sketched in figs. \ref{example1}(c,d), followed by the fusion of the  product $T'T$ of line operators to the ``identity,'' that is,
to a trivial line operator.  \label{example2}}
\end{figure}

In the analytic approach to geometric Langlands \cite{EFK,EFK2}, Hecke operators becomes ordinary operators acting on a Hilbert space of quantum states,
rather than more abstract functors acting on a category.   Not surprisingly, the gauge theory interpretation of Hecke operators in this sense is based on the
 same 't Hooft line operators as before, used somewhat differently.   In fig. \ref{example1}(b), we consider the same line operator as before, but now
running from left to right of the figure.   Some additional data must be supplied at the left and right endpoints where the line operator terminates on a
boundary of the strip.   Let us assume for the moment that this has been done.   Then the line operator becomes an ordinary operator acting on quantum states.
Reading the figure from bottom to top, an element of $\Hom(\B',\B)$ enters at the bottom and after the action of the line operator, a possibly different
element of $\Hom(\B',\B)$ emerges at the top.   (If we read the figure from top to bottom, we see the transpose operator acting on the dual
vector space $\Hom(\B,\B')$.)    It is because line operators that are supported on a one-manifold in space at a fixed time 
can act  in this way as ordinary quantum operators that they are traditionally\footnote{In traditional applications in particle physics, there
are no boundaries and the support of the line operator is taken to be a closed loop.   The operator is then often called a loop operator.} called
line ``operators.''

The purpose of figs. \ref{example1}(c,d) is to explain what is happening where the line operator of fig. \ref{example1}(b) ends on the left or right boundary.
In fig. \ref{example1}(c), we see that the left endpoint of the line operator corresponds to an element  $\alpha\in\Hom(\B,T\B)$, and in fig.
\ref{example1}(d), we see that the right endpoint corresponds to an element $\beta\in \Hom(T\B',\B')$.   Algebraically, the operator
$\h T:\Hom(\B',\B)\to \Hom(\B',\B)$ associated to a line operator $T$ with the additional data $\alpha,\beta$ can be described as follows.   For $\psi\in \Hom(\B',\B)$,
we have $\alpha\circ\psi\circ\beta\in \Hom(T\B',T\B)=\Hom(\B',T'T\B)$. Then using the fact that line operators
form an algebra and that the trivial line operator appears in the product $T'T$, we get a map $w:\Hom(\B',T'T\B)\to \Hom(\B' ,\B)$.  Finally
$\h T(\psi)=w\circ\alpha\circ\psi\circ\beta$.   This sequence of algebraic manipulations corresponds to the picture of fig. \ref{example2}.  
We will sometimes write $\h T_{\alpha,\beta}$ or $\h T_{R,p,\alpha,\beta}$  for the operator on $\H$ that is constructed from a line operator $T$ or $T_{R,p}$ with endpoint
data $\alpha,\beta$.

 \begin{figure}
 \begin{center}
   \includegraphics[width=3.4in]{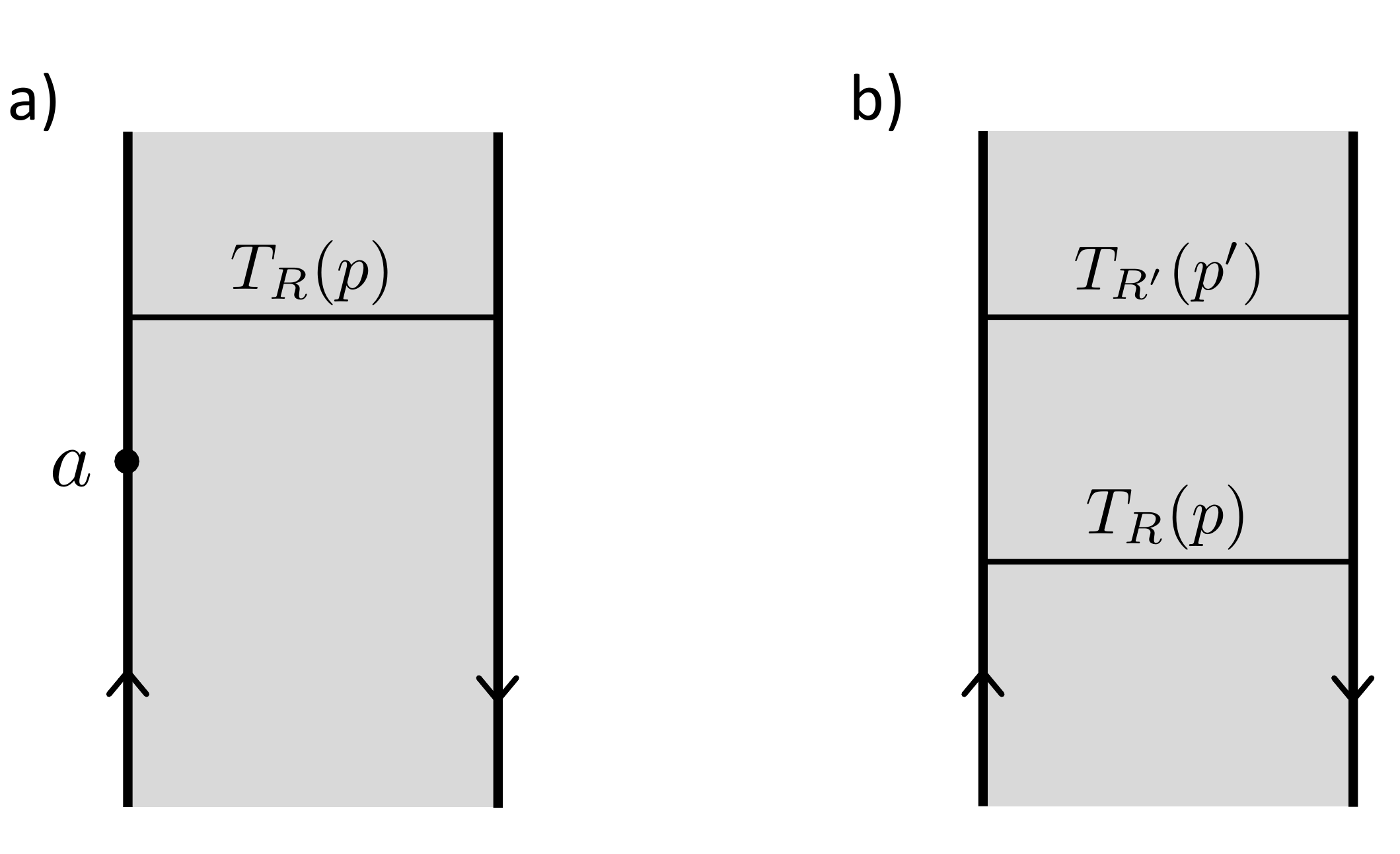}
 \end{center}
\caption{\small The argument showing that quantized Hitchin Hamiltonians commute can be adapted to show that line operators (viewed as actual operators
on a space of quantum states) commute with the quantized Hitchin Hamiltonians and with each other.   In each case, as sketched in (a) and (b) respectively,
the key point is that because of the existence of additional dimensions, the two operators can slide up and down past each other without singularity.    \label{example3}}
\end{figure}

 \begin{figure}
 \begin{center}
   \includegraphics[width=2.1in]{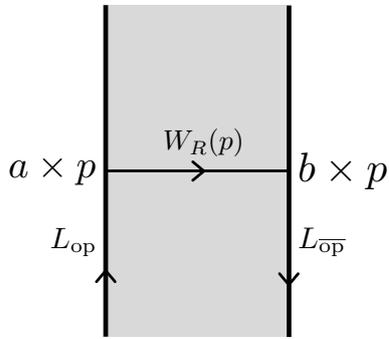}
 \end{center}
\caption{\small To compute the eigenvalues of the 't Hooft/Hecke operators, one considers dual Wilson operators that describe parallel transport from 
$a\times p$ to $b\times p$, where $p$ is a point in $C$ and $a,b$ are points on the left and right boundaries of $\Sigma$.   \label{example4}}
\end{figure} 

We will be particularly interested in elements $(\alpha,\beta)$ which originate from local endpoints of a four-dimensional line defect onto four-dimensional boundary conditions which lift $\B$ and $\B'$. As remarked in Section \ref{background}, the four-dimensional lifts of the boundary conditions we are considering are not topological. Instead, they are respectively holomorphic-topological and antiholomorphic-topological. As a consequence, 
the local endpoint lifting $\alpha$ will depend holomorphically on $p$ while the local endpoint lifting $\beta$ will depend antiholomorphically on $p$.
Notice that the actual path of the line defect in four-dimensions is immaterial, as long as it is topologically equivalent to a straight path. 
Only the positions of the endpoints in $C$ matter. 

One of the main properties of  't Hooft or Hecke operators, when regarded as in \cite{EFK,EFK2} as operators on quantum states, is that they commute with each
other and with the quantized Hitchin Hamiltonians.   This follows  from the same reasoning that we used to show that the Hitchin Hamiltonians commute
with each other.   An 't Hooft operator $T_{R,p}$ commutes with a Hitchin Hamiltonian $H_{\P,\alpha}$ because one can assume that the support
of $\alpha$ is disjoint from the point $p$, so that one can slide $T_{R,p}$ and $H_{\P,\alpha}$ up and down past each other (fig. \ref{example3}(a)) without
singularity.
Likewise, for distinct points $p,p'\in C$, 't Hooft operators $T_{R,p}$ and $T_{R',p'}$ commute (fig. \ref{example3}(b)).   Taking the limit $p'\to p$, it  follows that
$T_{R,p}$ and $T_{R',p}$ commute as well, even for $R\not= R'$.

\subsection{Wilson Operators And Their Eigenvalues}\label{wilop}

Since the 't Hooft operators commute with the Hitchin Hamiltonians, they can be 
diagonalized in the same basis, namely the basis of states $\psi_{u,\veps},$ $u\in \Upsilon,$
$\veps\in \T_u$,
where $\Upsilon=L_\op\cap L_\bop$ (Section \ref{inccenter}).   In fact, we can use electric-magnetic duality to determine the eigenvalues of the 't Hooft operators.   An 't Hooft operator
$T_{R,p}$ is dual to a Wilson operator $W_{R,p}$, labeled by the same representation $R$ of $G^\vee_\C$
and supported at the same point $p\in C$.   
While the 't Hooft operator is a ``disorder'' operator, whose microscopic definition involves a certain sort of singularity, Wilson operators are defined classically
in terms of holonomy, as follows.  In the relevant gauge theory on a four-manifold $M$ (for our purposes, $M=\Sigma\times C$), one has a 
$G^\vee_\C$ bundle $E^\vee_\C\to M$,
with connection $\cA=A+\i \phi$, to which we can associate a vector bundle $E^\vee_R=E^\vee_\C\times_{G^\vee_\C} R$.   We denote the induced connection
on this bundle simply as $\cA$.   The Wilson operator is constructed from the holonomy of the connection $\cA$ on $E^\vee_R$, integrated in general
along some oriented path $\gamma\subset M$.   If $\gamma$ is a closed loop, we take the trace of the holonomy around $\gamma$, and this gives
a version of the Wilson operator that is important in many physical applications.   

However,  to compute the eigenvalues of an operator defined by an 't Hooft line operator that stretches across the strip
(fig. \ref{example1}(b)), we need to consider a dual Wilson operator that similarly stretches across the strip (fig. \ref{example4}).   
  This is a Wilson operator supported, not on a closed loop, but  on a path
$\gamma\subset \Sigma\times p$ from $a\times p$ on the left boundary of the strip to $b\times p$ on the right boundary.

  In this case, the holonomy
is best understood as a linear transformation from the fiber $E^\vee_R$ at $a\times p$ to the fiber of this bundle at $b\times b$.   Thus  with an obvious
notation for these fibers, $W_{R,p}$ is a linear transformation
\be\label{lintr} W_{R,p}:E^\vee_{R,a\times p}\to E^\vee_{R, b\times p}.\ee
In order to treat the left and right edges of the strip more symmetrically, it is convenient to introduce the representation $R'$ dual to $R$ and view
$W_{R,p}$ as a linear function on a representation.  Then we have
\be\label{bintr} W_{R,p}\in \Hom(E^\vee_{R,a\times p}\otimes E^\vee_{R',b\times p},\C). \ee

So far, we have a linear function on a vector space, rather than a complex-valued function of connections, which could be quantized to get a quantum
operator.   To get a complex-valued function of connections, we need to supply vectors $v\in E^\vee_{R,a\times p}$, $w\in E^\vee_{R',b\times p}$.
Then $W_{R,p}(v\otimes w)$ is a complex valued function that can be quantized to get an operator.   

A natural construction of suitable vectors was described in Section \ref{localoper}.   In eqn. (\ref{linmap}), we described, for a holomorphic
oper with associated bundle $E^\vee_R$, a ``highest weight section'' $s_R:K_C^{(N-1)/2}\to E^\vee_R$.   Thus, if we are given a vector
$v\in K_{C,p}^{(N-1)/2}$, then we can define $s_R(v)\in E^\vee_{R,p}$.    Similarly, if $E^\vee_{R'}$ is an antiholomorphic oper, we have
$\bar s_{R'}:\bar K_C^{(N-1)/2}\to E^\vee_{R'}$, and hence, for $w\in K_{C,p}^{(N-1)/2}$, we have $\bar s_{R'}(w)\in E^\vee_{R',p}$.  Note
that a dual pair of representations $R,R'$ have the same value of $N$.

In the case of a bundle  $E^\vee_R\to \Sigma\times C$ that is a holomorphic oper on the left boundary and an antiholomorphic oper on the right
boundary, we can apply the holomorphic version of this construction on the left boundary and the antiholomorphic version on the right boundary, to get
\be\label{wintr}W_{R,p,v\otimes w}=W_{R,p}(s_{R}(v)\otimes \bar s_{R'}(w)). \ee
This finally is a complex-valued function of connections that can be quantized to get a Wilson operator on physical states.  We will call this
operator $W_{R,p,v\otimes w}$.   In the notation, we make use of the fact that the right hand side of eqn. (\ref{wintr}) depends on $v$ and $w$ only in the
combination $v\otimes w$.     Hopefully it will cause no serious confusion that we use the same notation for a classical holonomy (or its matrix element)
and the corresponding quantum operator.  

Because of the way the $B$-model localizes on flat connections, it is trivial to diagonalize this operator.   The flat connections that satisfy the boundary
conditions, with trivializations of the center on the boundary and modulo gauge transformations, are in one-to-one correspondence with 
 the usual basis of states $\psi_{u,\varepsilon}$, $u\in\Upsilon,$ $\varepsilon\in \T_u$ that diagonalize the Hitchin Hamiltonians.   
 The Wilson operators are diagonal in this basis.
  The eigenvalue of the quantum operator $ W_{R,p,v\otimes w}$ on a given
basis vector $\psi_{u,\varepsilon}$ is just the value of the  corresponding classical function  on the classical solution corresponding to $\psi_{u,\varepsilon}$.   That value
is simply the natural dual pairing $(s_R(v), \bar s_{R'}(w))$ of the vectors $s_{R}(v)$ and $\bar s_{R'}(w)$ in the dual vector spaces $E^\vee_{R,p}$ and
$E^\vee_{R',p}$.     This is true because a flat connection on $\Sigma\times C$ is actually a pullback from $C$.

The center $\zZ(G^\vee)$ acts on the physical Hilbert space $\H$ by $(s_\ell, s_r)\to (b s_\ell, s_r)$ (Section \ref{opergauge}).   This transforms
the eigenvalue of $\h W_{R,p}$ by a root of unity, which is simply the value of the central element $b$ in the representation $R$.   For example,
if $G^\vee=\SL(n,\C)$ and $R$ is the $n$-dimensional representation, eigenvectors of $\h W_{R,p}$ come in $n$-plets of the form
$\lambda\exp(2\pi \i k/n)$, $\lambda\in\C$, $k=0,1,\cdots, n-1$.   This is dual to the fact that on the $A$-model side, the Higgs bundle moduli space
has components labeled by a characteristic class $\zeta\in H^2(C,\pi_1(G))$.   For $G^\vee=\SL(n,\C)$, there are $n$ components,
which are cyclically permuted by the 't Hooft operator $\h T_{R,p}$ dual to $\h W_{R,p}$, leading to the same structure of the spectrum.   
In general, the action of $\zZ(G^\vee)$ on $R$ mirrors the way $\h T_{R,p}$ permutes the components of the moduli space. 

The holomorphic and antiholomorphic 
corner data needed to define the operators $W_{R,p}$ has consisted precisely of the vectors $v\in K_C^{(N-1)/2}$, $w\in \bar K_C^{(N-1)/2}$.  For the $n$-dimensional
representation of $\SL(n,\C)$, we have $N=n$.
We expect the same data to be needed to define holomorphic and antiholomorphic corner data for the dual 't Hooft operators.

These are arguably the simplest Wilson operators and we will call them principal Wilson
operators.   However, if the representation $R$ is reducible when restricted to a principal $\su(2)$ subalgebra of $\g^\vee$, then it is possible
to use the more general objects $s_{R,n}:Q_n\otimes K_C^{(n-1)/2}\to E^\vee_R$ (eqn. (\ref{holmap})).   So picking $v\in Q_n \otimes K_C^{(n-1)/2}$,
$w\in \bar Q_{n'}\otimes K_C^{(n'-1)/2}$ (where $n$ and $n'$ can be chosen independently), we can define $W_{R,p,n,n',v\otimes w}=W_{R,p}(s_{R,n}(v)\otimes
\bar s_{R',n'}(w))$, which can again be interpreted as a quantum mechanical operator.  Even more generally, we can consider holomorphic and antiholomorphic derivatives with respect to $p$ of $W_{R,p,n,n',v\otimes w} $.   It is enough to consider $n-1$ holomorphic derivatives and $n'-1$ antiholomorphic
ones; this suffices to define a complete set of Wilson operators,
since it amounts to applying the  linear form  $W_{R,p}:E^\vee_{R,a\times p}\otimes E^\vee_{R',b\times p}\to \C $ 
(eqn. (\ref{bintr})) to a set of vectors that according to the analysis 
in Section \ref{localoper}
form a basis of the finite-dimensional vector space on which $W_{R,p}$ is acting.

What happens if we continue to differentiate?
With $n$ holomorphic derivatives or $n'$ antiholomorphic ones, we will run into differential equations satisfied by the Wilson operators $W_{R,p,n,n',v\otimes w} $.
The holomorphic and antiholomorphic sections $ s_{R,r}$ and  $\bar s_{R',n'}$ that were used to define these Wilson operators
obey certain holomorphic and antiholomorphic 
differential equations (Section \ref{localoper} and Appendix \ref{somex}) as a function of $p$, and the corresponding Wilson operators obey the same equations.

Having defined operators that act on the physical Hilbert space $\H$, it is natural to ask what algebra they obey.   In bulk, the product of line operators
mimics the tensor product of representations of $G^\vee$.   Thus, if $R_i\otimes R_j\cong \oplus_k N^k_{ij}R_k$, with vector spaces $N^k_{ij}$, then
the corresponding decomposition of parallel Wilson operators is $W_{R_i} W_{R_j}=\oplus_k N^k_{ij} W_k$.     This is the appropriate relation for Wilson
operators understood as functors acting on boundary conditions, as illustrated in fig. \ref{example1}(a).   From this algebra, the structure of the nonabelian
group $G^\vee $ can be reconstructed, in principle.
However, for Wilson operators as operators on quantum states, as we are discussing here, the picture is different.
Matters are simplest if we multiply two principal Wilson operators, associated to highest weight vectors in the corresponding representations.
Since the tensor product of highest
weight vectors in two representations $R_1$ and $R_2$ is a highest weight vector in the tensor product $R_1\otimes R_2$,
the product of two principal Wilson line operators is another principal Wilson line operator, for a representation $R_3$ whose highest weight is the
sum of the highest weights of $R_1$ and $R_2$.   We will see the structure of the nonabelian group if we multiply the more general Wilson
operators $W_{R,p,n,n',v\otimes w}$ and their derivatives.

A dual 't Hooft operator  defined using the $S$-dual data will, of course, have the same eigenvalues as the Wilson
operators. The main challenge is to identify precisely the appropriate $A$-model endpoints.  We turn to that problem in Section \ref{dualhooft}.

We conclude this discussion of Wilson operators with the following remark.
The most illuminating realization of the oper boundary condition in four-dimensional gauge theory involves the Nahm pole \cite{GWknots}.   
Compared to a more direct approach that was assumed earlier \cite{KW} (in which the boundary condition is defined by just specifying the $(0,1)$ or
$(1,0)$ part of $\cA$ along the boundary), the Nahm pole description of the oper boundary condition has two
advantages:  it leads directly to the local constraints discussed in Section \ref{localoper}; and it also leads to a simple explanation of the duality
with the $A$-model description via $\B_\cc$.    If one uses the Nahm pole description of the oper boundary condition, then the complex connection $\cA$
is singular along the boundary, and some renormalization is involved in defining the classical holonomy $W_R$ across the strip and the corresponding
quantum operator.   The renormalization amounts to a complex gauge transformation that removes the Nahm pole singularity.

\subsection{The Dual 't Hooft Operators}\label{dualhooft}

 \begin{figure}
 \begin{center}
   \includegraphics[width=3.6in]{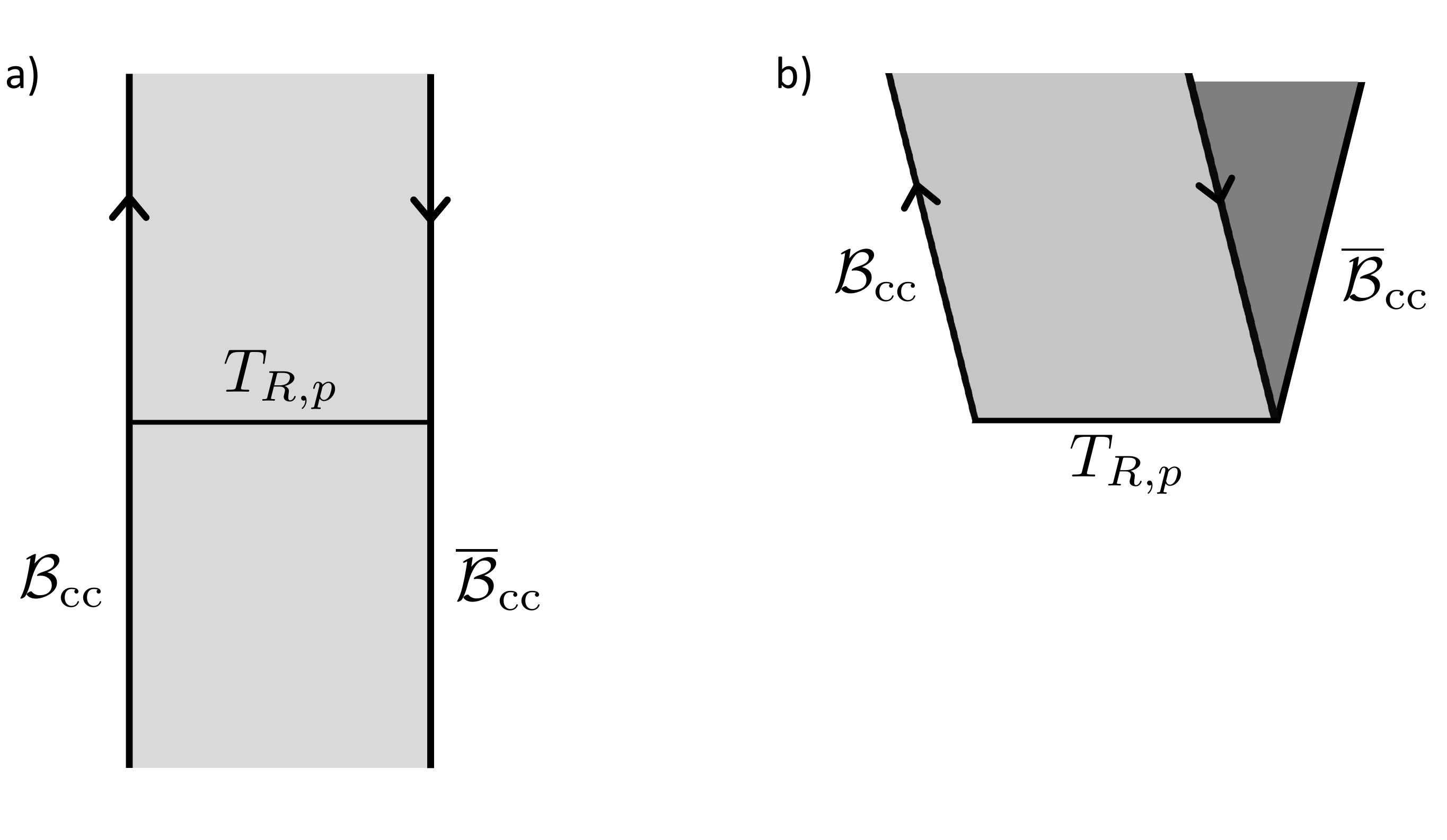}
 \end{center}
\caption{\small (a) An 't Hooft operator viewed as an interface from the $A$-model of $\M_H(G,C)$ to itself. (b) After  ``folding,'' the same 't Hooft
operator is interpreted as a brane defining a boundary condition in the product of two copies of $\M_H(G,C)$, with equal and opposite symplectic
form.    \label{example7}}
\end{figure} 

\subsubsection{'t Hooft Operators and Hecke Modifications}\label{THO}

An 't Hooft operator  $T_{R,p}$ (fig. \ref{example7}(a)) 
will produce a jump in the fields $(A,\phi)$ and  in the associated   Higgs bundle $(E,\varphi)$, in the sense that the Higgs bundle $(E,\varphi)$ just
below the 't Hooft operator is generically not isomorphic to the Higgs bundle $(E',\varphi')$ just above it.   They differ by what is known as
 a Hecke modification.   The  type of Hecke modification is 
 determined by the magnetic singularity of the 't Hooft operator, which is classified by a choice of
irreducible representation $R$ of the dual group $G^\vee$.  
Hecke modifications of bundles and Higgs bundles were described for physicists  in \cite {KW} and in more detail in  Section 4 of
\cite{More}.   Here we will give a very brief synopsis.

A Hecke modification of a holomorphic $G_\C$ 
bundle $E$ at a point $p$ is a new bundle $E'$ that is presented with an isomorphism to $E$ away from
$p$, but such that this isomorphism does not extend over $p$.   A section $w$ of $E'$ is a section of $E$ that is allowed to have poles of a specified
type at $p$, or that is constrained so that some components have zeroes of a specified type, or both.   

For example, if $E$ is a rank 2 holomorphic
vector bundle, trivialized near $p$ so that a section of $E$ is a pair of holomorphic functions $\begin{pmatrix}f\cr g\end{pmatrix}$, then an example of a Hecke modification of $E$ at $p$
is a new bundle $E'$ whose sections are a pair $\begin{pmatrix}f\cr g\end{pmatrix}$, 
where $f$ is holomorphic at $p$ but $g$ is allowed to have a simple pole at $p$.  This example
can be slightly generalized to give a family of Hecke modifications of $E$ at $p$ that are parametrized by $\bCP^1$.   We simply pick a pair of complex numbers
$u,v$, not both zero, representing a point in $\bCP^1$, and allow a section of $E'$ to have a polar part proportional to this pair:
\be\label{thep} w= \frac{1}{z}\begin{pmatrix}u\cr v\end{pmatrix}+{\mathrm{regular}}, \ee
where $z$ is a local parameter at $p$.  

 The relation between $E$ and $E'$ is reciprocal: instead of saying that we obtain $E'$ from $E$ by allowing a pole of a certain type, we could say
 that we obtain $E$ from $E'$ by requiring a certain type of zero.  If an 't Hooft operator $T_{R,p}$ can map $E$ to $E'$, then the dual 't Hooft operator
 (which is associated to the dual representation) can map $E'$ to $E$.

A Hecke modification of a Higgs bundle $(E,\varphi)$ actually does ``nothing'' to $\varphi$.    This means the following.   A Hecke modification of $(E,\varphi)$
is just a Hecke modification $E'$ of $E$ such that $\varphi:E\to E\otimes K_C$ is holomorphic
as a map $E'\to E'\otimes K_C$.   In the example of the rank 2 bundle, the Higgs field is locally $\varphi=\varphi_z\,\d z$, where $\varphi_z$ is a $2\times 2$ matrix of
holomorphic functions. 
   For $\varphi$ to be holomorphic as a map $E'\to E'\otimes K_C$, the necessary and sufficient condition is that, if $w$ is a section of $E'$
as characterized in eqn. (\ref{thep}), then the polar part of $\varphi w$  should be a multiple of $\begin{pmatrix}u\cr v\end{pmatrix}$; in other
words, $\begin{pmatrix}u\cr v\end{pmatrix}$ must be an eigenvector of the matrix $\varphi_z(p)$.
Generically, this condition is not satisfied and hence most Hecke modifications of $E$ are not valid as Hecke modifications of $(E,\varphi)$.   
If $\varphi_z(p)$ is not nilpotent,  it has  two distinct eigenvectors and 
the Higgs bundle $(E,\varphi)$ has two 
 possible Hecke modifications of this type  at the point $p$;
if $\varphi_z(p)$ is nilpotent but not zero, it has only one eigenvector and there is just one possible Hecke modification of this type; only if $\varphi_z(p)=0$ 
does  $(E,\varphi)$ has the same  $\bCP^1$ family of possible Hecke modifications of this type at $p$ that $E$ would have by itself.

Hecke modifications of the type just described can be viewed in two different ways.  
They are  dual to the two-dimensional representation of $G^\vee=\U(2)$.   This group is self-dual, so the  discussion
is applicable in gauge theory of $G=\U(2)$.    Alternatively, the same Hecke modification is dual to the two-dimensional representation of $G^\vee=\SU(2)$.
In this case, the dual group is $G=\SO(3)$.   In this application, since the rank two bundles $E$ and $E'$  do not have $\SO(3)$ structure group, the
preceding discussion should be restated in terms of the corresponding adjoint bundles $\ad(E)$ and $\ad(E')$

In this example, let  $\delta$ be the eigenvalue of $\varphi(p)$ acting on $\begin{pmatrix}u\cr v\end{pmatrix}$:
\be\label{defdelta} \varphi(p) \begin{pmatrix}u\cr v\end{pmatrix}=\delta \begin{pmatrix}u\cr v\end{pmatrix}.\ee
As we vary the choice of $E$ and $E'$, $\delta$ varies holomorphically. It defines a holomorphic function on the space of Hecke modifications of this type. Since
the eigenvalues of $\varphi(p)$ are $\pm \delta$, we have 
\be\label{delchar} \delta^2=-\det \varphi(p),\ee
 where $\det\varphi(p)$ is a linear combination of the Hitchin Hamiltonians.
The sign of $\delta$ distinguishes the two choices of 
Hecke modifications compatible with a given Higgs field.

\subsubsection{Hecke Correspondences}\label{HC}

The 't Hooft operator $T_{R,p}$ can be viewed as an interface between the $A$-model of $\M_H(G,C)$
and itself.  This is a tautology; in fig. \ref{example7}(a), we see the $A$-model of $\M_H(G,C)$ above and below $T_{R,p}$, so we can view
$T_{R,p}$ as an interface between two copies of this $A$-model.   In fact, this interface is of type $(B,A,A)$, because of the supersymmetric properties of $T_{R,p}$.
   It is convenient to use a folding trick similar to the one of fig. \ref{folding}.   Instead of associating to $T_{R,p}$
an interface in the $A$-model of $\M_H(G,C)$, we can associate to it a brane or boundary condition
 $\B_{R,p}$ of type $(B,A,A)$
 in the $A$-model of a product $\M_H(G,C)\times \M_H(G,C)$
(fig. \ref{example7}(b)).   Here the symplectic structure of $\M_H(G,C)\times \M_H(G,C)$ is $\omega_K\boxplus (-\omega_K)$, with a minus
sign in one factor because folding reverses the sign of the symplectic structure.         

Let us consider explicitly what $\B_{R,p}$ will look like for the basic example, described in Section \ref{THO},  that $R$ is the two-dimensional representation of $G^\vee=\SU(2)$.
What is its complex dimension?   If $C$ has genus $g$, then the choice of a $G_\C$ bundle $E$ depends on $3g-3$ complex parameters.   Choosing 
$E'$'s that can be made from $E$ by action of $T_{R,p}$ adds one more complex parameter.    But we have to constrain the Higgs field $\varphi$
so that $\begin{pmatrix}u\cr v\end{pmatrix}$ is an eigenvector of $\varphi(p)$.   So the choice of $\varphi$ involves $3g-4$ parameters,
not $3g-3$.    The upshot is that the support $Z_{R,p}$ of $\B_{R,p}$ has dimension $6g-6$, and thus  $Z_{R,p}$ is middle-dimensional in $\M_H(G,C)\times \M_H(G,C)$.
Because $Z_{R,p}$ is middle-dimensional and is the support of a brane of type $(B,A,A)$, it must be a complex Lagrangian submanifold.
The brane  $\B_{R,p}$ has rank 1 because the Hecke transformation by which $T_{R,p}$ produces $E'$ from $E$ is
generically unique, if it exists.   In short, $\B_{R,p}$ is a rank 1 Lagrangian brane of type $(B,A,A)$.

We can be more specific, because $\B_{R,p}$ is manifestly invariant under scaling of $\varphi$, which corresponds to 
the $\C^*$ symmetry of $\M_H(G,C)\times \M_H(G,C)$.   We can give a simple
description in the same sense in which $\M_H(G,C)$ can be approximated by its dense open set
 $T^*\M(G,C)$. (As in the general discussion of quantization, we expect that this approximation is sufficient in an $\lmark^2$ theory.)
   The intersection of $Z_{R,p}$ with $\M(G,C)\times \M(G,C)\subset T^*\M(G,C)\times T^*\M(G,C)$ is the variety $X_{R,p}$
 that parametrizes pairs $E,E'$ such that $E'$ can be reached from $E$ by a Hecke transformation of type $R$ at the point $p$.   $\C^*$ invariance
 of $Z_{R,p} $ means that it can identified as the conormal bundle\footnote{If a submanifold $U\subset M$ is defined locally by vanishing of 
 some coordinates $q_1,\cdots, q_r$, then its conormal bundle in $T^*M$ is defined by setting to zero those coordinates and the momenta that
 Poisson-commute with them.}    of $X_{R,p}$.   $X_{R,p}$ is called the Hecke correspondence in this situation, and $Z_{R,p}$ is the Hecke correspondence
 for Higgs bundles.
   
 The particular example of an 't Hooft operator dual to  the two-dimensional representation of $G^\vee=\SU(2)$ is relatively simple because the space
 of possible Hecke modifications of a given bundle $E$ at a given point $p$ is 1-dimensional.   In general, the space of Hecke modifications of $E$ that
 can be made at $p$ by $T_{R,p}$ has a dimension that depends on $R$ and becomes arbitrarily large if $R$ is a representation of $G^\vee$ of large highest weight.
 When the dimension is sufficiently large, every $E'\in \M(G,C)$ can be made from $E$ by $T_{R,p}$ and the ways to do so form a complex manifold $\Phi_{E',E,R,p}$
 of positive dimension.
 In such a situation, $Z_{R,p}$ is rather complicated.   It has a component on which the Higgs field vanishes (if $E'$ is produced from $E$ by a generic
 Hecke modification of very high weight, then no nonzero $\varphi:E\to E\otimes K$ is holomorphic as a map $E'\to E'\otimes K$), and other
 components with nonzero Higgs field (it is possible to pick a Hecke modification of $E$ of arbitrarily high weight 
 such that $\varphi:E'\to E'\otimes K$ is holomorphic). 
The $\CP$ bundle of $\B_{R,p}$ restricted to the various components is not just of rank 1.   For example, restricted to the component on which the
Higgs field vanishes, this $\CP$ bundle is formally  (that is,  modulo a proper treatment of singularities), the cohomology of $\Phi_{E',E,R,p}$.
 
\subsubsection{'t Hooft Line Operators as Operators on Quantum States}\label{lineq} 
   
 Under fairly general conditions, if a symplectic manifold $M$ can be quantized by branes to get a Hilbert space $\H$, an $A$-brane $\B$ in $M\times M$ (with
 additional data at the ``corners,'' as discussed presently) can be interpreted
 as a quantum operator on $\H$.  This is discussed in general in Section 4 of \cite{GW}.    In general it is difficult to get an explicit description of such
 an operator.   But here we are in a special situation with a simple answer.   That is because $M=\M_H(G,C)$ is effectively a cotangent bundle $T^*\M(G,C)$,
 and the brane of interest is supported on the conormal bundle of a subvariety $X_{R,p}\subset \M(G,C) \times \M(G,C)$.  
 
A fairly general operator $\O$ acting on the quantization of $T^*\M(G,C)$ can be represented   by an integral kernel $F(x,y)$ which is a half-density
on $\M(G,C)\times \M(G,C)$,   The action of $\O$ on a state $\Psi$ is
\be\label{actst} \O\Psi(x)=\int_{\M(G,C)}\d y \, F(x,y) \Psi(y). \ee
What sort of integral kernel should we expect for the quantum operator $\h T_{R,p}$ associated to the 
't Hooft operator $T_{R,p}$?   The points $y$ and $x$ in eqn. (\ref{actst})  correspond in fig. \ref{example7}(a) to the fields $(A,\phi)$ just below
and just above the line operator $T_{R,p}$.  So they correspond to bundles $E,$ $E'$, such that $E'$ can be reached from $E$ by a Hecke modification of type $R$
at $p$.   Hence in a classical limit, $F(x,y)$ is a distribution supported on $X_{R,p}$, which parametrizes such pairs $(E,E')$ .    The simplest case, which
we will consider first, 
is that $F(x,y)$ is a delta function in the directions normal to $X_{r,p}$.  More generally, in its dependence on the normal directions, $F(x,y)$ can be proportional
to arbitrary derivatives of a delta function in the normal variables.  

In general, one would expect
quantum corrections to the claim that $F(x,y)$ is supported on the underlying classical correspondence $X_{R,p}$.   However, in the present situation, there are no corrections, because the symplectic manifold
that is being quantized is a cotangent bundle $T^*\M(G,C)$, and the correspondence  $Z_{R,p}$ is a conormal bundle in $T^*\M(G,C)\times T^*\M(G,C)$.
The scale invariance of the cotangent bundle and the conormal bundle imply that the kernel $F(x,y)$ cannot depend on $\hbar$ and can be evaluated
in a semiclassical limit. 

We should add a note  on why it is valid here to argue based on scaling symmetry.
Brane quantization is based, as always, on studying $T^*\M(G,C)$ in the context of a suitable complexification.   Similarly to study $Z_{R,p}$ in brane
quantization involves 
complexifying it in a complexification of $T^*\M(G,C)\times T^*\M(G,C)$.
A scaling argument in brane quantization really involves scaling of the complexifications.   Such an argument is valid in the present setting because
the structure of $T^*\M(G,C)$ as a cotangent bundle and of $Z_{R,p}$ as a conormal bundle do extend to holomorphic stuctures of the same
type for their complexifications.    

We should also clarify what we mean by ``semiclassical limit.''  A one-loop correction is built into the assertion that a wavefunction is a half-density rather than a function and that $F(x,y)$ is correspondingly a half-density on $\M(G,C) \times \M(G,C)$.   The assertion that  $F(x,y)$ can be computed semiclassically 
means that there is no quantum correction beyond this fact.

According to Etingof, Frenkel, and Kazhdan \cite{EFK2},  the Hecke operator dual to the two-dimensional representation of $\SU(2)$ is defined by an integral
kernel that can be factored as the product of holomorphic and antiholomorphic factors.   Such a holomorphic factorization
is expected in the $A$-model.   A holomorphic factor $f$ will come from  
 the left endpoint of  $T_{R,p}$ in fig. \ref{example7}(a), or equivalently the lower left corner in fig. 
\ref{example7}(b),  and an antiholomorphic factor $\t f$ will come from the right endpoint or the lower right corner.
We view the 't Hooft operator as a rank 1 brane of type $(B,A,A)$ in $\M_H(G,C)\times \M_H(G,C)$, with trivial $\CP$ bundle.    In general, the space of corners between a brane of this type, supported on a Lagrangian submanifold $L$, and the canonical coisotropic $A$-brane is $H^0(L,K_L^{1/2})$.   The 't Hooft operator corresponds to a brane whose support
 is the Hecke correspondence $Z_{R,p}$.  So in this case, a holomorphic corner is a holomorphic section $f\in H^0(Z_{R,p},K^{1/2}_{Z_{R,p}})$.
  Similarly, an antiholomorphic corner is an element $\t f\in H^0(\bar Z_{R,p},\bar K^{1/2}_{\bar Z_{R,p}})$, where
  $\bar Z_{R,p}$ is $Z_{R,p}$ with opposite complex structure.

The product $\mu = f \t f$ of $f$ and $\t f$ will be a half-density on $Z_{R,p}$. We will  show that this data is precisely what is needed to define a 
distributional kernel $F(x,y)$. If $f$ and $\tilde f$ are pull-backs from $X_{R,p}$, we will get a delta function kernel. If they have a polynomial dependence on the fiber of $Z_{R,p} \to X_{R,p}$, we will get a linear combination of derivatives of a delta function. 

In order to describe a delta function distribution, we  pick some local coordinates. We parametrize the input to the Hecke transformation
by coordinates $\vec x=x_1,\cdots, x_{3g-3}$ on $\M(G,C)$.  We will write $|\d \vec x|$ for the half-density $(\d x_1 \cdots \d x_{3g-3} \d \bar x_1 
\cdots \d \bar x_{3g-3})^{1/2}$, and similarly for other variables introduced momentarily.  
For a given $\vec x$,  the output of the Hecke transformation ranges over a copy of $\bCP^1$ that we will call $\bCP^1_x$; we parametrize it 
 by a complex variable $z$.  $\bCP^1_x $ is of complex codimension $3g-4$ in $\M(G,C)$.  $\bCP^1_x$ can be defined locally by a condition $\vec n=0$, where  $\vec n=(n_1,\cdots, n_{3g-4})$ are local holomorphic coordinates  on the normal bundle $N$  to $\bCP^1_x$ in $\M(G,C)$. We can write the kernel as 
\be\label{nuff} 
F(x,y)= b(z,\bar z,x, \bar x)|\d\vec x ||\d z \d \vec n |\delta(\vec n,\vec{\bar n})
\ee
where $\d \vec n=\d n_1 \d n_2\cdots \d n_{g-4}$ and the delta function is defined by $\int \d \vec n \d \vec{ \bar n}\delta(\vec n,\vec{\bar n})=1$. 
In view of that last relation, the delta function transforms under a change of coordinates on the normal bundle as $(\d \vec n \d \vec{\bar n})^{-1}$, which
means that the possible kernels are in one-to-one correspondence with objects
\be\label{luff} \mu = b(z,\bar z,x, \bar x)|\d\vec x| |\d z(\d \vec n)^{-1}|. \ee
The Hecke correspondence $Z_{R,p}$ for Higgs bundles is parametrized locally by the coordinates $\vec x$ and $z$, introduced above, which parametrize the Hecke correspondence $X_{R,p}$ for bundles, and additional variables $\vec m$  that parametrize the choice of Higgs field.  Since $Z_{R,p}$ is the conormal bundle of $X_{R,p}$, the variables $\vec m$ are dual to the normal bundle coordinates $\vec n$ that appear in eqn. (\ref{nuff}). 
Therefore, we can replace $(\d\vec n)^{-1}$ with $\d \vec m$, and the possible delta function kernels are in one-to-one correspondence with half-densities 
\be\label{luff1} \mu = b(z,\bar z,x, \bar x)|\d\vec x \d z\d \vec m|^2. \ee
on $Z_{R,p}$ such that $b$ is independent of $\vec m$. 

This discussion is immediately generalized to linear combinations of normal derivatives of a delta function.   A kernel that involves normal derivatives of a delta
function
\be\label{nuff2} 
F(x,y)= b(x, \bar x,z,\bar z, \partial_{\vec n}, \partial_{\vec {\bar n}}) |\d\vec x | |\d z \d \vec n |\delta(\vec n,\vec{\bar n})
\ee
corresponds to a half-density 
\be\label{luff2} \mu = b(x, \bar x, z,\bar z,\vec m, \vec {\bar m})|\d\vec x \d z\d \vec m|. \ee
on $Z_{R,p}$ such that $b$ depends polynomially on $\vec m$. 

A holomorphically factorized kernel will  take the form $\mu = f \t f$ where $f$ and $\t f$ are respectively holomorphic and antiholomorphic.
We can now see what kind of holomorphic object $f$ must be.   $f$ must be a half-density on the Hecke correspondence $Z_{R,p}$
in the holomorphic sense: $f=v(\vec x, z,\vec m) (\d \vec x\d z\d\vec m)^{1/2}$.   In more standard language, $f$ must be
an element of $H^0(Z_{R,p}, K^{1/2}_{Z_{R,p}})$.   As explained earlier, this is the expected form of the answer in the $A$-model for a left endpoint of $T_{R,p}$.
Similarly, $\t f$ is an antiholomorphic section of the anticanonical bundle of $Z_{R,p}$, again in accord with the $A$-model expectation.

 If and only if $Z_{R,p}$ is a Calabi-Yau manifold,  there is a particular holomorphic section $\lambda_0$ of $K_{Z_{R,p}}^{1/2}$
  that is everywhere nonzero.   If such a $\lambda_0$
exists, then the data that defines any other holomorphic corner is $f= g \lambda_0$, where $g$ is a holomorphic function on $Z_{R,p}$.   In particular, such
a function $g$ (with only polynomial growth) 
is a polynomial in the Hitchin Hamiltonians $H_{\P,\alpha}$ and  the holomorphic function $\delta$ that was defined in eqn. (\ref{defdelta}), up to the relation
$\delta^2 = -\det \varphi(p)$, which expresses $\delta^2$ as a linear combination of the $H_{\P,\alpha}$.
   A nonconstant polynomial  $g(H_{\P,\alpha},\delta)$ has nontrivial zeroes, so $g\lambda_0$
is everywhere nonzero only if $g$ is a constant, showing that  $\lambda_0$ is unique, up to a constant multiple, if it exists.\footnote{More generally, if $Z$ is a complex manifold with $b_1(Z)=0$, then an everywhere nonzero
holomorphic function $g$ on $Z$, with no exponential growth, is constant.    Indeed, since $b_1(Z)=0$, the closed 1-form $\d g/g$ is exact, $\d g/g=\d w$
for some $w$, so $g=C e^w$ (with a nonzero constant $C$) and $g$ has exponential growth unless it is constant.   
The Hecke correspondence for $G_\C$-bundles satisfies 
$b_1(X_{R,p})=0$.   The Hecke correspondence for $G_\C$ Higgs bundles is the conormal bundle of $X_{R,p}$ and hence $b_1(Z_{R,p})=0$.}
Corners of the form $g \lambda_0$, where $g$ is a polynomial in the Hitchin Hamiltonians and $\delta$, precisely match
 the corners for Wilson lines, built from $s$, $Ds$ and 
polynomials in the observables dual to the Hitchin Hamiltonians.

Such a $\lambda_0$ does indeed exist, by virtue of a result of Beilinson and Drinfeld \cite{BD} that was important 
 in the work of Etingof, Frenkel, and Kazhdan \cite{EFK2}.  
 The properties of $\lambda_0$ mirror those of the simplest Wilson line corner $s$ of Section \ref{wilop}. In particular, according to the result of Beilinson and Drinfeld, $\lambda_0$, like $s$, varies with $p$ as a section  of   $K_p^{-1/2}$.  Of course, the existence of such a mirror of $s$ is expected from electric-magnetic duality.
 
  A slightly different formulation was useful in \cite{EFK2}.  To explain this, let us go back to 
 eqn. (\ref{luff}), from which we see that if $\mu$ can be holomorphically factorized, then the holomorphic
 factor is a holomorphic form $k = w(\vec x,z) (\d \vec x    \d z)^{1/2}  (\d \vec n)^{-1/2}$.    We can replace $\d z^{1/2} (\d\vec n)^{-1/2}$ with
 $\d z (\d \vec y)^{-1/2}$ where $\vec y=(z,\vec n)$ parametrizes the output of the Hecke transformation.   So in other words a holomorphically factorized kernel
 will come from a holomorphic object\footnote{Since $\vec y$ is determined by $\vec x$ and $z$, and reciprocally $\vec x$ is determined by $\vec y$ and $z$,
 we could equally well write $w(\vec y,z)$ instead of $w(\vec x,z)$.}
 \be\label{conf} k=w(\vec x,z) (\d \vec x)^{1/2} (\d\vec y)^{-1/2}\d z. \ee 
 If we multiply $k$ by its complex conjugate, we get
 \be\label{onf}|k|^2= |w(\vec x,z)|^2 |\d\vec x| |\d z\d\bar z| |\d\vec y|^{-1}.\ee
 This leads directly to the definition used in \cite{EFK2}.   The quantity $|k|^2$ can be regarded as a map from half-densities in $\vec y$ to half-densities in $\vec x$,
 valued in differential forms $|\d z \d\bar z|$ that can be integrated over $\bCP^1_x$.    That integral gives the Hecke operator at the point $p\in C$:
 \be\label{tonf} H_p=\int_{\bCP^1_x} |k|^2. \ee

In Section \ref{sec:chiral}, we will interpret some of these statements via two-dimensional chiral algebras.
In particular, the considerations about adjoint-valued chiral fermions in Section \ref{fft} are a physicist's interpretation of the original analysis of Beilinson
and Drinfeld.
The chiral algebra approach is local on $C$ and and can potentially be extended to 
situations where the space of Hecke modifications relating two given bundles has positive dimension.

\subsubsection{The Affine Grassmannian}\label{affgr}

So far, we have considered the simplest examples of Hecke modifications.   But to develop the theory
further, one wants a more systematic approach.   

 As motivation, we consider  first the case of a holomorphic vector bundle of rank $n$.
  Let $z$ be a local holomorphic parameter that vanishes at a point $p\in C$.  Let $U$ be a small neighborhood of  $p$.  $C$ has an open cover
  with two open sets, namely $U$ and $C'=C\backslash p$ ($C$ with $p$ removed). 
  Pick a trivialization of $E$ in a small neighborhood $U$ of the point $p$ and restrict $E$  to $C'$.  Let $E_0\to U$ be
  a trivial rank $n$ vector bundle.    We have an open cover of $C$ by open sets $C', U$ with vector bundles $E\to C'$ and $E_0\to U$.
  So we can define a new vector bundle $E'\to C$ by gluing together $E$ and $E_0$ over $U'=C'\cap U$
 via a gauge transformation.   For example, we can use the diagonal gauge transformation from $E_0$ to $E$
  \be\label{gtr} g(z) =\begin{pmatrix} z^{d_1} &&&\cr
                                                                      & z^{d_2} &&\cr
                                                                       &&\ddots & \cr
                                                                         &&& z^{d_n}\end{pmatrix},\ee with integers $d_1,\cdots, d_n$.
For $G=\U(n)$, with suitable choices of the $d_i$,  this gives an example of  a Hecke modification of a $G_\C$-bundle  dual to an arbitrary irreducible finite-dimensional
representation of $G=\U(n)$. 
 For $G=\SU(n)$, one modifies this by requiring $\sum_i d_i=0$ so that $g(z)$ is valued in $\SL(n,\C)$, and for $G=\PSU(n)$, one considers the
$d_i$ to be valued in $\frac{1}{n}\Z$, with $d_i-d_j\in \Z$ (this is equivalent to saying that if $g(z)$ is written in a representation of $\PSU(n)$, then only
integer powers of $z$ appear).     A constant shift of all $d_i$ by $d_i\to d_i+c$, where $c\in \Z$ (or $c\in \frac{1}{n}\Z$ for $G=\PSU(n)$) does not affect the space
of Hecke modifications, since one can compensate for it by $E'\to E'\otimes \O(p)^c$.

For any $G$, one can make a similar construction replacing diagonal matrices whose entries are powers of $z$ with a homomorphism $g(z)= z^\m :\C^*\to T_\C$ 
where $\C^*$ is the punctured $z$-plane, $T_\C$ is a complex maximal torus of $G_\C$, and $\m$, which generalizes the $d$-plet of integers $(d_1,\cdots, d_n)$
in the previous paragraph, is an integral weight of the dual group $G$, corresponding physically to the magnetic charge of an 't Hooft operator.

What we have described so far is a standard example of a Hecke transformation dual to an arbitrary finite-dimensional representation $R$ of $G$; $R$ is encoded
in the integers $d_i$ or the choice of $\m$.   
This construction depended  on the initial choice of a trivialization of $E$ over $U$.  By varying the choice of trivialization, one can
  obtain a whole space of Hecke modifications of the same type.  Once we pick a  reference trivialization, any other trivialization of $E$ 
  over the set $U$ would be obtained from the reference
  one by applying
 some gauge transformation $g'(z)$ that is holomorphic in $U$. The standard Hecke modification associated to this alternative trivialization of $E$ is described in the reference trivialization by the modified singular gauge transformation $g'(z) z^\m$. 
 
Two singular gauge transformations lead to the same $E'$ if they can be related by composition from the right with a gauge transformation $g''(z)$ defined on $U$. Two trivializations of $E$ will thus give the 
same standard Hecke modification if the corresponding gauge transformations $g_1'(z)$ and $g_2'(z)$ satisfy 
\begin{equation}
g_1'(z)z^\m = g_2'(z)z^\m g''(z)
\end{equation}
for some $g''(z)$. 

These relations can be formalized with the help of the {\it affine Grassmannian} $\Gr_{G_\C}$. It is customary to denote as $G_\C[{\cal K}]$ the (infinite-dimensional) space of $G_\C$-valued  gauge transformations  defined on $U'$ and as $G_\C[{\cal O}]\subset G_\C[{\cal K}]$ the subspace of such gauge transformations which extend holomorphically to $U$. The affine Grassmannian 
\begin{equation}
\Gr_{G_\C} \equiv G_\C[{\cal K}]/G_\C[{\cal O}]
\end{equation}
parameterizes equivalent singular gauge transformations.  We define $\Gr^\m_{G_\C}$ to be the orbit in $\Gr_{G_\C}$ of the standard
Hecke modification by the gluing function $z^\m$:
\begin{equation}
\Gr^\m_{G_\C} \equiv \left[G_\C[{\cal O}] z^\m \right].
\end{equation}
Every point in $\Gr^\m_{G_\C}$ is on such an orbit, for some $\m$.

For generic choices of $G$ and $\m$, one runs into a phenomenon of ``monopole 
bubbling'' in which a downward jump can occur in the magnetic charge of an  't Hooft or Hecke operator
(this was introduced in \cite{Kron,Pauly}; see Section 10.2 of \cite{KW} for a short introduction).   
Essentially, the orbit $\Gr^\m_{G_\C}$ is not closed in the affine Grassmannian, and the (possibly singular) closure of the orbit
includes other orbits $\Gr^{\m'}_{G_\C}$ with smaller charge. This can considerably complicate the analysis.

For $G=\U(n)$ or a related group $\SU(n)$ or $\mathrm{PSU}(n)$, the condition to avoid monopole bubbling is that $|d_i-d_j|\leq 1$ for all $i,j$.
Up to a constant shift of all the $d_i$ (which does not affect the space of Hecke modifications, as explained in the discussion of eqn. (\ref{gtr})), 
to avoid monopole bubbling we can assume that $k$ of the $d_i$ are $-1$ and the others 0.   
Explicitly, this corresponds to a Hecke modification of the following sort.
For $E\to C$ a rank $n$ holomorphic vector bundle and $p\in C$, one chooses a $k$-dimensional subspace $V\subset E_p$ and defines a new vector
bundle $E'\to C$ whose sections are sections of $E$ that are allowed to have a simple pole at $p$ with residue in $V$.   The space of Hecke
modifications of this type is parametrized by the choice of $V$, that is, by the Grassmannian of $k$-dimensional subspaces of $E_p$.
Such Hecke modifications are dual to the $k^{th}$ antisymmetric power of the fundamental representation of $G=\U(n)$ or $\SU(n)$.   
In this situation,  $\varphi$ will act as a $k \times k$ matrix on the polar part of $w$. The characteristic polynomial of the $k \times k$ restriction of $\varphi$ 
takes the general form $u(x) = x^k + \delta_1 x^{k-1} + \cdots \delta_k$, for some holomorphic functions $\delta_1,\cdots,\delta_k$.   These functions
 satisfy polynomial relationships with the Hitchin Hamiltonians $\H_{\P,\alpha}(\varphi)$ which encode the 
constraint that $u(x)$ divides the characteristic polynomial of $\varphi$.

\section{Quantum-Deformed WKB Condition}\label{wkb}

\subsection{Quantum States And The Hitchin Fibration}\label{hitchinwkb}
 \begin{figure}
 \begin{center}
   \includegraphics[width=4.7in]{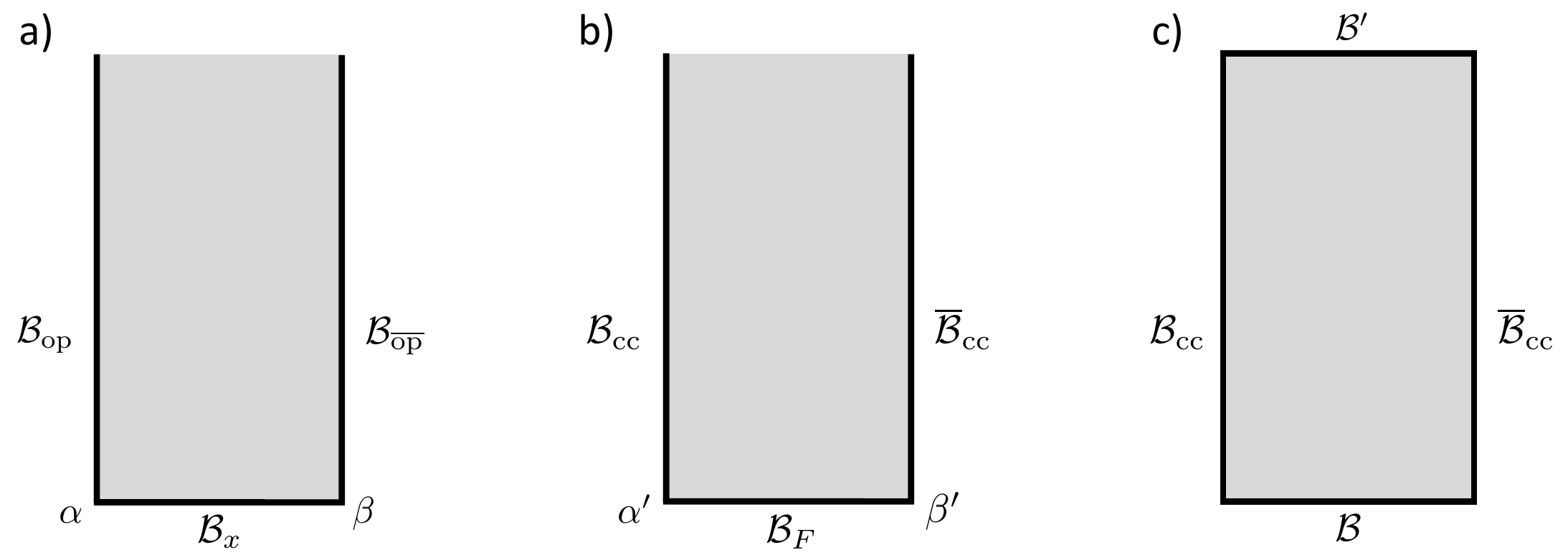}
 \end{center}
\caption{\small  (a) A  $B$-brane $\B_x$ supported at a point $x$, defining (with the help of
some data at the corners) a state
in $\H=\Hom(\B_{\bop},\B_{\op})$. (b) A dual $A$-brane $\B_F$ supported on a fiber $F$ of the
Hitchin fibration, defining a state in $\H=\Hom(\bar\B_\cc,\B_\cc)$.  (c) A pairing between states associated
to $A$-branes $\B$ and $\B'$ (together with data at corners).   If at least one of $\B$ and $\B'$ has compact support,
this pairing is well-defined; otherwise, it may not be.   \label{example6}}
\end{figure} 

A slightly different way to think about an eigenfunction of the Wilson operators and the Hitchin Hamiltonians is as follows.  Let $x$ be a point in
$\M_H(G^\vee,C)$ corresponding to a flat $G^\vee_\C$ bundle $E^\vee_x\to C$,
and let $\B_x$ be a $B$-brane supported at $x$, with a rank 1 (and inevitably trivial) $\CP$ bundle.   Then the part of $\Hom(\B_x,\B_\op)$ of degree  zero\footnote{The $B$-model has a conserved fermion number symmetry, with the differential $Q$ having fermion number or degree 1. When $x\in L_\op$,
$\Hom(\B_x,\B_\op)$ is also nonzero in positive degrees, but the positive degree states do not contribute in this discussion because $\Hom(\B_\bop,\B_\op)$ is
entirely in degree 0.   A similar remark applies later when we discuss the $A$-model.}
is a copy of $\C$ if $x\in L_\op$, that is if $E^\vee_x$  is a holomorphic oper.   Otherwise $\Hom(\B_x,\B_\op)=0$.
Similarly, the degree 0 part of $\Hom(\B_\bop,\B_x)$ is $\C$ if $x\in L_\bop$, that is if $E^\vee_x$  is an antiholomorphic oper, and otherwise zero.   If and only
if $E^\vee$ is both an oper and an anti-oper, we can pick nonzero elements $\alpha\in \Hom(\B_x,\B_\op)$, $\beta\in \Hom(\B_\bop,\B_x)$,
and then define the element $\alpha\circ\beta \in \H=\Hom(\B_\bop,\B_\op)$.   A picture representing this situation is fig. \ref{example6}(a). 
The brane $\B_x$ is used to provide a boundary condition at the bottom of the strip; $\alpha$ and $\beta$ provide the ``corner data'' needed to
define boundary conditions at the corners of the picture.   The path integral with the ``initial conditions'' set by $\B_x,\alpha,\beta$ defines a physical state of the system.
This state is the desired eigenstate of the Hitchin Hamiltonians and the Wilson operators.

Via electric-magnetic duality, we can get a dual picture.    After compactification on an oriented two-manifold $C$, electric-magnetic duality of
$\N=4$ super Yang-Mills theory reduces at low energies to a mirror symmetry of the Higgs bundle moduli space \cite{BJSV,JWS}.  As explained in
\cite{HT}, this is a rare instance in which the SYZ interpretation \cite{SYZ} of mirror symmetry as $T$-duality on the fibers of a family of Lagrangian tori can
be made very explicit.   The Hitchin fibration is the map that takes a Higgs bundle $(E,\varphi)$ to the characteristic poiynomial of $\varphi$.
The fibers of the map are abelian varieties that are
complex Lagrangian submanifolds in complex structure $I$.  This in particular means that they are Lagrangian submanifolds
from the point of view of the real symplectic structure $\omega_K=\Im\,\Omega_I$ of the Higgs bundle moduli space.   Hence, in the $A$-model
with symplectic structure $\omega_K$, the Hitchin fibration can be viewed as an SYZ fibration by Lagrangian submanifolds that generically are tori.
$T$-duality on the fibers of this fibration maps the $A$-model of symplectic structure $\omega_K$ to the $B$-model of complex structure $J$.   
The relation of electric-magnetic duality to this instance of mirror symmetry was an important input in \cite{KW}.   

In particular, mirror symmetry in this situation maps a rank 1 brane $\B_x$ supported at a point $x$ to a brane $\B_F$ supported on a fiber $F$ of the
Hitchin fibration, with a $\CP$ bundle that is a flat line bundle\footnote{\label{canon} In general, the $\CP$ bundle of a rank 1 brane is more canonically a $\Spinc$
structure rather than a line bundle.   In the present context, as $F$ is an abelian variety and so has a canonical spin structure, the distinction is  not important.}
$\S\to F$.  The duals of $\alpha\in \Hom(\B_x,\B_\op)$ and $\beta\in\Hom(\B_\bop,\B_x)$ are elements $\alpha'\in \Hom(\B_F,\B_\cc)$ and $\beta'\in \Hom(
\b\B_\cc,\B_F)$.   The element $\alpha'\circ\beta'\in \H$, which corresponds to the picture of fig. \ref{example6}(b), represents an element of $\H$ in the magnetic
description, in which the Hitchin Hamiltonians are differential operators acting on half-densities on $\M(G,C)$ and Wilson operators are replaced by
't Hooft operators.   

The branes $\B_x$ and $\B_F$ both have compact support, consisting of either a point $x$ or a fiber $F$ of the Hitchin fibration.      Compact support
makes it manifest that the states created by these branes (plus corner data)  are normalizable.   More than that,  compact support
means that these states have well-defined pairings with states that are constructed similarly using  an arbitrary brane $\B'$, possibly
with noncompact support, again with suitable corner data.   In other words, the pairing constructed from the rectangle of fig. \ref{example6}(c) is always
well-defined, regardless of the brane at the top of the rectangle, as long as the brane at the bottom of the rectangle has compact support.   This means roughly that an arbitrary brane, with a choice of corner data,   defines a distributional state, not
necessarily a normalizable vector in the Hilbert space $\H$,
 while a brane of compact support defines a vector  that can be paired with any distribution.  
 The Hilbert space is contained in the space of distributional states, and contains a subspace, roughly analogous to a Schwartz space, spanned by states
 associated to branes of compact support.
The eigenfunctions of the Hitchin Hamiltonians lie in this subspace.

The $B$-model gives a clear answer to the question of which  points $x\in \M_H(G^\vee,C)$ are associated to physical states: these are points in 
$\Upsilon=L_\op\cap L_\bop$.   
It is more difficult to extract directly from the $A$-model a prediction
for which pairs $F,\S$ are similarly associated to physical states.   We will only be able to get a sort of semiclassical answer, which we expect to be valid
asymptotically, in a sense that will be explained.  

We recall that the $\CP$ bundles of the branes $\B_\cc$ and $\b\B_\cc$ are dual line bundles $\L$ and $\L^{-1}$, and that the prequantum line bundle
of  $\M_H(G,C)$, in the sense of geometric quantization, is $\frL=\L^2$.    We want the condition on $F$ and $\S\to F$ such that corners
$\alpha'\in \Hom(\B_F,\B_\cc)$ and $\beta'\in \Hom(\b\B_\cc,\B_F)$ exist.    

We will describe the condition for $\alpha'$ and $\beta'$ to exist to lowest order in $\sigma$-model perturbation theory, and we will also explain to what
extent higher order corrections can or cannot change the picture, in the regime where they are small.   So let us first explain the regime in which $\sigma$-model
perturbation theory is valid.   This is the case that the Higgs field $\varphi$ is parametrically large and far away from the discriminant locus (on which
the spectral curve becomes singular).   Concretely if $(E,\varphi)$ is any Higgs bundle with a smooth spectral curve, and we rescale $\varphi$ by
a large factor 
\be\label{kcf} \varphi\to t\varphi,~~|t|\gg 1,\ee
with any fixed value of $\mathrm{Arg}\,t$,  then $\sigma$-model perturbation theory becomes valid.   For $t\to\infty$, the Higgs bundle
moduli space has  a concrete ``semi-flat'' description \cite{GMN}, leading to semiclassical results and asymptotic expansions as $t\to\infty$ for a variety of questions, including what we will consider here.   We will call the region $t\to\infty$ the WKB limit, for reasons that will emerge.

The brane $\B_F$ is of type  $(B,A,A)$.  For such a brane, with support $F$ and $\CP$ bundle $\S$, the leading $\sigma$-model approximation to
$\Hom(\B_F,\B_\cc)$, where $\B_\cc$ has $\CP$ bundle $\L$, is the $\bar\partial$ cohomology of $F$ with values in\footnote{As remarked in footnote \ref{canon}, $\S$ and likewise
$\S^{-1}$ is canonically  a $\Spin_c$ structure on $F$, not a line bundle.   Likewise, $K_F^{1/2}$ is not canonically defined as a line bundle (and for general may not
exist at all as a line bundle) but on any complex manifold, $K_F^{1/2}$ is canonically defined as a $\Spin_c$ structure.   Hence the product $\S^{-1}\otimes
K_F^{1/2}$ exists canonically as an ordinary line bundle and the answer stated in the text makes sense.  In our application,
 $K_F$ is trivial, and we can likewise take $K_F^{1/2}$ to be trivial and define $\S$ as a line bundle.}  $\L|_F\otimes \S^{-1}\otimes K_F^{1/2}$,
where $K_F^{1/2}$ is a square root of the canonical bundle of $F$  (see Appendix B of \cite{GW}).   
This comes about because the leading $\sigma$-model approximation to the $A$-model
differential $Q$ is
\be\label{jno}Q\sim \bar\partial+a , \ee
where $a$ is a $(0,1)$-form on $F$ that defines the complex structure of the line bundle $\L\otimes \S^{-1}\otimes K_F^{1/2}$.   
The line bundle $\L$ is flat when restricted to $F$, because $F$ is a complex Lagrangian submanifold.  The line bundle $\S$ is flat, because it is the
$\CP$ bundle of a Lagrangian brane.   And as $F$ is a complex torus, we can take $K_F^{1/2}$ to be trivial and omit this factor.   The $\bar\partial$ cohomology
of a complex torus with values in a flat line bundle $\L|_F\otimes \S^{-1}$ vanishes unless this line bundle is trivial.   So in the leading $\sigma$-model approximation,
$\Hom(\B_F,\B_\cc)$ vanishes unless we pick $\S\cong \L|_F$.  For $\S=\L|_F$,we can pick a nonzero $\alpha' \in H^0(F,\L|_F\otimes \S^{-1})\cong \C$.

Similarly, as the $\CP$ bundle of $\B_\bop$ is $\L^{-1}$,
the leading $\sigma$-model approximation to $\Hom(\B_\bop,\B_F)$ is the $\bar\partial$ cohomology of $F$ with values in $\S\otimes \L|_F\otimes K_F^{1/2}$, or,
taking $K_F^{1/2}$ to be trivial, just $\S\otimes \L|_F$.
This cohomology vanishes if $\S\otimes \L|_F$ is nontrivial; if it is trivial, which is so precisely if $\S=\L^{-1}|_F$, we can choose a nonzero $\beta'\in H^0(F, \S\otimes\L|_F)\cong \C$.

In short, we can use the brane $\B_F$ with suitable corner data to define a state in $\H$ if and only if we can choose $\S$ to be isomorphic
to both $\L|_F$ and $\L^{-1}|_F$.    In other words, the condition is that $\L^2|_F$ must be trivial.   As the prequantum line bundle over  $\M_H(G,C)$
is $\frL=\L^2$, the condition is that $\frL$ must be trivial when restricted to $F$.

This is actually the WKB condition of elementary quantum mechanics, which also is part of  the theory of geometric quantization.   To put the condition
in a more familiar form, recall that the symplectic form $\omega_K$ of $\M_H(G,C)$ is cohomologically trivial, so it can be written as $\omega_K=\d\lambda_K$
for a 1-form $\lambda_K$.  The prequantum line bundle $\frL$ is supposed to be a unitary line bundle with a connection of curvature $\omega_K$.
We can take $\frL$ to be a trivial line bundle with the connection $D=\d+\i\lambda_K$.   $\frL$ is flat when restricted to $F$ because $F$ is a Lagrangian
submanifold.   The condition that $\frL$ is trivial is that its global holonomies vanish.   In other words, the condition is that if $\gamma\subset F$ is a 1-cycle,
then the holonomy of $\frL$ around $\gamma$ must vanish:   $\exp(\i \oint_\gamma\lambda_K)=1$ or in other words $\oint_\gamma \lambda_K\in 2\pi\Z$.
To put this condition in a perhaps more familiar form, we can approximate $\M_H(G,C)$ as a cotangent bundle $T^*\M(G,C)$
 and choose $\lambda_K=\sum_i p_i\d q^i$, where $q^i$ are coordinates on the base of the cotangent bundle and $p_i$ are fiber coordinates.  Then
 the condition is that
 \be\label{weffo}\sum_i\oint_\gamma p_i \d q^i\in 2\pi \Z,\ee
 which may be recognizable as the WKB condition for associating a quantum state to the Lagrangian submanifold $F$.   
 
 We have reached this conclusion to lowest order in $\sigma$-model (or gauge theory) perturbation theory, and we do not claim that the result is exact.
 However, it is possible to argue that, at least sufficiently near the WKB limit, there is a quantum-corrected WKB condition that leads to qualitatively
 similar results.   Consider correcting the
 computation of $\Hom(\b\B_\cc,\B_F)$ or $\Hom(\B_F,\B_\cc)$ in perturbation theory in inverse powers of the parameter $t$ that was introduced in
 eqn. (\ref{kcf}).   This has the effect of shifting the $(0,1)$-form $a$ by a $(0,1)$-form $c_1/t+c_2/t^2+\cdots$, leading to
 \be\label{zeflo}Q=\bar\partial+a +\frac{c_1}{t}+\frac{c_2}{t^2}+\cdots .\ee
 Whatever the $c_k$ are, we can compensate for them by shifting $a$.   An arbitrary shift in $a$ can be interpreted as the sum of a $\bar\partial$-exact
 term, which does not affect the cohomology of $Q$, plus a term that can be interpreted as resulting from a shift in the line bundle $\S$.   Thus, instead of needing
 $\L|_F\otimes \S^{-1}$ and $\L|_F\otimes \S$ to be trivial in order to associate a quantum state with $F$, we need $\S$ to satisfy conditions
 that are asymptotically close to these.    Correspondingly, the classical WKB condition for $\frL|_F$ to be trivial is modified, at least for
 sufficiently large $t$, to a quantum WKB condition that determines which fibers of the Hitchin fibration are associated to quantum states.

\subsection{WKB Condition and Special Geometry}
In order to better understand the quantization condition, it is useful to recall some facts about the special geometry which 
governs the structure of $\M_H(G,C)$, as well as complex integrable systems that arise in Seiberg-Witten theory \cite{Seiberg:1994rs,GKMMM,DoW,Freed:1997dp, Gaiotto:2008cd}.
The basic ingredients of the geometry are 
\begin{itemize}
\item The base ${\cB}$ of the Hitchin fibration, of complex dimension $r$. We denote a point in the base as $u$ and the discriminant locus as ${\cal D}$.
\item A local system $\Gamma$ of lattices of rank $2r$ defined over ${\cal B}\backslash {\cal D}$ ($\cal B$ with $\cal D$ removed), equipped with a symplectic form $\langle\cdot,\cdot \rangle$.\footnote{In a true Seiberg-Witten geometry the symplectic form is integer-valued and $\Gamma$ is self-dual. The Seiberg-Witten geometry is self-mirror. Hitchin systems are related to Seiberg-Witten geometries by discrete orbifold operations which relax these conditions.} We will denote a charge (an
element of $\Gamma$) as $\gamma$. 
\item A collection of central charges $Z: \Gamma \to \C$ which vary holomorphically on ${\cal B}\backslash{\cal D}$. We will denote the central charge evaluated on a charge $\gamma$ as $Z_\gamma$. The $Z_\gamma$ are also identified with periods of the canonical 1-form on the spectral curve of the Hitchin system. 
\item Real angular coordinates $\theta: \Gamma \to S^1$ on the fibers of the complex integrable system. We will denote the coordinates evaluated on a charge $\gamma$ as $\theta_\gamma$. They are dual to the $Z_\gamma$ under the complex Poisson bracket 
\begin{equation}
\{Z_\gamma, \theta_{\gamma'} \} = \langle \gamma, \gamma' \rangle
\end{equation}
\end{itemize}

The complex symplectic form in complex structure $I$ is defined with the help of the inverse pairing:
\begin{equation}
\Omega = \langle \d Z, \d\theta \rangle
\end{equation}
Correspondingly, we have a 1-form 
\begin{equation}
\lambda  = \langle Z, \d\theta \rangle\equiv \lambda_J + \i \lambda_K
\end{equation}
satisfying $\d\lambda=\Omega$, with periods $2 \pi Z_\gamma$. 

We will use this special coordinate system for $\M_H(G^\vee,C)$. The parameters $u$ specify a fiber $F$ of the Hitchin fibration, which is an abelian
variety, and the angles $\theta_\gamma$ parameterize 
the choice of a flat line bundle $\S\to F$. The WKB conditions for the existence of corners are thus that $\theta_\gamma = \pm \mathrm{Im}\,Z_\gamma$ 
and the quantization condition becomes $\mathrm{Im}\, Z_\gamma \in \pi \mathbb{Z}$.

The $\B_F$ branes as $A$-branes are supposed to depend on the data $(u, \theta)$ holomorphically in complex structure $J$, 
as the $A_K$ twist on $\M_H(G,C)$ is mirror to the $B_J$ twist on $\M_H(G^\vee,C)$. Writing functions of $(u, \theta)$ which are holomorphic in complex structure $J$ is essentially as challenging as computing the hyper-K\"ahler metric on the moduli space. In the WKB region, though, 
the functions 
\begin{equation} \label{eq:semiflat}
X_\gamma = \exp \left(\mathrm{Re}\,Z_\gamma + i \theta_\gamma \right)
\end{equation}
 are an excellent ``semiflat'' approximation to $J$-holomorphic functions. The Cauchy-Riemann equations fail by corrections suppressed exponentially in the WKB region \cite{Gaiotto:2008cd}.\footnote{One can define locally some corrected $X_\gamma$ which are truly $J$-holomorphic, but non-trivial ``wall crossing'' coordinate transformations are required in different patches
 \cite{GMN}. This subtlety will not be important here. These statements  have a transparent $A$-model interpretation. The semiclassical $A$-brane  moduli combine the $\CP$ data with the deformation data associated to the same 1-forms on the brane support to give the $(\C^*)^{2r}$ coordinates $X_\gamma$. The only corrections are non-perturbative and due to disk instantons ending on a cycle $\gamma$. These only exist at codimension one loci where $Z_\gamma$ is real 
and lead to the wall-crossing transformations. } The complex symplectic form in complex structure $J$ is approximately 
\begin{equation}
\Omega_J = \langle \d \log X, \d \log X \rangle
\end{equation}

Stated in this language, the WKB condition for the existence of a corner becomes
\begin{equation} \label{eq:wkb}
X_\gamma = e^{Z_\gamma(u)}
\end{equation}
This can be interpreted as the parametric definition of an $r$-dimensional complex Lagrangian submanifold, as expected. 
Surprisingly, the parameters coincide with the coordinates $u$ on ${\cal B}\backslash{\cal D}$ and thus we get an approximate holomorphic identification 
between ${\cal B}\backslash{\cal D}$ and the space of branes equipped with a corner. 

This identification is only valid in a semiclassical approximation.  To make such an expansion, as in Section \ref{hitchinwkb}, we replace $u$ with $tu$, and
take $t$ to be large.   The corner condition receives perturbative corrections, which 
will take a systematic form 
 \begin{equation}
X_\gamma = e^{t Z_\gamma(\tilde u) + t^{-1} c_{1,\gamma}(\tilde u) + \cdots}
\end{equation}
We introduced parameters $\tilde u$  that cannot be taken to coincide with 
the  coordinates $u$ that are holomorphic in complex structure $I$. The relation between the two follows from the comparison of (\ref{eq:wkb}) and (\ref{eq:semiflat}):
\begin{equation}
\mathrm{Re}\,Z_\gamma(u) = \mathrm{Re}\,Z_\gamma(\tilde u) + t^{-2} \mathrm{Re}\,c_{1,\gamma}(\tilde u) + \cdots
\end{equation}
The perturbatively-corrected quantization condition to have both types of corners will become 
\begin{equation}
\mathrm{Im}\,Z_\gamma(\tilde u) + t^{-2} \mathrm{Im}\,c_{1,\gamma}(\tilde u) + \cdots \in \pi \mathbb{Z}.
\end{equation}
 
\subsection{Match with real WKB opers}
There is another natural occurrence of the periods $Z_\gamma$ of the canonical 1-form on the spectral curve. If we attempt a WKB analysis on $C$ of the oper differential equation, the monodromy data of the oper flat connection will be computed at the leading order in terms of the exponentiated periods $e^{Z_\gamma(\tilde u) }$ \cite{Voros1,Voros2,Voros3,Voros4,Gaiotto:2014bza}. Here we employed a non-canonical identification between $L_{\op}$ and ${\B}$. For example, for $G^\vee = \SU(2)$ we would write the oper differential operator as a reference operator deformed by a large quadratic differential $U(z;\tilde u)$ 
\begin{equation}
\partial^2_z - t_2(z) \equiv \partial^2 - t^2 U(z;\tilde u)- t_2^{0}(z) 
\end{equation}
and the leading WKB approximation would involve the periods $Z_\gamma(\tilde u)$ of the WKB 1-form $\sqrt{U(z;\tilde u)} \d z$. Here $t_2(z)$ is the classical stress tensor, $t_2^0(z)$ a reference choice of classical stress tensor and $t$ is the scaling parameter.

The WKB calculation is really a combination of a topological and an analytic problem. The topological problem involves a careful Stokes analysis of the asymptotic behaviour of the flat sections of the connection. The analytic problem involves the computation of the Voros symbols, which are periods of the all-orders WKB 1-form. Remarkably, the Voros symbols precisely compute the corrected $X_\gamma$ coordinates of the oper in the space of flat $G^\vee$ connections \cite{Gaiotto:2014bza}:
\begin{equation}
X_\gamma = e^{t Z_\gamma(\tilde u) + t^{-1} c_{1,\gamma}(\tilde u)  + \cdots}
\end{equation}
The WKB analysis  of the oper differential equation thus gives directly the $B$-model analogue of the WKB corner condition in the mirror $A$ model. 

\section{Real Bundles}\label{realb}

\subsection{The Setup}\label{setup}
So far we have studied the quantization of $\M_H(G,C)$ as a real symplectic manifold.   
An alternative is to view $\M_H(G,C) $ as a complex symplectic manifold with complex structure $I$ and holomorphic symplectic structure 
$\Omega_I=\omega_J+\i \omega_K$, and quantize a real symplectic submanifold $\M_H^\R \subset \M_H$.    For this, as explained in Section \ref{background}, we look for an antiholomorphic involution $\tau$ of $\M_H(G,C)$ 
that satisfies $\tau^*\Omega_I=\bar\Omega_I$.   In our application, $\M_H^\R$ will be a cotangent bundle, leading to a simple description of the Hilbert space
and of the action of the Hecke operators, and  a real integrable system, equipped with a real version of the Hitchin fibration.

The classification of the possible antiholomorphic involutions is the same whether one considers holomorphic $G_\C$ bundles as in \cite{Hurtu,Hurtu2}
or Higgs bundles, as in 
 \cite{BaSc,BaSc2,BG}.   Those references provide much more detail than we will explain here.   We also note that 
 the three-manifold $U_\tau=(\Sigma\times \hat I)/\{1,\tau\}$ that we will use in studying the duality was  introduced in Section 11 of \cite{BaSc}.  That paper also contains a duality proposal based on the structure of $\M_H^\R(G,C)$ as a real integrable system.   

A suitable involution of $\M_H$ can be constructed starting with an antiholomorphic involution $\tau$ of the Riemann surface $C$, which exists
for suitable choices of $C$.    An antiholomorphic map reverses
the orientation of $C$; conversely, if $\tau$ is an orientation-reversing smooth involution of an oriented two-manifold $C$, then one can pick a complex
structure on $C$ such that $\tau$ acts antiholomorphically.\footnote{Pick any Riemannian metric $\mathrm g$ on $C$.   Then $\mathrm g'
=({\mathrm g}+\tau^*({\mathrm g}))/2$ is  a $\tau$-invariant metric, and $\tau$ acts antiholomorphically in the complex structure determined by $\mathrm g'$.}
Topologically, the  possible choices of $\tau$ can be classified as follows.  If $C$ has genus $g$, then its Euler characteristic
is $2-2g$ and the quotient $C'=C/\{1,\tau\}$ will have Euler characteristic $1-g$.  $C'$ can be any possibly unorientable two-manifold, possibly with boundary,
of Euler characteristic $1-g$.  The boundary of $C'$ comes from the fixed points of $\tau:C\to C$.  These fixed points make up a certain number of circles;
any integer number of circles from $0$ to $g+1$ is possible.
For example, if $g=0$, $C'$ can be a disc or $\Bbb{RP}^2$; if $g=1$, $C'$ can be a cylinder, a Mobius strip, or a Klein bottle. 
Having fixed $\tau$, we then choose the topological type of a lift of $\tau$ to act on a smooth $G$-bundle $E\to C$.   We choose the lift
to preserve $\tau^2=1$.   If $s$ is a fixed point of $\tau:C\to C$,  then $\tau$ acts on the fiber $E_s$ of $E$ over $s$
by an automorphism $x_s$ of $G$.   Here $x_s$ can be an inner automorphism, that is, conjugation by an element $g_s$ of $G$,
but more generally, if $G$ has non-trivial outer automorphisms, $x_s$ may be an outer automorphism.
The conjugacy class of $x_s$ is constant on each fixed circle $S$, and we denote it as $x_S$.
  The condition  $\tau^2=1$ places a condition on $x_S^2$.
Since a Higgs bundle is defined by fields $(A,\phi)$ that are adjoint-valued, purely to define an involution of the Higgs bundle moduli space
$\M_H(G,C)$, we require only that $x_S^2=1$ in the adjoint form of $G$; as an element of $G$, $x_S^2$ might be a central element not equal to 1.
The precise condition that should be imposed on $x_S^2$ depends on topological subtleties that were reviewed in Section \ref{toposubt}; we will return to this point in Section \ref{nonfree}.   
If $\tau$ acts freely and $G$ is not simply-connected, other issues
come into play in lifting $\tau$ to act on $E$. For example, for $G=\SO(3)$, a lift of $\tau$ to act on $E$ only exists if $\int_C w_2(E)=0$.   
In that case, there are two topologically
inequivalent lifts, parametrized by $\int_{C'}w_2(E')$, where $E'\to C'$ is the quotient of $E\to C$ by $\{1,\tau\}$.
  
 A Riemann surface $C$ endowed with 
antiholomorphic involution $\tau$ can be viewed as an algebraic curve  defined over $\R$.   The boundary points of $C'$, if any,
correspond to the real points of $C$ over $\R$. Once it has been lifted to act on the smooth bundle $E$,
  $\tau$ also acts on $\M(G,C)$, the moduli space of flat connections on $E$, and on the
corresponding Higgs bundle moduli space $\M_H(G,C)$.  Whether or not there are fixed points in the action of $\tau$ on $C$, there always
are fixed points in the action of $\tau$ on\footnote{If $A$ is any connection on $E$, then $\frac{1}{2}(A+\tau^*(A))$ is a $\tau$-invariant connection.
Generically, the $(0,1)$ part of this connection defines a stable $G_\C$-bundle, which corresponds to a $\tau$-invariant point in $\M(G,C)$.   The
same connection with zero Higgs field defines a $\tau$-invariant point in $\M_H(G,C)$.}
$\M(G,C)$ and on $\M_H(G,C)$.    Each component of the fixed point set of $\tau$ on $\M(G,C)$ or $\M_H(G,C)$ is middle-dimensional.   This
is a general property of antiholomorphic involutions of complex manifolds.   
 
 We will write $\M^\R(G,C)$ or $\M_H^\R(G,C)$ for a component of the fixed point set of $\tau$ acting on $\M(G,C)$ or $\M_H(G,C)$, respectively.
 The components in general are classified by additional data that we have not introduced so far.  In particular, viewing $\M(G,C)$ or $\M_H(G,C)$
 as moduli spaces of flat $G$-valued  or $G_\C$-valued connections, 
  the monodromy $h_S$ of a flat bundle around $S$ will be invariant under $x_S$,
  so it will lie in the $x_S$-invariant subgroup $H_S$ or $H_{S,\C}$ of $G$ or $G_\C$.   $H_S$ and $H_{S,\C}$ are
   connected if $G$ is simply-connected, but in general not otherwise
  (for example, if $G=\SO(3)$ and $g_S=\mathrm{diag}(1,-1,-1)$, then $H_S$ contains the component of the identity and another component
  that contains the element $\mathrm{diag}(-1,-1,1)$).   So in general, to specify a component of the fixed point sets requires an additional choice for each $S$.
  
  Associated to each fixed circle $S$ is a real form $G_{\R,S}$ of the complex Lie group $G_\C$.   Writing $g\to \bar g$ for the antiholomorphic
  involution of $G_\C$ that leaves fixed the compact form $G$,  $G_{\R,S}$ is defined by the condition
  \be\label{ralform} g=x_S(\bar g). \ee
  In case $x_S$ is conjugation by $g_S\in G$, the condition becomes 
  \be\label{realform} g=g_S \bar g g_S^{-1}.\ee

$\M_H^\R(G,C)$, with symplectic structure $\omega_J$, is the real symplectic manifold  that we want to quantize.  The definition of  $\M_H^\R(G,C)$ has
depended on various choices which we are not indicating in the notation, but regardless of those choices, $\M_H^\R(G,C)$ has  $\M_H(G,C)$  as a complexification
and this complexification has the appropriate properties for brane quantization of $\M_H^\R(G,C)$.
$\M_H^\R(G,C)$ has a real polarization that is defined
as follows: a leaf of the polarization consists of $\tau$-invariant Higgs pairs $(A,\phi)$ with fixed $A$, but varying $\phi$.   The same condition, without
the requirement of $\tau$-invariance, defines a holomorphic polarization of the complexification $\M_H(G,C)$ of $\M^\R_H(G,C)$.  Hence brane quantization
of $\M^\R(G,C)$ is equivalent to its quantization using this real polarization.   
Concretely, in this real polarization, $\M_H^\R(G,C)$ can be approximated by the cotangent bundle $T^*\M^\R(G,C)$
for the same reason that such a statement holds for the full Higgs bundle moduli space: if $(A,\phi)$ is a $\tau$-invariant
Higgs bundle representing a point in $\M_H^\R(G,C)$, then generically $\phi$ represents a cotangent vector to $\M^\R(G,C)$ at the point in $\M^\R(G,C)$
corresponding to $A$.   
Therefore, the Hilbert space $\H_\tau$ that arises in quantization of $\M^\R_H(G,C)$ is the space of $\lmark^2$ half-densities on $\M^\R(G,C)$,
the usual answer for geometric quantization of a cotangent bundle.   

We can also restrict the Hitchin fibration to the $\tau$-invariant locus.   $\tau$ acts on the base ${\cB}$ of the Hitchin fibration with a fixed point
set ${\cB}^\R$. Functions on ${\cB}^\R$ are Poisson-commuting (with respect to the real symplectic structure $\omega_J$ of $\M_H^\R(G,C)$).  
So $\M_H^\R(G,C)$ is a  real integrable system.
The generic fiber of the map $\M_H^\R(G,C)\to {\cB}^\R$ is a real torus.   
Baraglia and Schaposnik \cite{BaSc} proposed to define a mirror symmetry between $\M_H^\R(G,C)$ and a similar moduli space with $G$ replaced by
$G^\vee$ by $T$-duality on the fibers of the
real integrable system.   In general, one would expect such a definition to give reliable results at least asymptotically, far from the discriminant locus of the
real integrable system.  At least when $\tau$ acts freely, one can be more precise, as we discuss next.

\subsection{Four-Dimensional Picture And Duality}\label{fourdpic}

To learn something interesting about this construction, one wants to apply duality, and for this, as usual, a four-dimensional picture is helpful.

For a four-dimensional picture, we start with the usual four-manifold $M=\Sigma\times C$, where $\Sigma=\R\times I$ is a strip.   Here $\R$ labels the ``time''
and $I$ is an interval, say the interval $0\leq w\leq 1$.   We then modify   $M$ by imposing an equivalence relation on the right boundary at $w=1$:
we declare that, for $p\in C$, points $(t,1,p)$ and $(t,1,\tau(p))$ are equivalent. (No equivalence is imposed except at $w=1$.)  
Imposing this equivalence relation amounts to requiring that the data at $w=1$ should be $\tau$-invariant (up to a gauge transformation).  So this
is a way to implement in four dimensions what in the two-dimensional picture of  Section \ref{setup} was the brane $\B_\R$ at the right boundary.
We write $M_\tau$ for the quotient of $M$ by this equivalence relation.   It is convenient to factor out the time and define $M_\tau=\R\times U_\tau$.

There is an alternative construction of $U_\tau$ as a quotient by the $\Z_2$ group generated by $\tau$.  For this, we start with a doubled interval
$\h I: 0\leq w\leq 2$.   Then we divide $\h U= \h I\times C$ by the symmetry that acts by $(w,p)\to (2-w,\tau(p))$.   A fundamental domain is the region
$0\leq w\leq 1$, so the quotient is the same as before.  

Simplest is the case that $\tau$ acts freely on $C$.   $U_\tau$ is then actually an orientable manifold,
whose boundary is a single copy of $C$, at $w=0$.  $U_\tau$ is the total space of a real line bundle over $C'=C/\{1,\tau\}$.
$C'$ itself and the real line bundle are both unorientable, but the total space $U_\tau$ is orientable.   

The four-dimensional picture associated to quantization of $\M_H^\R(G,C)$ is just the $A$-model on $M_\tau=\R\times U_\tau$, with $\B_\cc$ boundary
conditions  at $w=0$.   There is no need for an explicit boundary condition at $w=1$, as there is no boundary there.  
What in the two-dimensional description was a Lagrangian brane $\B_\R$ supported on $\M_H^\R(G,C)$ has been absorbed into the geometry of $M_\tau$.

Therefore, without further ado we can describe a dual description.   The dual is just the $B$-model with gauge group $G^\vee$
on the same four-manifold $M_\tau$, but now with
$\B_\op$ boundary conditions at $w=0$.  In a two-dimensional language, the brane $\B^\vee_\R$ dual to $\B_\R$ is supported on $\M_H^\R(G^\vee,C)$.
In particular, it is supported on the same locus ${\cB}_\R$ on the base of the Hitchin fibration.   This is in accord with the duality proposal of
Baraglia and Schaposnik \cite{BaSc}, which appears to be valid at least when $\tau$ acts freely on $C$.

As usual, it is relatively easy to describe the physical states of the $B$-model in quantization on $U_\tau$ and the eigenvalues of the Hitchin Hamiltonians.
The first step is localization of the $B$-model on complex-valued flat connections, which here means $G^\vee_\C$-valued flat connections on $U_\tau$
that satisfy the oper boundary condition.   
To put it differently, the localization is on flat $G_\C$ bundles $E^\vee\to U_\tau$ whose restriction to $C=\partial U_\tau$ is
a holomorphic oper.   Let $\Upsilon_\tau$ be the set of isomorphism classes of such bundles.    
For positivity of the Hilbert space inner product, we expect $\Upsilon_\tau$ to be a discrete set  of nondegenerate points (see
Appendix \ref{bmodel}),
similarly to the analogous set $\Upsilon=L_\op\cap L_\bop$ encountered in the quantization of the full Higgs bundle moduli space.
Assuming this, the Hilbert space $\H_\tau$ has a basis $\psi_u$ labeled by $u\in \Upsilon_\tau$.
The $\psi_u$ are eigenfunctions of the Hitchin Hamiltonians.  The eigenvalue of a Hitchin Hamiltonian $H_{\P,\alpha}$ on $\psi_u$ is the value
of the corresponding function $f_{\P,\alpha}:L_\op\to \C$ at the point in $L_\op$ that corresponds to the boundary values of the flat bundle
associated to $u$.

If an oper bundle over $C=\partial U_\tau$ extends over $U_\tau$ as a flat bundle, then in particular this
implies that  the antihomomorphic involution $\tau:C\to C$ lifts to an action on $E^\vee$
and therefore that $E^\vee$ is an antiholomorphic oper, as well as a holomorphic one.   So the set $\Upsilon_\tau$ 
 has a natural map to  $\Upsilon=L_\op\cap L_\bop$.    However, in general the map from $\Upsilon_\tau$
to $\Upsilon$ is not an embedding, since a $\tau$-invariant flat bundle on the boundary of $U_\tau$ that is a holomorphic oper may have more
than one extension as a flat bundle over the interior of $U_\tau$.   To analyze this situation, note that flat $G^\vee_\C$ bundles over $U_\tau$ correspond
to homomorphisms  $\varrho:\pi_1(U_\tau)\to G^\vee_\C$, up to conjugation.   As $U_\tau$ is contractible to $C'=C/\{1,\tau\}$, we can equally well
consider $\varrho:\pi_1(C')\to G^\vee_\C$.    The group $\pi_1(C')$ has an index 2  subgroup $\pi_{1,+}(C')$  consisting of orientation-preserving loops in $C'$.
Such loops can be deformed to the boundary of $U_\tau$, so once a flat $G^\vee_\C$ bundle is given on the boundary of $U_\tau$, the restriction of the corresponding
homomorphism $\varrho$ to $\pi_{1,+}(C')$ is uniquely determined.   However, there is some freedom in the extension of $\varrho$ to the rest of $\pi_1(C')$.
$\pi_1(C')$ is generated by the index 2 subgroup $\pi_{1,+}(C')$ together with any  orientation-reversing element $\sigma\in \pi_1(C')$.   To complete the description of
$\varrho$, we need to specify $\varrho(\sigma)$.     $\varrho$ is supposed to be a homomorphism, so we require $\varrho(\sigma)^2=\varrho(\sigma^2)$,
where, since $\sigma^2\in \pi_{1,+}(C')$, $\varrho(\sigma^2)$ is determined by the boundary data.    If $\varrho(\sigma^2)$ is a regular element of $G^\vee_\C$,
then there are only finitely many choices for $\varrho(\sigma)$, but  $\varrho(\sigma)$ is not uniquely determined.    In particular, we are always free
to transform $\varrho(\sigma)\to \veps \varrho(\sigma)$, where $\veps$ is an element of order 2 of the center of $G^\vee$.   Thus the subgroup
$\zZ_2(G^\vee)$ of the center of $G^\vee$ consisting of elements of order 2 acts freely on $\Upsilon_\tau$.  (For simple $G^\vee$, $\zZ_2(G^\vee)$ is
1, $\Z_2$, or $\Z_2\times \Z_2$, depending on $G^\vee$.)    
 
 The $A$-model dual of the action of  $\zZ_2(G^\vee)$  is the following.   Topologically, for simple $G$, a $G$-bundle on the three-manifold
 $U_\tau$ is classified by $H^2(U_\tau,\pi_1(G))$.   In the present instance, $U_\tau$ is contractible to $C'=C/\tau$ so $H^2(U_\tau,\pi_1(G))=H^2(C',\pi_1(G))$.
 As $C'$ is unorientable, one has $H^2(C',\pi_1(G))=\pi_{1;2}(G)$, where $\pi_{1;2}(G)$ is the subgroup of $\pi_1(G)$ consisting of elements of
 order 2.    Thus on the $A$-model side, there is a grading of the Hilbert space by $\pi_{1;2}(G)$.  As $\pi_{1;2}(G)=\zZ_2(G^\vee)$, this 
matches the $\zZ_2(G^\vee)$ action in the $B$-model.   
 
 Finally let us discuss the natural line operators in this problem. 
As in Section \ref{wtw}, we could in principle consider Wilson and 't Hooft line operators supported on an arbitrary 1-manifold $\gamma\subset M_\tau$.
But a natural special case is the following.  Pick a point $p\in C$.   In $\h I\times C $, there is a natural path $\h\gamma_p$ between the points $0\times p$ and $2\times p$
on the left and right boundaries: we simply set $\h\gamma_p = \h I\times p$.    Upon dividing by the group generated by $\tau$, 
$\h\gamma_p$ descends to a path  $\gamma_p\subset
U_\tau$,  between the boundary points $0\times p$ and $0\times \tau(p)$.  In a purely two-dimensional description, one would have needed endpoint or
corner data associated to the real brane ${\B}_\R$, but in the four-dimensional description, this is not needed as $\gamma$ has no endpoint at $w=1$.

As in Section \ref{wilop}, we can choose an arbitrary representation $R$ of $G^\vee$
and consider the holonomy $W_{R,\gamma}$
of the bundle $E^\vee_R=E^\vee\times_{G^\vee_\C}R$ on the curve $\gamma_p$.   However, to turn this holonomy into a quantum operator $ W_{R,p}$,
we need, as before, to supply  endpoint data. 
For example, if we choose canonical endpoints $s_{R,n}$ at $0\times p$ and 
$s_{\bar R, m}$ at $0\times \tau(p)$, the Wilson operator will take the form of the inner product between $s_{R,n}$ transported to $\tau(p)$ and $s_{\bar R, m}$. As 
in Section \ref{wilop}, the Wilson operators $W_{R,p}$ constructed this way are diagonal
on the basis of states $\psi_u$, with eigenvalues given by evaluating the inner product on the concrete canonical sections in the flat bundle corresponding to $u$.  

The duality predicts that a dual 't Hooft operator associated to the same representation $R$, supported on the same curve 
$\gamma_p$, and with $S$-dual endpoints, has the same eigenvalues.   The dual 't Hooft operator is associated to a real Hecke correspondence $Z^\R_{R,p}$ of $\M_H^\R(G,C)$ with itself;
$Z^\R_{R,p}$ is just the $\tau$-invariant locus of the ordinary Hecke correspondence $Z_{R,p;\bar R,\tau(p)}$  of $\M_H(G,C)$ with itself, 
for a $\tau$-conjugate pair of points labeled by the conjugate (or dual) 
pair of representations $R,\bar R$.   
As in Section \ref{lineq}, because of the scaling symmetry of the cotangent bundle, the quantum operator associated to the
Hecke correspondence can be defined by a semiclassical formula.   Beyond the ingredients that were used in Section \ref{lineq}, one needs one further
fact.  Suppose that $X$ is a complex manifold with canonical bundle $K_X$; let $\tau:X\to X$ be an antiholomorphic involution  with fixed point set $X^\R$,
and assume thast $X^\R$ is orientable.
Then a holomorphic section of $K_X^{1/2}\to X$ restricts on $X^\R$ to a complex-valued half-density.   Hence a holomorphic endpoint or corner that one
would use (together with an antiholomorphic one) 
in defining an 't Hooft operator in the quantization of $\M_H$ restricts on the real locus to a half-density that defines an 't Hooft
operator in the quantization of $\M_H^\R$.   In our application, $Z^\R_{R,p}$ is orientable because $Z_{R,p;\bar R,\tau(p)}$ is a Calabi-Yau manifold,
as discussed in Section \ref{lineq}; the Calabi-Yau form of a Calabi-Yau manifold $X$ that has a real structure can be chosen to be real and restricts
on the real locus to a top-degree differential form that defines an orientation of $X^\R$.

\subsection{The Case That $\tau$ Does Not Act Freely}\label{nonfree}

Now we consider  the case that $\tau$ does not act freely on $C$.      Suppose that the action of $\tau$ leaves fixed a circle $S\subset C$.
Then $U_\tau=\h U/\Z_2$ contains $S$ as a locus of $\Z_2$ orbifold fixed points.    The local behavior near $S$ looks like
\be\label{wofo} S\times \R^2/\{1,\tau\}, \ee
where $\tau$ acts on $\R^2$ as a $\pi$ rotation.
  In the four-manifold $M_\tau=\R\times U_\tau$,
the fixed point set is $\R\times S$. 

We recall that in general the $\tau$ action on $S$ is accompanied by an action of an automorphism $x_S$ that satisfies $x_S^2=1$ at least
in the adjoint form of $G$.    In $G$ gauge theory, one might expect to require $x_S^2=1$ in $G$, but one has to take into account
the topological subtleties that were reviewed in Section \ref{toposubt}.    

To illustrate the issues, we consider the case that $G$ has rank 1 and thus is $\SU(2)$ or $\SO(3)$. 
In this case, $G$ has no outer automorphisms and $x_S$ is conjugation by an element
$g_S$ of $G$.
As discussed in Section \ref{toposubt}, there
 are two versions of $\SU(2)$ gauge theory.  In ordinary $\SU(2)$
gauge theory, the most obvious condition is to require $g_S^2=1$ acting on an $\SU(2)$ bundle $E\to M$.   Then the  only options are $g_S=1$ and $g_S=-1$. 
On the other hand, in
$\Spin\cdot \SU(2)$ gauge theory, the most obvious condition is to ask for 
$g_S^2=1$ acting on $\cS\otimes E$, where $\cS$ is a $\tau$-invariant spin bundle on $M$ (defined
at least locally near the fixed point set) and $E$ is an $\SU(2)$ bundle (defined wherever $\cS$ is).   Since $\tau$ acts as a $\pi$ rotation of the normal plane at the fixed 
point set, $\tau^2$ is a $2\pi$ rotation and acts as $-1$ on $\cS$.   This would suggest that we require
 $\tau^2=-1$ on $E$, leading to $g_S=\mathrm{diag}(\i,-\i)$,
up to conjugation.\footnote{A role for this conjugacy class was suggested by D. Baraglia.} 

However, we believe that it may also be possible to reverse these choices.    For example, in $\SU(2)$, before trying to divide by $\tau$, we could assume
that there is a monodromy defect with monodromy $-1$ along the fixed point locus of $\tau$.  Then in taking the quotient we would want $g_S^2=-1$.
Similarly, including such a monodromy defect before taking the quotient would motivate $g_S^2=+1$ for $\Spin\cdot \SU(2)$.

Similarly,  there are two versions of $\SO(3)$ gauge theory, with or without a factor 
$\Delta=(-1)^{\int_M w_2(M) w_2(E)}$ in the integrand of the path integral.    With or without this factor, we want $g_S^2=1$, which gives two
possibilities, namely $g_S=1$ and $g_S=\mathrm{diag}(-1,-1,1)$.   In the presence of a codimension 2 singularity with $g_S=\mathrm{diag}(-1,-1,1)$,
$\Delta$ is not well-defined topologically, so it appears that in this case we want $g_S=1$.   Without the factor $\Delta$, both possibilities for $g_S$ are viable.
For $\SO(3)$, as the center is trivial, we do not have the option of including a defect with central monodromy, but we can include the dual of this, which is
a defect that senses the topology of the gauge bundle $E$ restricted to the fixed point set.  For $\SO(3)$, this defect is a factor in the path integral of 
$(-1)^{\int_W  w_2(E)}$, where $W$ is the fixed point set.

In general, it is a subtle question to find the dual of a singularity of this nature.  Part of the reason for the subtlety is that one {\it cannot} assume
that the dual of an $\R^2/\Z_2$ orbifold singularity, defined by a condition of $\tau$-invariance, is another $\R^2/\Z_2$
orbifold singularity, defined by a dual condition of $\tau$-invariance.   In general, one only knows that the
dual of an $\R^2/\Z_2$ orbifold singularity is a codimension 2 defect that preserves the same supersymmetry as the $\R^2/\Z_2$ orbifold singularity.
    In a somewhat similar problem of rigid surface defects, it proved difficult to get a general understanding of the action
of duality \cite{GWr}.

It is tempting to claim a simple answer if $g_S$ is central for all $S$, on the following grounds.
 Suppose that $g_S=1$.   Then the singularity is only in the geometry, not the gauge field.
The $\R^2/\Z_2$ orbifold singularity is not a singularity at all topologically, as $\R^2/\Z_2$ is equivalent topologically to $\R^2$.   So if $g_S=1$ for all
fixed circles, $U_\tau$ is actually a manifold topologically, and one can ``round off'' the orbifold singularities to give it a smooth geometry.  Let us call the rounded
version $\t U_\tau$.   If it is correct in the $A$-model to replace the orbifold with the smooth manifold $\t U_\tau$, then the dual is the $B$-model on the same
manifold.   This reasoning has a potential analog for the more general case that $g_S$ is central but not equal to the identity.
We can still round off the defect to get the smooth
manifold $\t U_\tau$, but now  $\t U_\tau$ contains a defect with central monodromy, supported on the orbifold locus $W$.   
As remarked earlier, the dual of this defect is a defect that senses
the topology of the bundle, for example a defect defined by a factor $(-1)^{\int_W
 w_2(E)}$ in the case of gauge group $\SO(3)$.

The following is a  strategy, in principle, to analyze the general case.   Surround the fixed point  locus by a two-torus. Then the orbifold defect defines 
a boundary condition for the 2d theory which arises from  $T^2$ compactification of four-dimensional super Yang-Mills theory. 
This is not quite a $\sigma$-model, because of the 
large unbroken gauge symmetry. If we can identify the mirror of the boundary condition, it will provide boundary conditions for the 
$G^\vee_\C$ flat connection restricted to the two-torus on the dual side. This would be sufficient to characterize how the oper flat connection 
at the boundary of $U_\tau$ can extend to the interior and thus determine the spectrum on the $B$-model side of the duality.

\subsection{Real Hecke Operators}\label{rho}

The most important novelty of the case that $\tau$ does not act freely on $C$ may be the existence of line operators supported on a real point in $C$,
as opposed to the line operators considered in Section \ref{fourdpic} that are supported on a $\tau$-conjugate pair of points.

First we describe a gauge theory picture for  an 't Hooft operator supported at a real point  $p\in C$.   We work on the four-manifold $M=\R\times \h I\times C$,
with $\R$ parametrized by $t$, and $\h I$ by $w$.   A local picture suffices, so we take $C$ to be simply the complex $z$-plane $\C$, and we consider the involution $\tau$
that acts by $(t,w,z)\to (-t,w,\bar z)$.  Acting just on $\C$, $\tau$ has the fixed line $S$ defined by ${\mathrm{Im}\, z}=0$; we acccompany $\tau$ with
an automorphism $x_S$  satisfying $x_S^2=1$ (or a slightly more general condition discussed in Section \ref{nonfree}).      An 't Hooft operator supported at $z=t=0$
is described by a Dirac monopole solution of the $G$ gauge theory.  The solution has a structure group that reduces to a maximal torus $T\subset G$
and can be characterized by its curvature:
\be\label{curv} F=\frac{\m}{2}\star_3 \d \frac{1}{(t^2+|z|^2)^{1/2}}, \ee
where $\star_3 $ is the Hodge star for the metric $\d t^2+|\d z|^2$ on $\R\times\C$, and $\m$ is a constant element of the Lie algebra $\frak t$ of $T$.
For a connection with this curvature to exist, $\m$ must be an integral coweight, dual to a representation $R$ of $G^\vee$ (it coincides with the
object that was called $\m$ in Section \ref{affgr}).   Now we ask whether
this solution is invariant under $\tau$, accompanied by the automorphism $x_S$.   Since $\star_3$ is odd under $\tau$, a necessary
and sufficient condition is that $\m$ should be odd under $x_S$:
\be\label{oddurv} x_S( \m )=-\m. \ee
When and only when it is possible to choose $\m$ in its conjugacy class so that it is odd under $x_S$, the solution constructed this way is $\tau$-invariant
and descends to a solution on $M_\tau=M/\{1,\tau\}$.   It describes a real 't Hooft operator, supported on a real point in $C$ and associated to the representation
$R$ of $G^\vee$.  

Using this model solution, we can define a space of ``real'' Hecke modifications which can be implemented by a real 't Hooft operator.
There are real versions $G_{\R,S}({\cal K})$ and $G_{\R,S}({\cal O})$ of  $G_{\C}({\cal K})$ and $G_{\C}({\cal O})$ which consist of 
gauge transformations which lie in $G_{\R,S}$ along $S$. We can thus define a real version 
\begin{equation}
\Gr_{G_{\R,S}} = G_{\R,S}({\cal K})/G_{\R,S}({\cal O})
\end{equation}
of the affine Grassmannian and the orbits $\left[G_{\R,S}({\cal O}) z^\m\right]$ of real Hecke modifications of type $\m$. 

A knowledge of which real 't Hooft operators are possible for a given $x_S$ puts a very strong constraint on the dual of a $\Z_2$ orbifold singularity with a given $x_S$.

Geometrically, a real 't Hooft operator stretches from the boundary  of $U_\tau$ to a fixed point $p$ in the interior.
A full analysis of real 't Hooft operators, which we will not attempt here, would include a discussion of the possible endpoints of the 't Hooft operator 
on a fixed point and a derivation of the corresponding integral operators. 
 
On the $S$-dual side, we will have a Wilson operator stretched from the boundary to $p$.  A specific endpoint of the Wilson line will give a vector $v_{\bar R,m}$ in the space of flat sections of the gauge bundle in a neighborhood of $p$. We cannot characterize this vector more precisely without knowing the $S$-dual of the orbifold
singularity;  in particular, the flat bundle in the $B$-model may not extend over the fixed point set.
The Wilson operator expectation value will take the form of a pairing $(v_{\bar R,m}, s_{R,n})$,
selecting a specific solution of the oper differential equations. 

More generally, one can consider an arbitrary oriented three-manifold $U$ with boundary
$C$, and study the dual $A$- and $B$-models on $\R\times U$.   Modulo technical difficulties (the moduli space of flat bundles on $U$ may be very
singular), one can hope to define a space $\H_U$ of physical states, with an action of the Hitchin Hamiltonians on the $A$-model side and a prediction
for their eigenvalues in terms of classical data on the $B$-model side. One can also study line operators, though in general there will be no close
analogs of the ones that we have considered in this article.   The state space  $\H_U$  will have a hermitian inner product, but it is not
clear that this inner product will be positive-definite in general, since it is no longer obtained by quantizing a cotangent bundle.

\section{Four-Dimensional Avatars of BAA Boundary Conditions and Corners}\label{avatar}

Four-dimensional super Yang-Mills theory admits many half-BPS boundary conditions \cite{Gaiotto:2008sa} 
which are topological in the four-dimensional $A$-twist and descend to BAA boundary conditions upon twisted compactification 
on $C$ \cite{KW,Gaiotto:2016hvd,Gaiotto:2016wcv}. 

Such a boundary condition, along with its ``corners'' with $\B_\cc$ and $\b\B_\cc$, can be used to define a quantum state in
$\H=\Hom(\b\B_\cc,\B_\cc)$.   We have already made use of this construction; see fig. \ref{example6} of Section \ref{wkb}.
Apart from studying additional examples, what we will add in the present section is the use of two-dimensional chiral algebras to study the
corners and the associated quantum states. 

In four dimensions, a junction or corner between two boundary conditions occurs on a two-manifold, which for our purposes is a copy of
the Riemann surface $C$.   In the case of a brane of type BAA that can be engineered in four-dimensional gauge theory, its corners
with $\B_\cc$, if they can likewise be engineered in four-dimensional gauge theory, are frequently holomorphic-topological and support
holomorphic chiral algebras. Chiral algebras that arise this way were studied in   \cite{Gaiotto:2016hvd,FG}.
  Corners with $\b\B_\cc$, if they can be engineered in four dimensions, likewise typically support antiholomorphic
chiral algebras.   This is the situation that we will study in the present section.

\subsection{The Analytic Continuation Perspective}\label{ACP}
We begin by recalling a construction that simplifies the analysis of the relevant junctions.   The brane $\B_\cc$ can be derived
from a deformed Neumann boundary condition in four dimensions.\footnote{\label{deformed} Ordinary Neumann boundary conditions for a gauge field assert that 
$n^i F_{ij}=0$, where $F$ is the Yang-Mills curvature and $n$ is the normal vector to the boundary.   Deformed Neumann boundary conditions
express $n^i F_{ij}$ in terms of the boundary values of some other fields.  Note that a different, undeformed, Neumann boundary condition,
with a different extension to the rest of the supermultiplet, will enter the story in Section \ref{Enriched}.}    We will call this the deformed Neumann or $\B_\cc$ boundary condition.
The path integral of the $A$-twisted 4d gauge theory in the presence of such a deformed Neumann  boundary can be interpreted as 
a slightly exotic path integral for a three-dimensional theory defined on the boundary   \cite{Witten:2010zr,Witten:2011zz}.
  The action of this three-dimensional theory  is a holomorphic function of complex
variables, and the path integral is taken on a middle-dimensional integration cycle in the space of  fields.   The integration cycle is defined
by $A$-model localization in four dimensions, but the details of this are not important for our purposes.   We will only discuss properties that
do not depend on the choice of integration cycle.  We will call this type of path integral loosely a contour path integral.

The relevant three-dimensional auxiliary theory is most familiar not in the case  of  conventional geometric
Langlands but for what is known mathematically as ``quantum'' geometric Langlands.   This corresponds in gauge theory to working at a generic
value of the canonical parameter $\Psi$ that was introduced in \cite{KW}.   

At generic $\Psi$, the boundary theory associated to suitably deformed Neumann boundary conditions is a Chern-Simons theory with a complex
connection $\cA$, with curvature $\cF=\d \cA+\cA\wedge \cA$, and  action
\be\label{csact} I_\CS =\frac{\Psi}{4\pi}\int_N \left( \Tr\,\cA\wedge \d \cA +\frac{2}{3}\cA^3\right). \ee
Here $N$ is the boundary or a portion of the boundary of a four-manifold $M$.    The boundary condition that leads to the theory $I_\CS$ on $N$ is
``topological'' in the sense that the only structure of $N$ that is required to define it is an orientation.   If $N=\partial M$, then the $A$-model path integral
on $M$ with boundary condition that leads to the boundary coupling $I_\CS$ is a Chern-Simons path integral on $N$ with a non-standard integration cycle (which depends on $M$).
More generally, $N$ itself may have a boundary; along $\partial N$, we consider a junction or corner between the deformed Neumann boundary condition that leads
to $I_\CS$ and some other boundary condition.   In this case, as in conventional Chern-Simons theory on a three-manifold with boundary, for suitable
choices of the second boundary condition,  a current algebra or Kac-Moody symmetry will appear along $\partial N$.   The level of these currents is $\Psi-h$, where $h$ is the dual Coxeter number of $G$.
The contribution $\Psi$ to the level can be computed classically from the failure of $I_\CS$ to be gauge-invariant on a manifold with boundary, and the $-h$ is a 1-loop correction, which will be described in Section \ref{Dbc}.

For the present article, we are interested in ``ordinary'' geometric Langlands at $\Psi=0$.   We will not get anything sensible if we simply set $\Psi=0$
in the action (\ref{csact}), since a contour path integral with zero action will not make sense,
and in fact there is no way to take the limit $\Psi\to 0$ while preserving topological invariance along $N$.   That is one way
to understand the fact that the $\B_\cc$ boundary condition that has been important in the present article is holomorphic-topological rather than
topological.   To take the limit $\Psi\to 0$ to get a holomorphic-topological boundary condition, we can do the following.  Let $C$ be a complex Riemann
surface with local complex coordinate $z$,  and assume that $N=S\times C$, where $S$ is a 1-manifold parametrized by $t$.
Then take the limit $\Psi\to 0$ keeping fixed $\varphi=\frac{\Psi}{4\pi} \cA_z\d z$.
The Chern-Simons action goes over to 
\be\label{degac}
I_{\varphi\cF}= 
\int _N\Tr\, \varphi_z \cF_{t \bar z} \d t \d^2z.
\ee
The degeneration of the Chern-Simons action (\ref{csact}) to the action of eqn. (\ref{degac}) is somewhat analogous to the degeneration from a complex
flat connection to a Higgs bundle as the complex structure of the Higgs bundle moduli space $\M_H(G,C)$ is varied.
The action $I_{\varphi\cF} $ describes what can be interpreted as a topological gauged quantum mechanics on the cotangent bundle to the space of $(0,1)$ connections on $C$. 
The Higgs field $\varphi$ is the momentum conjugate to $\cA_{\bar z}$. The analogous statement in two-dimensional terms
is that the $A$-model of $\M_H(G,C)$, with a $\B_\cc$ boundary, is related to 
an analytically continued  quantum mechanics on 
$\M_H(G,C)$   \cite{Witten:2010zr}.

This formulation makes it obvious that the deformed Neumann boundary condition is not topological in the $A$-twist. Instead, 
it depends on a choice of complex structure on $C$ and it admits local operators which vary holomorphically along $C$ and topologically along $S$: it is a holomorphic-topological boundary condition. The equations of motion derived
from $I_{\varphi\cF}$ imply that $\varphi$ is holomorphic in $z$ and independent of $t$. 
Away from the boundary of $N$, the gauge-invariant  local operators on $N$ are the 
gauge-invariant polynomials  ${\cal P}[\varphi](z)$ of $\varphi$, which descend to the Hitchin Hamiltonians in the 2d $A$-model, and fermionic partners
of these operators, which are described in Appendix \ref{coho}.  (At generic $\Psi$, there are no
gauge-invariant local operators on a boundary characterized by the Chern-Simons action $I_\CS$.)  

What happens if $N$ has a boundary?  
A junction between the deformed Neumann boundary condition that leads to the $\varphi\cF$ theory and a topological 4d boundary condition can in many cases be described by a boundary condition in the $\varphi\cF$ theory,  encoding both the topological boundary condition and the choice of junction.  
In many important examples, the topological boundary condition is a half-BPS boundary condition of type BAA.
 With suitable choices, as we discuss further in Section \ref{Dbc}, $\varphi$ can behave  along $\partial N$ as a holomorphic current generating a Kac-Moody symmetry.  
  But now the level of the Kac-Moody symmetry comes entirely from the 1-loop 
correction and is $-h$.

In general, the boundary conditions which appear in the $\varphi\cF$
 auxiliary gauge theory are holomorphic as well and may support holomorphic local operators. This reflects the same property of the corresponding junctions between the deformed Neuman boundary and the half-BPS boundary: they are holomorphic in the $A$-twisted theory. The appearance of holomorphic junctions in the GL-twisted theory at general $\Psi$ and the relation to the Chern-Simons level were analyzed in \cite{Gaiotto:2017euk}. 

\subsection{The Role of Chiral Algebras} \label{RoleChiral}
A holomorphic junction between a holomorphic-topological boundary condition and a topological one
may support local operators which depend holomorphically on their position on $C$. Essentially by definition, these operators define a chiral algebra. Although the chiral algebra depends on the choice both of the topological boundary condition and of the junction, we will suppress that dependence for notational convenience and denote the
chiral algebra simply as ${\cal V}$. The chiral algebra is akin to the chiral algebras of holomorphic local operators which can be 
found in a 2d CFT, or at the boundary of a 3d topological field theory (TFT) 
 such as Chern-Simons theory. There are some differences  \cite{Costello:2020ndc}  due to the fact that the $\varphi \cF$ theory is holomorphic-topological. 

A single junction in the $A$-twisted theory can be used to build a variety of different corners in the 2d $A$-model, depending on the choice of 
local operators $O_i(z_i)$ placed at points $z_i$ in $C$. These corners are not all independent: they depend on the $z_i$ holomorphically,  with singularities as $z_i \to z_j$ controlled by the OPE of the chiral algebra.  The OPE or the associated Ward identities imply  recursion relations between different corners.\footnote{A solution of the Ward identities for a chiral algebra is usually called a {\it conformal block}. In a physical 2d CFT, conformal blocks can usually be obtained by some sewing procedure on the Riemann surface. This may not be possible for a general chiral algebra such as ${\cal V}$, but the notion of conformal blocks is still available and can be used to characterize the space of $A$-model corners which can be produced from a given junction.}
 
Recall the gauge-invariant local operators ${\cal P}[\varphi](z)$ on the deformed Neumann boundary condition which 
give rise to the Hitchin Hamiltonians. These operators can be brought to the junction along the topological direction $t$ along the boundary. The resulting boundary-to-junction OPE is not singular and produces a collection of operators
$S_{\cal P}(z)$ in ${\cal V}$. These operators are {\it central}, i.e.  they have non-singular OPE with the other operators in ${\cal V}$, 
for the same reason that the ${\cal P}[\varphi](z)$ have non-singular OPE with each other: they can be freely displaced along the $t$ direction. 

That property has an important corollary: $A$-model corners labelled by a collection of operators $S_{\cal P}(z) O_1(z_1) \cdots O_n(z_n)$
satisfy the same Ward identities as a function of the $z_i$ as corners labelled only by $O_1(z_1) \cdots O_n(z_n)$. This insures that the 
the action of ${\cal P}[\varphi](z)$ on this space of $A$-model corners is well-defined.

 \begin{figure}
 \begin{center}
   \includegraphics[width=1.7in]{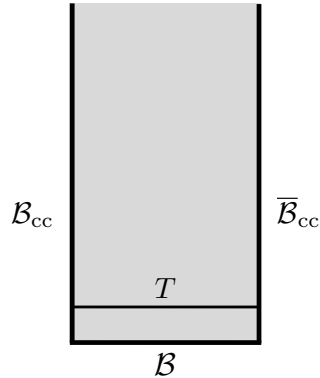}
 \end{center}
\caption{\small An 't Hooft operator $T$, of type BAA, acting on the state created by a brane $\B$ of type BAA, together with suitable chiral corners.   
We can fuse $T$ with $\B$ to
make a new brane $T\B$, again of type BAA.   This involves the same action of line operators on branes that we started with in fig. \ref{example1}(a).  In addition
we have to consider the composition of the chiral corners at the left and right endpoints or boundaries  of $T$ and $\B$.  \label{example8}}
\end{figure} 

We can also discuss the chiral algebra ingredients which occur in a four-dimensional description of the action of 't Hooft operators on the states created by branes
with chiral algebra corners.  In fig. \ref{example8}, we sketch an 't Hooft operator $T$, of type BAA, acting on the state created by a brane $\B$ of the same type, with
suitable chiral and antichiral corners.  
The action of $T$ on the state can be described as the fusion of the  interface represented by $T$  with the boundary $\B$, accompanied by a composition of the corresponding corners both with $\B_\cc$ and with $\bar \B_\cc$. 
In  four dimensions, in contrast to the two-dimensional picture of fig. \ref{example8},  there is an obvious difference between the 't Hooft line defect and the half-BPS boundary: the former is supported at a point $p\in C$ while the latter wraps the whole $C$. 
So the composition $T\B$ coincides with $\B$ away from $p$; it can be described as the brane $\B$ enriched with a line defect.
In order to describe the action of 't Hooft operators we should thus first generalize the discussion in this section to allow for
boundary line defects ending on the junction at some point $p\in C$. As usual, this setup will depend holomorphically on $p$. It depends only on $p$ because in the A-twisted theory, the boundary line defect itself is the image of a topological 't Hooft line defect and is thus also topological. 

The presence of a boundary line defect $\ell$ and its endpoint does not affect the choice of chiral algebra operators available elsewhere on $C$, nor the Ward identities they satisfy away from $p$. It affects, though, their behavior near  $p$. For each $\ell$, there is  a vector space of possible endpoints $O_{\ell,i}(p)$ and they form a module ${\cal V}_\ell$ for ${\cal V}$: the module structure encodes the OPE between chiral algebra 
operators and endpoints and controls the Ward identities satisfied at $p$ by the ${\cal V}_\ell$ insertions.  

As we bring an 't Hooft operator with endpoint $\alpha_{R,n}(p)$ to the half-BPS boundary, we will produce a boundary defect $\ell$ as well as 
a specific endpoint $S_{R,n}(p) \in {\cal V}_\ell$ at the junction. We can predict two general properties of such endpoints: they will have non-singular OPE with the chiral algebra ${\cal V}$, just like the  $S_{\cal P}(z)$ associated to Hitchin Hamiltonians, and they will satisfy the same differential equations in $p$ as $\alpha_{R,n}(p)$ do, with coefficients controlled by the $S_{\cal P}(z)$. These properties can be derived immediately by separating the 't Hooft line from the boundary along the topological direction and transforming $S_{R,n}(p)$ back to $\alpha_{R,n}(p)$.

Again, the centrality property guarantees that $A$-model corners labelled by a collection of operators $S_{R,n}(p) O_1(z_1) \cdots O_n(z_n)$
satisfy the same Ward identities as a function of the $z_i$ as corners labelled only by $O_1(z_1) \cdots O_n(z_n)$. This insures that the 
the action of the 't Hooft operator on the state created by the brane is well-defined. 

We conclude this general discussion with some comments on the description of the 't Hooft lines in the auxiliary $\varphi F$ theory. 
The 't Hooft lines perpendicular to $N$ can be interpreted as monopole local operators in the $\varphi F$ theory. Operators such as  ${\cal P}[\varphi](z)$
appear as local operators both in the $\varphi F$ theory and in the four-dimensional theory because they are polynomial of the elementary fields and
their definition does not depend on a choice of contour for the path integral. The definition of disorder operators such as monopole operators, instead, 
requires one to modify the space of field configurations allowed in the path integral and thus affects the possible choices of integration contours. This modification 
is encoded in the presence of an actual line defect in the four-dimensional theory, ending on $p$ in $N$.  These considerations also apply to boundary disorder operators 
at $\partial N$, which will appear as $O_{\ell,i}(p)$ endpoints of a boundary line defect in the four-dimensional theory. 

It would be interesting to analyze directly the space of monopole operators available in the $\varphi F$ theory, as well as their images at the boundary. 
Some of the tools were developed in \cite{Gaiotto:2016wcv} and applied to Chern-Simons theory there. They involve cohomology calculations on the 
affine Grassmanian which are likely to give a local version of the analysis in Section \ref{lineq}. We leave this exercise to an enthusiastic reader. 

\subsection{From Corners to States} \label{fcs}

Now consider a strip with $\B_{\cc}$ and $\bar \B_{\cc}$ at the two ends, and with boundary conditions set at the bottom of the strip
by some other brane $\B$ (fig. \ref{example6}).   At the ``corners,'' the construction that we have just described produces chiral algebras $\V$ and $\bar \V$.
With suitable operator insertions at the corners, the path integral on the strip  produces a (possibly distributional) state
\be\label{chidef}
\chi = \left|O_1(z_1) \cdots O_n(z_n) \bar O_{n+1}(\bar z_{n+1}) \cdots \bar O_{n+\bar n}(\bar z_{n+ \bar n})\right\rangle
\ee
 in the usual  Hilbert space $\H=\Hom(\b\B_\cc,\B_\cc)$.
An important special case is that the chiral algebras $\V$ and $\bar\V$ at the two corners are complex conjugate,
though we are not restricted to this case.  Operators sitting on different corners cannot have OPE singularities, so the $\bar O_j(\bar z_j)$ 
operators at the $\bar \B_{\cc}$ corner do not affect the Ward identities for the $O_i(z_i)$ and vice  versa. 

We can also consider the pairing of $\chi$ with a test state $\Psi \in {\cal H}$.
 The resulting inner product may in general be ill-defined, but it is well-defined if  $\Psi$ is sufficiently nice, for example
 if $\Psi$ is created in a similar way using a brane of compact support at the top of the strip (fig. \ref{example6}(c)).
If the inner product is well-defined for all choices of the operator insertions, it gives a collection of correlation functions 
\be
\langle \Psi| O_1(z_1) \cdots O_n(z_n) \bar O_{n+1}(\bar z_{n+1}) \cdots \bar O_{n+\bar n}(\bar z_{n+ \bar n}) \rangle
\ee
on $C$ which satisfy the Ward identities for ${\cal V}$ and $\bar {\cal V}$.

Half-BPS boundary conditions decorated by boundary line defects will produce a larger collection of states
\be
\left|\prod_i O_i(z_i) \prod_j \bar O_{j}(\bar z_{j}) \prod_k O_{\ell_k,k}(p_k) \bar O_{\ell_k,k}(\bar p_k) \right\rangle,
\ee 
with corner data including endpoints for the boundary line defects.

In the remainder of this section we will make these structures explicit for some basic examples of half-BPS boundary conditions. 

\subsection{Dirichlet Boundary Conditions}\label{Dbc}
Four-dimensional half-BPS Dirichlet boundary conditions fix the gauge connection  $A$ to vanish at the boundary, or more generally to take some specified
value at the boundary, 
extended to other fields in a supersymmetric fashion.   In particular, three of the six scalar fields satisfy Dirichlet boundary conditions,
and other three, which include the Higgs field $\phi$
of the 2d $\sigma$-model,  satisfy Neumann boundary conditions.

From the perspective of the auxiliary $\varphi \cF$ theory on a portion $N$ of $\partial M$, a junction between the half-BPS Dirichlet boundary condition
and the deformed Neumann boundary is represented by a Dirichlet boundary condition on the field $\cA_{\bar z}$ along $\partial N$.
 Gauge transformations are restricted to be trivial  at the boundary; otherwise it would not be possible to specify the value of $\cA_{\bar z}$ along
 the boundary. We do not impose any boundary condition on  $\varphi=\varphi_z\d z$, the conjugate of $\cA_{\bar z}$. 

Since gauge transformations are constrained to be trivial along $\partial N$, local operators on $\partial N$, that is, on the junction, 
are not required to be gauge-invariant. Instead, there is a $G$ global symmetry\footnote{\label{defglobal} A global symmetry transformation by an element $g\in G$
is a gauge transformation $g(x):N\to G$ whose restriction to $\partial N$ is constant, $g(x)|_{\partial N}=g$.   This preserves the condition  $\cA_{\bar z}|_{\partial N}=0$,
so it is a  symmetry of the theory defined with that boundary condition.  How the constant $g$ is extended over the interior of $N$ as a gauge transformation
does not matter.   Any two choices differ by 
a gauge transformation that is trivial on the boundary and acts trivially  on physical observables.}   
acting on operators on $\partial N$; the boundary value $a$ of $\cA$ can be interpreted as a background connection for that global $G$ symmetry. In the 4d setup, the $G$ symmetry acts on the whole half-BPS Dirichlet boundary condition, but in the $A$-twist there are no local operators it can act on at interior points of $N$. The global
symmetry acts on junction operators only. 

The  scalar field $\varphi$ is a valid local operator at the junction, identified with the same operator in the auxiliary $\varphi \cF$ theory. 
In the $A$-twisted theory, the operator $\varphi_z$ is actually the conserved current associated to the boundary $G$ symmetry. 
One way to derive this result is as follows.   If we vary the action $I_{\varphi\cF}=\int_N \Tr\,\varphi_z F_{t\bar z}\d t \d^2z$ with ``free''
boundary conditions, we find that the variation of $I_{\varphi \cF}$ contains a boundary term $\int_{\partial N} \Tr\,\varphi_z\delta A_{\bar z}$,
and therefore the Euler-Lagrange equations include a boundary condition $\varphi_z|_{\partial N}=0$, with no restriction on $A_{\bar z}$.   If instead
we want the boundary condition to be $A_{\bar z}=a_{\bar z}$ (where $a$ is some specified connection on $C$), we can add a boundary
term to the action, so that the full action becomes
\be\label{fullact} I'_{\varphi\cF}=I_{\varphi\cF}-\int_{\partial N}\Tr\,\varphi_z(A_{\bar z}-a_{\bar z}).\ee
Then the Euler-Lagrange equations give a boundary condition $(A_{\bar z}-a_{\bar z})|_{\partial N}=0$, with no constraint on $\varphi$.
But now the current is $J_z=\partial I'_{\varphi\cF}/\partial a_{\bar z}=\varphi_z$, as claimed.

As explained in Section \ref{ACP} and in \cite{Gaiotto:2017euk}, in a generalization of this problem with generic $\Psi$, the fact that $J_z$ is a Kac-Moody
current can be seen classically, though a 1-loop computation similar to what we are about to describe is needed to show that the level is $\Psi-h$ rather than
$\Psi$.   At $\Psi=0$, the Kac-Moody level comes entirely from a 1-loop calculation.   The necessary computation can be done very easily with the help
of  
 a simple shortcut, which has an analogue in gauge theory in any dimension.
 We place the $\varphi\cF$ theory on a slab $I\times C$, where $I$ is an interval $[0,L]$ with the same Dirichlet boundary conditions at each end.   
 Anomalies can always be computed from the low energy limit of a theory, so in this case we can drop the modes that are nonconstant along $I$
 and reduce to a purely two-dimensional theory.   This is equivalent to taking a naive $L\to 0$ limit.
The resulting 2d theory has action 
\be\label{belaction}
\int \Tr \, \varphi_z D_{\bar z} \cA_t\ee
and is a bosonic $\beta \gamma$ system with fields $\varphi_z$ and $\cA_t$ (which now depend  only on $z$ and $\bar z$, not on $t$) of spins 1 and 0, valued in the adjoint representation. 
A similar fermionic chiral $bc$ system with adjoint-valued fields  has  an anomaly coefficient $2h$, so the bosonic $\beta\gamma$ system has anomaly  $-2h$.
In gauge theory on the slab $I\times C$, the anomaly is localized on the boundaries of the slab since the bulk theory is not anomalous.
By symmetry,  half the anomaly comes from one end of the slab and half from the other end, so the anomaly coefficient at either end is $-h$.   This value of the level
has very special properties, as we will review momentarily.   The value $-h$ of the Kac-Moody level is often called the critical level and denoted as $\kappa_c$.

We will now determine the image in ${\cal V}$ of the bulk local operators 
${\cal P}[\varphi](z)$ 
of the $\varphi\cF$ gauge theory.
 Naively, the image would just be 
${\cal P}[J](z)$. This expression, though, is ill-defined because of the OPE singularity of $J_z$ with itself. We can regularize this expression by point splitting, carefully subtracting singular terms. This is actually a familiar exercise in 2d CFT. 

The simplest example is the image of the quadratic Hamiltonian $\Tr\, \varphi_z^2$. The regularized version of the operator 
is the Sugawara operator $S_2(z)$. Recall that for general level $\kappa$, the Sugawara operator is proportional to the stress tensor: $T(z) = \frac{1}{\kappa + h} S_2(z)$. In particular, the singular part of the OPE of $S_2(z)$ 
\begin{equation}
S_2(z) J(w) \sim (\kappa + h)\frac{J(w)}{(z-w)^2} + (\kappa + h) \frac{D_w J(w)}{z-w} + \cdots
\end{equation}
is proportional to the critically-shifted level $\kappa+h$. When $\kappa=\kappa_c=-h$, $S_2(z)$ has non-singular OPE with the Kac-Moody currents, i.e. it is central. This is precisely the property we expect  for the image of $\Tr \varphi_z^2$ in ${\cal V}$ and fully characterizes it 
among operators with the same scaling dimension.   So $S_2(z)$ corresponds to the quadratic Hitchin Hamiltonians.

The regularization of higher Hamiltonians takes more work, but we can invoke a general theorem \cite{BD}: the center of the critical Kac-Moody algebra for $G$ is generated by a collection of central elements $S_{\P}(z)$ which regularize ${\cal P}[J](z)$. As an algebra, the center of the critical Kac-Moody algebra is isomorphic to the space of holomorphic functions on the oper manifold $L_\op$ for the Langlands dual group $G^\vee$. We conclude that $S_{\P}(z)$ is the image at the junction of the 3d operators ${\cal P}[\varphi](z)$, which are also identified via $S$-duality with the generators of the algebra of holomorphic functions on $L_\op$. 

The presence of a global $G$ symmetry at a Dirichlet boundary adds an extra ingredient to the construction of holomorphic-topological BAA branes.   As already
remarked, we can generalize
Dirichlet boundary conditions by setting the boundary value of the connection to any fixed connection $a$ along $C$, rather than setting it to zero.
We thus produce a whole family of 
BAA branes $\mathrm{Dir}(a)$ parameterized by the choice of background $G$ connection $a$. In the $A$-twisted theory, the 
system only depends on $a_{\bar z}$ at the $\B_\cc$ corner and on $a_z$ at the $\bar \B_\cc$ corner. The dependence is encoded respectively in the holomorphic currents $J_z$ at the $\B_\cc$ corner and  anti-holomorphic currents $\bar J_{\bar z}$ at the $\bar \B_\cc$ corner.

In the absence of local operator insertions at the junctions, Dirichlet boundary conditions thus define a family of (distributional) states 
$\Delta(a) \in \H$. The insertion of Kac-Moody currents $J_z(z_i)$ or $\bar J_{\bar z}(\bar z_j)$ at the two corners gives functional derivatives
\be
\prod_i \frac{\delta}{\delta a_{\bar z}}(z_i) \prod_j \frac{\delta}{\delta a_{ z}}(\bar z_j) \Delta(a)
\ee 
of the state with respect to the background connection. 
 
The Kac-Moody nature of the holomorphic and anti-holomorphic connections has an important consequence: 
the two currents are separately conserved. Recall the anomalous conservation laws:
\begin{align}
D_{\bar z} J_z &= - \frac{\kappa_c}{2 \pi} f_{z \bar z} \cr
D_z J_{\bar z} &= \frac{\kappa_c}{2 \pi} f_{z \bar z}
\end{align}
where $f$ is the curvature of $a$. These imply that the state $\Delta(a)$ transforms covariantly but anomalously under infinitesimal
complexified gauge transformations of $a$.  To first order in $\lambda$,
\be 
\Delta(a_z + D_z \bar \lambda, a_{\bar z} + D_{\bar z} \lambda )=\left[1+\frac{\kappa_c}{2 \pi} \int_C \Tr\left(\bar \lambda - \lambda \right)f_{z \bar z}  \right] \Delta(a_z,a_{\bar z})
\ee
A nice enough abstract state $\Psi \in \H$ paired with Dirichlet boundary conditions  at $t=0$ gives a functional 
$\Psi(a)$ that transforms similarly under infinitesimal complexified gauge transformations of $a$:
\be \label{eq:cogauge}
\Psi(a_z + D_z \bar \lambda, a_{\bar z} + D_{\bar z} \lambda )=\left[1+\frac{\kappa_c}{2 \pi} \int_C \Tr\left(\bar \lambda - \lambda \right)f_{z \bar z}  \right] \Psi(a)
\ee  Note that $\Psi$ is invariant under real gauge transformations, which correspond to the special case $\lambda=\bar\lambda$.
Functional derivatives of $\Psi(a)$ with respect to $a$ give  correlation functions of critical Kac-Moody currents coupled to $a$. 

\subsection{Connections vs. Bundles}\label{CB}

Let $a$ be a connection on a Riemann surface $C$ with structure group the compact gauge group $G$.  Consider a function $\Psi(a)$ which
is invariant not just under $G$-valued gauge transformations, but under $G_\C$-valued gauge transformations, acting at the infinitesimal level by
\be\label{actfor}\delta a_{\bar z}=-D_{\bar z}\lambda, ~~\delta a_z=-D_z\bar\lambda. \ee
Such a function determines a function on $\M(G,C)$, because $\M(G,C)$ can be viewed as the quotient of the space of all $G$-valued connections by
the group of complex gauge transformations.\footnote{This description is slightly imprecise as one needs to take account of considerations of stability
to realize $\M(G,C)$ as a quotient.   But actually, there is no difficulty: as long as the function $\Psi(A)$ is continuous as well as invariant under $G_\C$-valued
gauge transformations, it does descend to  a function on $\M(G,C)$.   To get from the space of all connections to $\M(G,C)$, one throws away connections
that define unstable holomorphic bundles, imposes an equivalence relation on the semistable ones, and then takes the quotient.   A function $\Psi(a)$
that is continuous and $G_\C$-invariant is always invariant under the equivalence relation.   In any event, these considerations are unimportant for an $\lmark^2$
theory. Somewhat similar remarks apply in the next paragraph.}

Conversely, given a function $f$ on $\M(G,C)$, to define a  function $\Psi(a)$ on the space of connections, we simply declare $\Psi(a)$, for a given $a$,
to equal $f$ at the point in $\M(G,C)$ that is associated to the holomorphic bundle $E\to C$ that is determined by the $(0,1)$ part of $a$.   The function $\Psi(a)$
defined this way is automatically invariant under $G_\C$-valued gauge transformations.  

This correspondence between functions on $\M(G,C)$ and $G_\C$-invariant functions of connections can be extended
to functions $\Psi(a)$ that are not $G_\C$-invariant, but rather transform covariantly under $G_\C$-valued gauge transformations, with an anomaly.
These functionals can be identified with sections of some line bundle over $\M(G,C)$.
For us, the most important case is a function $\Psi(a)$ that transforms  with holomorphic and antiholomorphic anomaly coefficients $\kappa_c=-h$,
as in 
eqn. (\ref{eq:cogauge}). 
In this case, the line bundle  is actually the bundle of half-densities on $\M(G,C)$, as we will show momentarily.
More generally, if $\kappa_c$ is replaced by some other level $\kappa$ (the same both holomorphically and antiholomorphically),
$\Psi(a)$  would represent a section of $|K_{\M(G,C)}|^{\frac{\kappa}{\kappa_c}}$. 

There are many ways to demonstrate that a function that transforms with holomorphic and antiholomorphic
 anomaly coefficient $\kappa_c$  correponds to a half-density on $\M(G,C)$ (as shown originally by Beilinson and Drinfeld \cite{BD}).
We will proceed by showing that a function that transforms with the anomaly coefficient $2\kappa_c$ is a density on $\M(G,C)$.   A density
on $\M(G,C)$ is something that can be integrated over $\M(G,C)$ in a natural way, without using any structure of $\M(G,C)$ beyond the fact
that it is the quotient of the space of connections by the group of $G_\C$-valued gauge transformations.
To decide what kind of object $\Psi(a)$ can be integrated over $\M(G,C)$, we will use
a construction which is somewhat analogous to the definition of the bosonic string path integral. 

Consider a two-dimensional $G$ gauge theory with  connection $a$, coupled to some matter system with holomorphic and anti-holomorphic Kac-Moody symmetry at levels $(\kappa, \kappa)$. As the path integral of this theory is formally invariant under 
complexified gauge transformations, we may hope to gauge-fix the path integral to an integral over $\M(G,C)$. In order to do so, we need a family of  gauge-fixing conditions. We simply pick a representative 2d connection $a[m, \bar m] = (a_z[\bar m], a_{\bar z}[m])$ for every point $m$ in $\M(G,C)$ and gauge-fix $a = a[m, \bar m]$.
The aim is to reduce the integral over $a$ to an integral over $m,\bar m$.

We  introduce Faddeev-Popov ghosts for this gauge-fixing in the customary manner. For this, we introduce adjoint-valued ghosts $c$, $\bar c$ associated to complexified gauge parameters $\lambda$ and $\bar \lambda$, and  an adjoint-valued 1-form $(b_z, \bar b_{\bar z})$
associated to the gauge-fixing condition. The ghost action is 
\begin{equation}
\int\left( \Tr \,b_z D_{\bar z} c +  \Tr \,b_{\bar z} D_z c\right)\d^2z.
\end{equation}  
This ghost system has holomorphic and anti-holomorphic Kac-Moody symmetries, with levels $(2h,2h)=(-2\kappa_c, -2\kappa_c)$.
The BRST current is known to be nilpotent if and only if  the total anomaly of matter plus ghosts vanishes, that is, if and only if $\kappa - 2 \kappa_c=0$.

The integrand over $\M(G,C)$ is prepared with the help of the $b$ zero modes. If we denote the matter partition function as $\Psi(a)$, the gauge-fixed path integral becomes 
\begin{equation}
\int_{\M(G,C)} \Psi\left(a[m,\bar m]\right) \left\langle \prod_i \left[\int_C b_z \frac{\partial a_{\bar z}[m]}{\partial m_i} \d m_i \right] \left[\int_C b_{\bar z} \frac{\partial a_{z}[\bar m]}{\partial \bar m_i} \d \bar m_i \right] \right\rangle
\end{equation}   Here we have simply imitated the usual definition of the path integral of the bosonic string coupled to a conformal field theory of
holomorphic and antiholomorphic central charge $c=26$.
We thus learn that gauge-covariant functionals with level $2 \kappa_c$ correspond to densities on $\M(G,C)$. 

Gauge-covariant functionals with level $\kappa_c$, as in (\ref{eq:cogauge}), thus correspond to half-densities on $\M(G,C)$. We can write down explicitly the Hilbert space inner product in this presentation:
\begin{equation}\label{eq:innergauge}
\left( \Psi', \Psi \right) \equiv \int_{\M(G,C)} \overline{\Psi}'\left(a[m,\bar m]\right) \Psi\left(a[m,\bar m]\right) \left\langle \prod_i \left[\int_C b_z \frac{\partial a_{\bar z}[m]}{\partial m_i} \d m_i \right] \left[\int_C b_{\bar z} \frac{\partial a_{z}[\bar m]}{\partial \bar m_i} \d \bar m_i \right] \right\rangle
\end{equation}

We now have two ways to associate a functional $\Psi(a)$ to a (nice enough) state $\Psi \in \H$: we identify an abstract state with a half-density on $\M(G,C)$ and promote 
it to a functional of connections, or we contract $\Psi$ with the states $\Delta(a)$ produced by the shifted Dirichlet boundary condition $\mathrm{Dir}(a)$.  To show
that these two procedures are equivalent, we can reason as follows.
Classically, the BAA brane $\mathrm{Dir}(a)$ is supported on the fiber of $T^*\M(G,C)$ at the bundle $E$ defined by $a_{\bar z}$. 
This is the simplest type of conormal Lagrangian submanifold  and the corresponding state $\Delta(a)$ has delta function support
at $E$. So the pairing  $ (\Delta(a), \Psi)$ 
just evaluates the functional corresponding to $\Psi$ at the connection $a$.

As a final exercise, we can return to the definition of the quantum Hitchin Hamiltonians. We would 
like to derive a formula expressing the functional $[\D_\P(z) \circ \Psi](a)$ resulting from the action of a
Hamiltonian on some $\Psi$ in terms of  the functional $\Psi(a)$ associated to $\Psi$. We consider a strip with $\B_\cc$ boundary conditions on the left boundary and some initial condition at the
bottom of strip that, together with data at the corners, defines the state $\Psi$.  (This setup was sketched in fig. \ref{example6}(b), where the brane at the bottom
of the strip is called $\B_F$.)
By definition, the functional 
$[\D_\P(z) \circ \Psi](a)$ is computed by inserting $\P[\varphi_z]$ along the left boundary and moving it to the lower left corner of
the strip.    When we do this,  $\P[\varphi_z]$ is converted to 
the central element $S_{\P}(z)$ of the chiral algebra, which is a regularized polynomial in $J_z$ and its derivatives. In turn, 
the $J_z$ insertions can be traded for functional derivatives with respect to $a_{\bar z}$. 
As a result, $[\D_\P(z) \circ \Psi](a)$ is expressed as a certain differential operator $\D^{(a)}_\P(z)$ acting on $\Psi(a)$. The operator 
 $\D^{(a)}_\P(z)$ is a regularization of $\P\left[\frac{\delta}{\delta a_{\bar z}} \right]$. It maps gauge-covariant functionals to 
 gauge-covariant functionals precisely because $S_{\P}(z)$ is central: the $S_{\P}(z)$ insertion does not modify the 
 Ward identities of the currents and thus the differential operator ${\cal D}^{(a)}_{\P}(z)$ commutes with the 
 gauge-covariance constraints (\ref{eq:cogauge}). This is actually how the quantum Hitchin Hamiltonians 
 are defined mathematically \cite{BD}: they encode the effect of an $S_{\P}(z)$ insertion in a conformal block
 for the critical Kac-Moody algebra. 

A similar presentation of Hecke operators requires a discussion of boundary 't Hooft operators 
at Dirichlet boundary conditions and  their endpoints at the junction. In the auxiliary 3d perspective, 
the boundary 't Hooft operators map to boundary monopole operators. The classical moduli space of such disorder operators 
was discussed in a similar setting in \cite{Costello:2020ndc}: it coincides with the affine Grassmannian $\Gr_{G_\C}$. 
In the presence of the disorder operator, the gauge bundle at some small distance from the boundary is a specific Hecke modification 
of whatever fixed bundle is determined by the boundary value of the connection. Correspondingly, in order for $\varphi_z$  
to be non-singular at some distance from the boundary it must have some prescribed poles and zeroes at the boundary. 
This bare boundary monopole configuration can be dressed by local functionals of $\varphi_z$. 

In Section \ref{sec:chiral} we will discuss the ``spectral flow operators'' $\Sigma_g$ in the chiral algebra, which are labelled by a point in 
$\Gr_{G_\C}$ and enforce an appropriate version of the constraint on $J_z$. The spectral flow operators and their Kac-Moody descendants 
can play the role of endpoints of boundary 't Hooft operators. We will observe the existence of certain (continuous) linear combinations of 
spectral flow operators which are central and can thus play the role of the images $S_{R,n}(z)$ of endpoints of bulk 't Hooft operators. 
This will allow us to formulate Hecke operators in a 2d chiral algebra language.

\subsection{Nahm Pole Boundary Conditions}\label{nbc}

Half-BPS Dirichlet boundary conditions can be generalized to a larger collection of Nahm pole boundary conditions labelled by an embedding 
$\rho:\mathfrak{su}(2)\to\g$. These boundary conditions allow for a choice of background connection $a_\rho$ whose structure group commutes with $\rho$. 

Following the analogy with the $\Psi \neq 0$ results in \cite{Gaiotto:2017euk}, we will tentatively identify the chiral algebra 
associated to these boundary conditions with the critical level limit of the ${\cal W}^G_{\rho;\kappa}$ chiral algebras, 
which are in turn defined as the Drinfeld-Sokolov reduction associated to $\rho$ of a $G$ Kac-Moody algebra at level $\kappa$. 
As a basic check of this proposal, we observe that ${\cal W}_{\rho;\kappa_c}$ has the same large center as critical Kac-Moody, 
generated by appropriate $S_{\cal P}(z)$.

A particularly interesting case is the Nahm pole associated to a regular embedding $\rho$. The corresponding chiral algebra is 
the classical limit of a $W$-algebra and is completely central. It is generated by the $S_{\cal P}(z)$: all local operators on the junction  
are specializations of local operators on the deformed Neumann boundary. 

The regular Nahm pole is associated to a BAA brane $\B_N$ supported on the Hitchin section of the Hitchin fibration. This section is a complex Lagrangian
submanifold of $\M_H$, of type BAA, but it  lies completely outside $T^*\M(G,C)$.
If an eigenstate $\Psi$ of the Hitchin Hamiltonians is viewed purely as a square-integrable half-density on $\M(G,C)$, then it would appear not to make any sense to compute
the inner product of $\Psi$ with a state created by $\B_N$ (with appropriate corners), as the space of square-integrable half-densities comes by quantization of
$T^*\M(G,C)$, which is completely disjoint from the support of $\B_N$.    However, as discussed in Section \ref{wkb}, $\Psi$ is actually associated to a brane in $\M_H$
of compact support, and therefore should have a well-defined pairing with the state created by any brane.   In fact, in the dual $B$-model, the computation is
straightforward.   We return to this point at the end of Appendix \ref{bmodel}.

The regular Nahm pole boundary supports boundary 't Hooft lines which were studied in \cite{Witten:2011zz}. They are in natural correspondence 
with bulk 't Hooft lines, and they are indeed the image of bulk 't Hooft lines brought to the boundary. Nahm pole boundary conditions decorated by boundary 
't Hooft operators thus give rise to BAA branes supported on the Hecke modification of the Hitchin section 
at a collection of points. If the number of points is large enough, these BAA branes are nice submanifolds in $T^*\M(G,C)$. The associated 
states should play a role in the separation of variables analysis of \cite{T}.

\subsection{Enriched Neumann Boundary Conditions}\label{Enriched}

A basic BAA boundary condition in the 2d $\sigma$-model of $\M_H(G,C)$ is the Lagrangian boundary condition associated
to the Lagrangian submanifold $\M(G,C)\subset \M_H(G,C)$.    In 4d terms, this comes 
from a half-BPS boundary condition of type BAA  in  in which the gauge field $A$ satisfies Neumann boundary conditions and the Higgs field  $\phi$
satisfies Dirichlet boundary conditions.\footnote{The brane $\B_\cc$ comes instead from  a deformation of Neumann boundary conditions for $A$, in a
sense described in footnote \ref{deformed}, extended to the rest
of the supermultiplet in a different fashion and preserving a different symmetry (ABA rather than BAA).}   We will refer to this boundary condition as BAA Neumann.

In terms of the $\varphi\cF$ theory, this is simply the boundary condition defined by $\varphi|_{\partial N}=0$, with no constraint on $A_{\bar z}|_{\partial N}$.
We showed in Section \ref{Dbc} that this is the boundary condition one gets from the Euler-Lagrange equations of the action $I_{\varphi\cF}$, with ``free''
variations of all fields.

With this boundary condition, no restriction is placed on a gauge transformation on the boundary.   That is consistent, because the boundary
condition $\varphi|_{\partial N}=0$ is gauge-invariant.   However, there is a gauge anomaly on a boundary of $N$ that has this boundary condition.
The anomaly coefficient is $+h$.

An easy way to see this is to consider the $\varphi\cF$ theory on a slab $N=I\times C$,  with the  $\cA_{\bar z}=0$ boundary condition at the left end
of the slab and the $\varphi_z=0$ boundary condition at the right end.   These boundary conditions are invariant under constant gauge transformations
by an element $g\in G$.  At the left end of the slab, a constant gauge transformation is interpreted as a global symmetry.    
This gives an action of $G$ as a group of global symmetries of the theory on the slab.  
 The theory with $\varphi_z=0$
at one end of the slab and $\cA_{\bar z}=0$ at the other end is completely trivial: up to a gauge transformation, the only classical solution
is $\varphi_z=\cA_{\bar z}=0$ everywhere, and there are no low energy excitations.  So the $G$ action is anomaly free. As it acts by a constant gauge
transformation, its anomaly coefficient is the sum of the anomaly coefficient of the global symmetry at the left end of the slab and of the gauge symmetry at the right end.
We learned in Section \ref{Dbc} that the global symmetry has an anomaly coefficient $-h$ at the left end of the slab.  So
the gauge symmetry must have an anomaly coefficient $+h$ at the right end.   

So in short, the $\varphi_z=0$ boundary condition in the $\varphi\cF$ theory has anomaly $+h$.
A possible cure for the anomaly is to add extra degrees of freedom at the junction with anomaly $-h$, the critical level.  Unitary degrees of freedom at the junction
will not help, as they have a positive anomaly coefficient.
Instead, we can do the following.   
BAA Neumann boundary conditions can be enriched, preserving the supersymmetry of type BAA, by adding to the boundary
 3d matter degrees of freedom that make a 3d superconformal quantum
 field theory (SQFT) with $\N=4$ supersymmetry. 
  We will call the resulting boundary condition
   an enriched Neumann boundary condition (of type BAA, if it is necessary to specify this).   The interesting case is that the SQFT has
  $G$ symmetry and is coupled to the gauge field $A$ of the bulk ${\cal N}=4$ theory; this is possible, because with  Neumann boundary conditions, $A$
  is unconstrained on the boundary.  
 The $A$-twist of the bulk 4d theory induces an $A$-twist of the  boundary
3d SQFT.   The twisted boundary theory can contribute a negative amount to the anomaly at the junction.  For our application, we want  holomorphic boundary conditions 
for the 3d SQFT that support a $G$ Kac-Moody algebra at critical level $\kappa_c=-h$.

For this purpose, we can employ one of the holomorphic boundary conditions defined in \cite{Costello:2018fnz}.   The effect of ``enrichment'' is
that the boundary condition for the $\varphi\cF$ theory ending on an enriched Neumann boundary is no longer $\varphi_z|_{\partial N}=0$. Rather, $\varphi_z|_{\partial N}$ 
 equals the critical Kac-Moody currents of the boundary chiral algebra of the SQFT. In particular, the images $S_{\P}(z)$ of $\P[\varphi_z]$ are identified with the central elements built from the critical Kac-Moody currents for the matter. 

The simplest example is the case that the SQFT is a theory of free 3d hypermultiplets  transforming in a symplectic representation $R$ of $G$. Let $Z$ be
the bosonic field in the hypermultiplets.  Twisting turns the components of 
 $Z$  into spinors, still valued in the representation $R$.
As analyzed in  \cite{Gaiotto:2016hvd}, with the appropriate sort of boundary condition, the twisted hypermultiplet path integral on a three-manifold
with boundary is a 2d contour path integral, with a holomorphic action, of the general sort described in Section \ref{ACP}.   In this case, the holomorphic action is
\begin{equation}
S[Z,\cA] = \int_C \langle Z, \bar \partial_\cA Z \rangle .
\end{equation} 
where  $\langle \cdot,\cdot \rangle$ denotes the symplectic pairing on the representation $R$, and we have included a coupling to the complex gauge
field $\cA$. Fields $Z$ with such an action are sometimes called symplectic bosons and do have a negative Kac-Moody level; see Section \ref{sb} for more about them. The simplest possibility is to select an $R$ for which the level is precisely$-h$. It is also possible to select an $R$
for which the level is more negative and make up the difference with some extra 2d chiral fermions in a real representation $R_f$ of $G$ placed at the junction. We will discuss the simplest possibility here and briefly comment on the general case at the end. 

 Now consider a junction between the deformed Neumann boundary condition that supports the $\varphi\cF $ theory and the
BAA Neumann boundary condition enriched by hypermultiplets.   The appropriate holomorphic action is the sum of $I_{\varphi\cF}$ and $S[Z,\cA]$:
\be\label{welli}\h I=\int_N \Tr\,\varphi \cF +\int_{C=\partial N}  \langle Z, \bar \partial_\cA Z \rangle .\ee
The Euler-Lagrange equation for $\cA_{\bar z}$ gives a boundary condition
\be\label{nell}\varphi_z|_{\partial N}=\mu(Z), \ee
where $\mu(Z)$ is the holomorphic moment map for the action of $G_\C$ on the representation $R$.  Here the components of $\mu(Z)$ become
(after quantization)  the Kac-Moody
currents of the matter system, so this formula illustrates the statement that after enrichment, the appropriate boundary condition sets $\varphi$ equal to the Kac-Moody currents.  

We also have the classical equations of motion 
\begin{align}
0=\bar\partial_\cA\varphi =\bar\partial_\cA Z.
\end{align}
Triples $(\cA,\varphi,Z)$ satisfying these conditions along with eqn. (\ref{nell}) describe a brane over $\M_H(G,C)$ of type BAA.   The simplest case is that for given $\cA,\varphi$, there is at most one $Z$ satisfying the conditions. If so, the pairs $(\cA,\varphi)$ for which such a $Z$ does exist furnish a complex Lagrangian submanifold of $\M_H(G,C)$, in complex structure $I$, corresponding to a brane of type BAA.
These Lagrangian submanifolds are  of conormal type, since if a suitable $Z$ exists for one Higgs pair  $(\cA,\varphi)$, then a suitable $Z$ likewise exists after any rescaling of $\varphi$.   

The natural quantization of these BAA branes is a path integral over $Z$ \cite{Gaiotto:2016wcv}:
\begin{equation}\label{eq:ZPsi}
\Psi(a) = \int DZ D\bar Z \,e^{\int_C \left[\langle Z, \bar \partial_a Z \rangle -  \langle 
\bar Z, \partial_a \bar Z \rangle\right]}
\end{equation} 
possibly modified by the insertion of a non-trivial corner in the form of a collection of $Z$ and $\bar Z$ insertions in the path integral \cite{Costello:2018fnz}. 
From the point of view of the present paper, the meaning of this formula is as follows.  We place the enriched Neumann brane at the bottom of a strip,
playing the role of the brane denoted as $\B_x$ in fig. \ref{example6}(b).    Assuming no operator insertions are made at the bottom corners  of the strip,
the state in $\H=\Hom(\b\B_\cc,\B_\cc)$ defined by this picture is $\Psi(a)$.   The statement makes sense, because the $Z$ and $\bar Z$ fields support
current algebras at critical level $\kappa_c=-h$, so that the path integral of these fields does indeed define a half-density on $\M(G,C)$.  

The chiral algebra at the junction in this construction consists of the 
subalgebra of gauge-invariant operators within the boundary chiral algebra of the 3d matter theory, i.e. it consists of operators built from the $Z$'s and their derivatives which have trivial OPE with the critical Kac-Moody currents.   One can modify the construction just described by including chiral and antichiral operators at the
bottom corners of the strip; to describe the resulting state, one just includes the corresponding factors in eqn. (\ref{eq:ZPsi}).

We will discuss this construction  further  in Section \ref{sec:symp}. 

 \begin{figure}
 \begin{center}
   \includegraphics[width=3.2in]{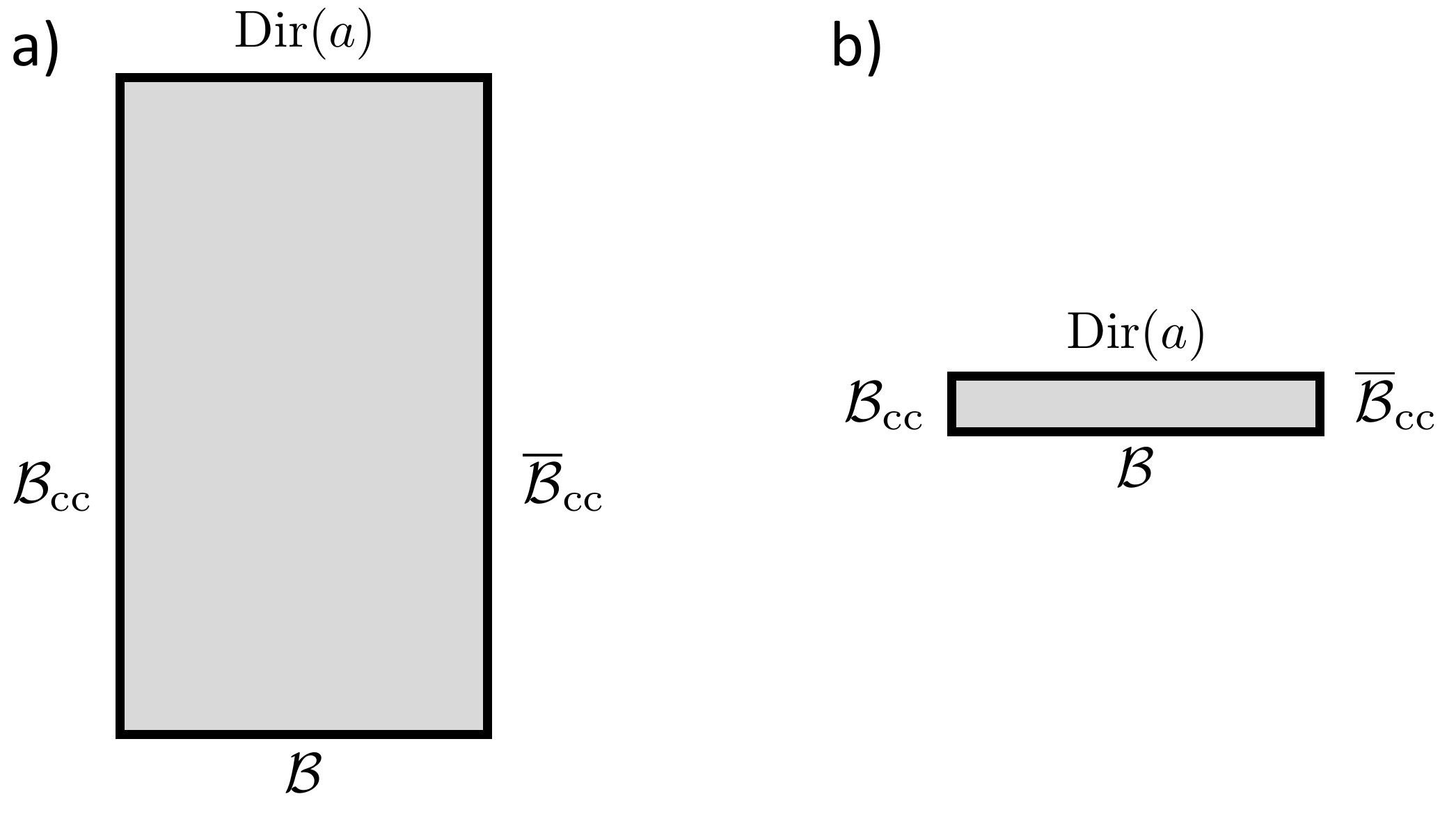}
 \end{center}
\caption{\small  (a)  A rectangle with an enriched Neumann brane $\B$ at the bottom and a generalized Dirichlet brane $\B'$ at the top.   (b)  In topological field
theory, the ``height'' and ``width'' of the rectangle are arbitrary.   In a limit in which the height is small, we reduce to a purely three-dimensional computation on
a product $I\times C$. As always, $C$ is not drawn. \label{example9}}
\end{figure} 

There is an alternative way to understand (\ref{eq:ZPsi}) directly in 4d. The alternative perspective can be applied as well to a more general
 situation where the corresponding BAA brane has a non-trivial $\CP$ bundle or where extra chiral fermions are added at the junctions. In order to read off $\Psi(a)$, we can contract the state $\Psi$ created by the enriched Neumann boundary with the state $\Delta(a)$  created by a Dirichlet boundary condition $\mathrm{Dir}(a)$.   The inner product between these two states is represented by
the path integral on the rectangle of fig. \ref{example9}(a) with $\mathrm{Dir}(a)$  boundary conditions at the top and enriched Neumann at the bottom.  In two-dimensional topological field theory, the ``height'' and ``width'' of the rectangle are arbitrary.
Take the limit that the height is much less than the width (fig. \ref{example9}(b)).   In this limit, the path integral reduces to a path integral in a 3d theory on $I\times C$.
The 3d theory is produced by compactification from four to three dimensions on an interval  with Dirichlet boundary conditions 
at one end and enriched Neumann boundary conditions at the other end.
   This compactification gives a simple answer, because the 4d fields are all frozen at one boundary or the other: one just gets back the 
 same 3d theory which was employed to construct the enriched Neumann boundary conditions.  In our example, this is the theory of the same free hypermultiplets
 that we started with, with the global $G$ symmetry now identified 
 with the $G$ symmetry that acts at the Dirichlet boundary. The answer of (\ref{eq:ZPsi}) is just the partition function of the 3d theory on $I\times C $, with the boundary conditions which give rise to the symplectic bosons or  their complex conjugates. We can thus apply (\ref{eq:ZPsi}) to a situation where the corresponding BAA brane is complicated, bypassing the 2d derivation. Any (anti)chiral fermions added at the junctions would just contribute their partition function, i.e.
\begin{equation}\label{eq:ZPsifer}
\Psi(a) = \int DZ D\bar Z D\psi D\bar \psi \,e^{\int_C \left[\langle Z, \bar \partial_a Z \rangle -  \langle 
\bar Z, \partial_a \bar Z \rangle\right]+\left[( \psi, \bar \partial_a \psi ) - ( 
\bar \psi, \partial_a \bar \psi)\right]}
\end{equation} 
As long as the combined level of the symplectic bosons and fermions is $-h$, this represents a half-density on $\M(G,C)$.

In this section we described states associated to elementary boundary conditions. The construction can be easily generalized to describe operators associated to analogous elementary interfaces. The composition of elementary interfaces can produce a vast collection of BAA boundary conditions and interfaces, which are associated to the composition of the corresponding operators. This would allow, among other things, the calculation of $C \times [0,1]$ partition functions for A-twisted 3d ${\cal N}=4$ gauge theories with chiral and antichiral boundary conditions at the two ends of the segment. We leave a detailed analysis of this problem, as well as the B-model analogue, 
to future work.

\section{Hecke Operators and Spectral Flow Modules}\label{sec:chiral}

\subsection{Preliminaries}

The quantization of BAA branes associated to enriched Neumann boundary conditions has given us examples (\ref{eq:ZPsi}) of wavefunctions which are defined as partition functions of 2d CFTs with chiral and antichiral critical Kac-Moody symmetry. In this section we describe how to compute the action of quantum Hitchin Hamiltonians and Hecke operators on such partition functions, directly in a 2d CFT language. At the same time, we will gain a better appreciation of the mathematical results we invoked in Section \ref{wtw} to define the Hecke operators.

We have already discussed briefly the 2d CFT interpretation of the quantum Hitchin Hamiltonians. The critical Kac-Moody chiral algebra 
has a large center, generated by certain local operators $S_{\cal P}(z)$ which have non-singular OPE with the currents.  
The transformation of a correlation function under complexified gauge transformations is described
by the Ward identities for the currents.
The statement that $S_{\cal P}(z)$ has non-singular OPE with the currents means that a correlation function with insertions of such operators only
\begin{equation}
\langle S_{{\cal P}_1}(z_1) \cdots S_{{\cal P}_n}(z_n) \rangle_a
\end{equation}
satisfies  the same transformation properties (\ref{eq:cogauge}) as a partition function. It thus also defines a half-density on $\M(C,G)$. 

Furthermore, the $S_{\cal P}(z)$ are assembled from Kac-Moody currents, which can be traded for functional derivatives with respect to 
the connection. We can thus expand recursively 
\begin{equation}
\langle S_{{\cal P}_1}(z_1) \cdots S_{{\cal P}_n}(z_n) \rangle_a = {\cal D}^{(a)}_{\P_1}(z_1 )\langle S_{{\cal P}_2}(z_2) \cdots S_{{\cal P}_n}(z_n) \rangle_a
\end{equation}
and the final answer will be independent of the order of the operators to which we apply the recursion. The differential operators ${\cal D}^{(a)}_{\P_1}(z_1 )$ thus commute. 

Although here we referred to correlation functions of some 2d CFT, this is unnecessary: given a half-density on $\M(C,G)$ represented by 
a gauge-covariant functional $\Psi(a)$ on the space of connections, the functional derivatives with respect to $a$ behave just as Kac-Moody currents.
The differential operators ${\cal D}^{(a)}_{\P}(z)$ represent in a gauge-covariant manner the action of the quantum Hitchin Hamiltonians 
${\cal D}_{\P}(z)$ on the half-density $\Psi$. 

When doing calculations in a neighborhood $U$ of a point $p$ in $C$, it is usually helpful to choose a representative connection for the bundle 
which vanishes on $U$. This is always possible because $\cA_{\bar z}$ can be set to zero locally by a complex-valued 
gauge transformation. That amounts to trivializing the bundle over $U$, as we did in discussing general Hecke transformations in Section
\ref{affgr}.  Then  the Kac-Moody currents are meromorphic on $U$ and satisfy the Kac-Moody OPE in a standard form
\begin{equation}
J^a(z) J^b(w) \sim \frac{\kappa_c \delta^{ab}}{(z-w)^2} + \frac{f^{ab}_d J^d(w)}{z-w}.
\end{equation}

Recall the definition of the Fourier modes of the Kac-Moody algebra
\begin{equation}
J^a_n \equiv \oint_{|z|=\epsilon} \frac{\d z}{2 \pi \i} z^n J^a(z)
\end{equation}
The insertion of such a Fourier mode represents an infinitesimal deformation of $a_{\bar z}$ supported on the loop $|z|=\epsilon$,
or a deformation of the bundle which modifies the gluing of a bundle over $U$ to a bundle over the rest of the surface by an infinitesimal gauge transformation in $U'=U\backslash p$.  

In the absence of other operator insertions in the disk $|z|<\epsilon$ (or in the presence of central operator insertions) correlation functions with insertions of
the non-negative Fourier modes vanish. The corresponding infinitesimal gauge transformations can be extended to $U$ and do not change the bundle. 
They represent changes in the original trivialization over $U$. In the presence of a generic operator insertion in the disk, 
the non-negative modes act non-trivially: the insertion of a general local operator requires some choice of trivialization of the bundle and 
the result depends on the choice. 

The negative Fourier modes can act non-trivially even in the absence of other operator insertions 
and represent infinitesimal gauge transformations which can change the bundle. 
Repeated action of the negative modes builds the image at $z=0$ of the vacuum module for the Kac-Moody algebra.
The operator 
$S_{\cal P}(0)$ and other central elements in the chiral algebra correspond by the operator-state correspondence to the vectors in the 
vacuum module that are annihilated by all the non-negative Fourier modes of the currents. For example, the Sugawara vector  is 
\be
|S_2\rangle \equiv \Tr \,J_{-1} J_{-1} |0 \rangle,
\ee with similar formulas for other central elements.

\subsection{Hecke Operators as Central Vertex Operators}\label{cvo}

The  Hecke integral operators can also be analyzed with 2d chiral algebra technology. We would like to lift the Hecke operators to operators acting on gauge-covariant functionals and give them a 2d chiral algebra interpretation in terms of the insertion of  local operators which have trivial OPE with the Kac-Moody currents. Such a formulation  immediately guarantees that the Hecke operators commute with the quantum Hitchin Hamiltonians
and with other Hecke operators. 

We can follow verbatim the definition of Hecke modifications from Section \ref{affgr}. First, we trivialize the bundle $E$ on a  small neighborhood $U$ of a point $p$. We can then think of $E\to C$
as built by gluing a trivial bundle over $U$ to the bundle $E$ over $C\backslash p$ with a trivial gluing map.  Then we produce a new bundle $E'$ by modifying the gluing map to $z^\m$ (where $\m$ is an integral weight of the dual group and $z$ is a local parameter at $p$). The bundles $E$ and $E'$ can be described by the same connection away from $U$. The connection $a$ which describes $E$ vanishes on $U$, while the connection $a'$ which describes $E'$ coincides with $a$ outside of $U$
and can be taken in $U$ to be some specific reference connection supported on an annulus in $U'$, and proportional to  $\m$.  

Take the functional $\Psi$ which represents the input wavefunction, and evaluate it on $a'$. This gives a new functional $\Psi_{\m}(a)$. 
Crucially, $\Psi_\m(a)$ is not covariant under complexified gauge transformations: the new bundle $E'$ depends on the original choice 
of trivialization of $E$. Formally, $\Psi_\m(a)$ and its functional derivatives can be interpreted as correlation functions of Kac-Moody currents in the presence of a ``spectral flow operator'' $\Sigma_{\m}(0)$.   The relation between Hecke operators and spectral flow automorphisms of Kac-Moody algebras was previously observed
from a different but related point of view in Section 8.5 of \cite{Teschner:2010je}.

The term ``spectral flow'' refers to a certain automorphism of the Kac-Moody algebra:
\begin{align}
J^\alpha_n &\to J^{\alpha}_{n+(\m, \alpha)} \cr
J^h_n &\to J^h_n - \m \kappa \delta_{n,0}
\end{align}
where $J^\alpha$ is the current associated to a root $\alpha$ and $J^h$ are the Cartan currents. 
This is precisely the effect of a $z^\m$ gauge transformation on the Fourier modes of the currents. 

By definition, a spectral flow module is the image of the vacuum module under the spectral flow. In particular, the image of the vacuum vector under spectral flow is annihilated by $J^{\alpha}_{n+(\m, \alpha)}$ with non-negative $n$ and is an eigenvector of  $J^h_0$ with a nontrivial eigenvalue. 
Correspondingly, a spectral flow operator $\Sigma_{\m}(0)$ is a local operator such that the OPE with the Kac-Moody currents become non-singular after a $z^\m$ gauge transformation. The $J^{\alpha}(z)$ will have a pole/zero of order $(\m, \alpha)$ at $z=0$ and $J^h(z)$ 
will have a simple pole of residue\footnote{Bosonization offers a convenient way to describe $\Sigma_{\m}$.
Schematically, if the Cartan currents are bosonized as $J^h = \partial \phi^h$ and the remaining currents as vertex operators $J^\alpha = e^{\frac{\alpha}{\kappa} \cdot \varphi}$, then the spectral flow operator can be represented by a vertex operator $\Sigma_{\m} = e^{\m \cdot \varphi}$ as well. This representation can be useful for some calculations, but behaves poorly under general $G_\C$ gauge transformations.} $\m\kappa \Sigma_{\m}(0)$.

We stress again that the functional $\Psi_{\m}(a)$ does not represent a half-density on $\M(G,C)$, as it depends on the 
choice of trivialization of $E$. The properties of the spectral flow operator $\Sigma_{\m}(0)$ characterize the precise failure of the 
gauge-covariance constraints (\ref{eq:cogauge}). Our objective is to build from $\Sigma_{\m}(0)$ some local operator insertion which 
is central and can thus represent the action of a Hecke operator on $\Psi(a)$. 

Before continuing with the general discussion, we  present the reference example of $G=\SO(3)$ and minimal charge. The basic spectral flow automorphism is 
\begin{equation}
J^\pm_n \to J^{\pm}_{n \pm 1} \qquad \qquad J^0_n \to J^0_n - \delta_{n,0}
\end{equation}
The spectral flow module is built from a vector $|1\rangle$ which satisfies 
\begin{align}\label{laterref}
J^{\pm}_{n \pm 1} |1\rangle &=0  \qquad \qquad n\geq 0 \cr
J^0_n |1\rangle &=0  \qquad \qquad n> 0 \cr
J^0_0|1\rangle &= |1\rangle.
\end{align}
For this case of the basic ``charge 1'' spectral flow operator, we will write $\Sigma_1(0)$ for $\Sigma_\m(0)$.

Following our discussion of the affine Grassmannian in Section \ref{affgr}, we can 
replace the gluing map $z^\m$ by another gluing map $g$ in the same orbit $\Gr^\m$.
The same construction with $z^\m$ replaced by $g$ produces a functional $\Psi_{g}(a)$. The functional derivatives of $\Psi_{g}(a)$ can be interpreted as correlation functions in the presence of a modified spectral flow operator $\Sigma_{g}(0)$. As a change of trivialization is implemented by the non-negative modes of the currents, we can express the action of these modes on $\Sigma_{g}(0)$ as certain differential operators along $\Gr^\m$.

We should stress that the definition of $\Psi_{g}(a)$ really requires a choice of reference connection supported within $U'=U\backslash p$ which realizes the gluing map $g$. Different connections describing the same $g$ are related by complex gauge transformations and 
thus may lead to a different normalization for $\Psi_{g}(a)$ and $\Sigma_{g}(0)$. As a result, $\Psi_{g}(a)$ and $\Sigma_{g}(0)$ are 
actually sections of a certain line bundle on $\Gr^\m$. We will indentify this line bundle in Section \ref{fft}.
The non-negative modes of the currents will act as vector fields on sections of
this line bundle. This is the chiral algebra manifestation of the mismatch between the bundles of half-densities before and after the Hecke modification. 

The line bundle on $\Gr^\m$ is controlled by the level of the Kac-Moody algebra. In the next section, we will show that
at critical level, this line bundle coincides with the bundle of densities on $\Gr^\m$. This means that in a theory that has holomorphic and antiholomorphic
Kac-Moody levels that are both critical, the spectral flow operator is a density on $\Gr^\m$ and can be naturally integrated:
 \begin{equation}\label{tendef}
\h\Sigma_\m(0) \equiv \int_{\Gr^\m}\Sigma_{g}(0) |\d g|^2.
 \end{equation}
$\h\Sigma_\m(0)$  has the appropriate properties for the Hecke operator of charge $\m$ dual to a Wilson operator
with minimal corners $s_{R}$ in the language of Section \ref{wilop} (that is, a Wilson
operator defined using wavefunctions built from highest weight vectors).  The integral over $\Gr^\m$ generalizes the integral over $\bCP^1_x$ in eqn. (\ref{tonf}).
 
The action of a non-negative mode of the currents on $\h\Sigma_\m(0)$ can be traded for a 
Lie derivative of $\Sigma_{g}(0)$ along the corresponding vector field on $\Gr^\m$. As long as no boundary terms appear upon integration by parts (this may
require a technical analysis when monopole bubbling is possible), $\h\Sigma_\m(0)$ will be annihilated by the non-negative modes of the Kac-Moody algebra and is thus central. Correspondingly, the averaged functional 
 \begin{equation} \label{eq:chirhecke}
\int_{\Gr^\m}\Psi_{g}(a)|\d g|^2
 \end{equation}
 obtained by acting with $\h\Sigma_\m(0)$ on $\Psi(a)$ 
is gauge-covariant and can represent the action of the principal Hecke operator of charge $\m$.

We can readily apply this construction to our illustrative example of $G=\SO(3)$ and minimal charge. 
We can define a $\bCP^1$ family of spectral flow operators $\Sigma_{1;\mu}(0)$ as a global $\SO(3,\C)$ rotation of $\Sigma_1(0)$.
Formally, we can write the corresponding states as 
\be 
|1;\mu\rangle = e^{\mu J^+_0 }|1\rangle
\ee
It is straightforward to express the action of the non-negative Fourier modes on $|1;\mu\rangle$ as differential operators in $\mu$
and verify that they are total derivatives. It is clear that 
\be 
J^+_0 |1;\mu\rangle = \partial_\mu e^{\mu J^+_0 }|1\rangle=  \partial_\mu |1;\mu\rangle
\ee
The action of $J^0_0$ is also straightforward
\be 
J^0_0 |1;\mu\rangle = e^{\mu J^+_0 }(J^0_0 + \mu J^+_0)|1\rangle  =  \partial_\mu \left(\mu |1;\mu\rangle\right)
\ee
Computing the action of $J^-_0$ requires only a bit more work:
\be 
J^-_0 |1;\mu\rangle = e^{\mu J^+_0 }(J^-_0 + 2 \mu J^0_0 + \mu^2 J^+_0)|1\rangle  =  \partial_\mu \left(\mu^2 |1;\mu\rangle\right)
\ee

 This makes the insertion of 
\be \label{eq:averso}
\int \Sigma_{1;\mu, \bar \mu}(0)|\d \mu|^2
\ee
central, as long as boundary terms for the integration by parts vanish. 
The natural way to show that boundary terms vanish is to show that (\ref{eq:averso}) is really the integral of a density on 
$\bCP^1$. In order to do so, we need to cover $\bCP^1$ with a second patch, starting from the opposite spectral flow operator 
$\Sigma_{-1}(0)$ and deforming it to $\Sigma_{-1;\mu}(0)$ as 
\be 
|-1;\mu\rangle = e^{-\mu^{-1} J^-_0 }|-1\rangle
\ee
The action of the non-negative Fourier modes on this family involves the same differential operators in $\mu$ as for $\mu^2 |1;\mu\rangle$. 
Including the antichiral modes we find that we can consistently identify 
\begin{equation}
\Sigma_{-1;\mu}(0) = |\mu|^4 \Sigma_{1;\mu}(0)
\end{equation}
and combine them into a density $\Sigma_{1;\mu}(0)$ defined on the whole $\bCP^1$.
We explain a different and more general approach to this result in Section \ref{fft}.

The $\Sigma_{1;\mu}(0)$ insertion, by construction, 
corresponds to a very specific modification $a \to a'[a;\mu]$ of the background connection.
Recall that we work in a gauge where $a$ vanishes inside the open patch $U$ and 
$a'[a;0]$ differs from $a$ by some reference connection supported on an annular region in $U'$. The insertion of the exponentiated Fourier mode $e^{\mu J^+_0}$ adds a further specific modification to the connection on a wider annular region, producing $a'[a;\mu]$.

The integral operator (\ref{eq:chirhecke}) corresponding to (\ref{eq:averso}) is thus 
 \begin{equation}\label{eq:chirhesutwo}
\int_{\bCP^1}\Psi(a'[a;\mu])|\d \mu|^2
 \end{equation}
 Compare this with (\ref{actst}). We should write $\Psi(y)$ there as $\Psi(a(y))$ here, with $a(x)$ denoting our gauge-fixing 
 choice of a representative connection for every bundle $x$. There is no reason for $a'[a(x);\mu]$ 
 to be already in a gauge-fixed form. A complexified gauge transformation will be needed to bring it to the gauge-fixed form $a(x_\mu)$ for the modified bundle $x_\mu$. The anomaly will give some rescaling factor which we can write as the absolute value of a holomorphic quantity $\omega$:
  \begin{equation} \label{eq:omega}
\Psi(a'[a(x);\mu]) = |\omega(x;\mu)|^2 \Psi(a(x_\mu))
 \end{equation} 
 The integral operator becomes 
  \begin{equation}
\int_{\bCP^1}|\omega(x;\mu)|^2 \Psi(a(x_\mu))|\d \mu|^2
 \end{equation}
We obtain:
  \begin{equation}
F(x,y) = \int_{\bCP^1} |\omega(x;\mu)|\delta(y;x_\mu)|\d \mu|^2
 \end{equation}
where $\delta(y;x)$ is a delta function supported on the diagonal in $\M \times \M$. 

 The left hand side of (\ref{eq:omega}) is a half-density 
 in $x$ and a density in $\mu$. The right hand side involves a half-density in $y=x_\mu$. 
 We can thus identify $\omega(x;\mu)$ with $w(\vec x;\mu)$ in $(\ref{conf})$. 
The factor $\omega(x;\mu)$  encodes the anomalous rescaling of $\Psi$ under a complexified gauge transformation
and is thus non-vanishing. This allows us to identify it with the holomorphic factor $k$ introduced by \cite{BD}.\footnote{We will see momentarily that $\omega(x;\mu)$ could be computed in the theory of adjoint free fermions.} We identify (\ref{eq:chirhesutwo}) with the Hecke operator $H_{p=0}$ associated to the two-dimensional representation of $G^\vee$ with a minimal choice of corners corresponding to $|k|^2$, as in eqn. (\ref{tonf}).

\subsection{Free Fermion Trick}\label{fft}

In the last section, we observed that in a CFT with Kac-Moody symmetry, the operator $\Sigma_g(0)$, where $g$ is a gauge transformation associated to a Hecke transformation of weight
$\m$, is a section of a line bundle over $\Gr^\m$.   This line bundle, since it is determined by the anomaly, depends only on the central charge $\kappa$ of the CFT.   
We would like to compute this line bundle for a CFT of critical level $\kappa_c=-h$, but it turns out that it is particularly simple to compute it for a CFT
whose level is $-\kappa_c=+h$.  This will give us the inverse of the line bundle over $\Gr^\m$ that we actually want.
 
After picking a spin structure on $C$, or equivalently a choice of $K_C^{1/2}$, we consider a system of chiral (Majorana-Weyl) fermions  $\psi^a$ of spin 1/2 
valued in the adjoint representation of the gauge group.  The Kac-Moody currents are 
constructed as normal ordered fermion bilinears and have anomalous gauge transformation due to 
the normal ordering. The central charge is exactly $h=-\kappa_c$. 

We claim that the spectral flow operators in the theory of adjoint free fermions are sections of $K^{-1}_{\Gr^\m}$.   This means that the spectral flow
operators in a CFT at the critical level $\kappa_c$ are sections of the inverse of this or $K_{\Gr^\m}$.
We will illustrate the case of $G=\SO(3)$ and minimal $\m$, and briefly indicate the generalization to other $G$ and $\m$.

After picking a Cartan subalgebra of $\SO(3)$, we have
 chiral fermions $\psi^\pm$ and $\psi^0$. The basic Hecke modification at $z=0$ results in $\psi^+$ having a pole at $z=0$ and $\psi^-$ having a zero. This is implemented simply by a $\psi^-$ insertion at $z=0$. A Hecke modification associated to a point
$(u,v)\in\bCP^1$ is implemented by an $\SO(3,\C)$ rotation of $\psi^-$, i.e. by $\psi^-_{(u,v)} \equiv u^2 \psi^-  + 2 u v \psi^0 +  v^2 \psi^+$. An
insertion of $\psi^-_{(y,v)}(0)$ imposes the vanishing of  $\psi^-_{(u,v)}$ at $z=0$, while giving a pole to other linear combinations of the components of $\psi$.

In the theory of adjoint fermions, we thus have $\Sigma_{1;(u,v)}(0) = \psi^-_{(u,v)}$. 
This is quadratic in homogeneous coordinates of $\bCP^1=\Gr^\m$, 
so it is  a global section of $\O(2) = K_{\Gr^\m}^{-1}$, as claimed. 
Hence at critical level, the spectral flow operator in this example is a section of $K_{\Gr^\m}$, a fact that was exploited in Section \ref{cvo}.

For a general gauge group and charge, the reference Hecke modification results in the fermions labelled by a root $\alpha$ having extra poles or zeroes of order $(\lambda, \alpha)$ at $z=0$. This is implemented by a very simple vertex operator:
\begin{equation}\label{opspin}
\prod_{\alpha | (\lambda, \alpha)<0} \prod_{n_\alpha =0}^{-(\lambda, \alpha)-1}\partial_z^{n_{\alpha}}\psi^{\alpha}
\end{equation}
It is straightforward to see that this product transforms as a section of $K_{\Gr^\m}^{-1}$: each fermion derivative in the product matches one of the 
non-negative Fourier modes $J_{n_\alpha}^\alpha$ which act non-trivially on $\Sigma_\m$; these modes provide a basis of the tangent bundle to $\Gr^\m$.

This computation could be expressed as a comparison of the Pfaffian of the Dirac operator acting on $\psi$, before and after
the Hecke modification.  This Pfaffian is analyzed in detail in \cite{BD}.

The operator $\Sigma_{g}(z,\bar z)$ fails to be a true function of $z$ because of the gauge anomaly. Indeed, even a rescaling of the local coordinate $z \to \lambda z$ changes the singular gauge transformation from $z^\m$ to   $(\lambda z)^\m$ and thus results in the action of the Cartan zero modes $\m \cdot J_0$ on the spectral flow operator, resulting in a non-trivial scaling dimension proportional to $(\m,\m)$. 
We can study this anomalous dependence on $z$ with the help of the free fermion trick. 
For example, for $\SO(3)$ and minimal $\m$ we have an insertion $\psi^-$ which behaves as a section of $K_C^{1/2}$. 
Correspondingly, for critical level the spectral flow operator is a section of $K_C^{-1/2}\otimes \bar K_C^{-1/2}=|K_C|^{-1}$.
This remains true for the averaged Hecke operator $\h\Sigma_\m(0)$ because in this example, the coordinates on $\Gr^\m=\bCP^1$ have scaling dimension $0$ and thus the measure $\d \mu$ does not contribute to the scaling dimension.    The fact that $\h\Sigma_\m(0)$ is a section of $|K_C|^{-1}$ is expected from $S$-duality.  It matches
the fact that the holomorphic and antiholomorphic sections $s$ and $\bar s$ used to define the dual Wilson operator (Section \ref{wilop}) are sections of $K_C^{-1/2}$
and $\bar K_C^{-1/2}$, respectively.

For general groups and representations, matching the scaling dimension of $\h\Sigma_\m(0)$ with the behavior of the corresponding Wilson
operator is more subtle.    The scaling dimension of the fermionic insertion grows quadratically in the charge, but so does the 
negative scaling dimension of the measure $\d\mu$ on $\Gr^\m$. There is a nice cancellation between the derivatives on the fermions and the scaling 
dimension of the measure, so that the scaling dimension of $\h\Sigma_\m$ is linear in $\m$.   The scaling dimensions of dual Wilson operators were 
described in Section \ref{wilop}.

\subsection{Integral-differential Hecke Operators and the Oper Differential Equation}\label{opde}

In Section \ref{lineq}, as well as integral Hecke operators, whose kernel has delta function support on the Hecke correspondence, 
we considered integral-differential Hecke operators,
whose kernel is a derivative of a delta function. 
The natural way to build such more general Hecke operators is to consider the insertion of Kac-Moody descendants of 
$\Sigma_{g}$, which represent functional derivatives $\frac{\delta}{\delta a}$ taken in a neighbourhood of the 
location of the Hecke modification. We can restrict ourselves to descendants by the negative modes of the Kac-Moody currents, 
as the non-negative modes can be traded for $g$ derivatives which would be integrated by parts. 

The action of non-negative modes on a descendant of $\Sigma_{g}$ will produce some linear combination of $g$ derivatives of other descendants. We need some $g$-dependent combination of descendants which transform as a density on $\mathrm{Gr}^\m$ and such that the action of non-negative Kac-Moody modes will produce total $g$ derivatives. We can produce a simple example of that: $\partial_z \h\Sigma_\m(z, \bar z)$. Indeed, the $z$ derivative of a basic spectral flow operator $\Sigma_\m(z)$ coincides with the Cartan Kac-Moody descendant :
\begin{equation}\label{firstone} \partial_z \Sigma_\m(z)= \frac{1}{2 \pi \i} \oint \frac{\d w}{w-z} \m \cdot J (w) \Sigma_\m(z, \bar z)\equiv  \m \cdot J_{-1} \circ \Sigma_\m(z, \bar z). \end{equation}
To demonstrate this relation, recall that the insertion of $\Sigma_\m(z)$ in a correlation function represents a specific modification of the background connection in a neighbourhood $U$ of $z$. As we vary $z$, the modified connection changes. The change is supported in the annular region $U'$ and can be described by the insertion of a current integrated against the variation of the background connection. 
The entire comparison occurs within the $\U(1)$ subgroup of the gauge group determined by $\m$. 
As a small shortcut, we can compare the effect of the $z$ derivative and of the integrated current insertion at the level of the bundle modifications they implement. If $\Sigma_\m(z)$ implements the gauge transformation $g(w) = (w-z)^\m$ on $U'$, the $z$ derivative $\partial_z \Sigma_\m(z)$ implements $\partial_z  (w-z)^\m = \frac{\m}{z-w} (w-z)^\m$. The $\frac{\m}{z-w}$ part is identified with the gauge transformation produced by the 
$\m \cdot J_{-1}$ Fourier mode and $(w-z)^\m$ represents $\Sigma_\m(z)$ again. As we are working at the level of the bundle modification instead of the connection, we could be missing effects due to the anomaly.  A simple check in the free fermion theory can exclude that.\footnote{The bosonized description of $\Sigma_\m(z)$ is also an effective way to verify the computation.}

Inserting this relation into the definition of $\h\Sigma_\m(z, \bar z)$, we find that $\partial_z \h\Sigma_\m(z, \bar z)$ can be written as an integral over $\Gr^\m$ of a specific Kac-Moody descendant of $\Sigma_{g}(z, \bar z)$. 

Another natural way to produce well-defined integral-differential operators of this type is to consider descendants of 
$\h\Sigma_\m$ by modes of the Sugawara vector or other central elements. The resulting local operators are clearly gauge-invariant. 
We expect that the classification of $g$-dependent combinations of Kac-Moody descendants of $\Sigma_{g}(z, \bar z)$ which are a total derivative on $\mathrm{Gr}^\m$ will match the corresponding classification of 't Hooft line endpoints $\alpha_{R_\m,n}$. 

%

\subsection{Wakimoto Realization}
We will give here an alternative derivation of the properties of $\Sigma_1$ with the help of the Wakimoto construction at critical level. As a bonus, we will recover in a different way the oper differential equation.

The critical-level Wakimoto construction presents the Kac-Moody currents as the symmetry currents for a twisted $\beta\gamma$ system for $\bCP^1$:
\begin{align}\label{waki}
J^+ &= \beta \cr
J^0 &= - \beta \gamma + \partial \alpha \cr
J^- &= -\beta \gamma^2 - 2 \partial \gamma + \partial \alpha \gamma \cr
\end{align}
where $\alpha$ is a locally-defined holomorphic function. The Sugawara vector simplifies to a Miura form $S_2 = (\partial\alpha)^2 + \partial^2 \alpha$ and is thus manifestly a multiple of the identity operator, with trivial OPE with the currents. 

The spectral flow automorphism extends naturally to the $\beta\gamma$ system, so that the spectral flow operator gives a zero to $\gamma$ and a pole to $\beta$. An
operator that does this is usually indicated as $\delta(\gamma)$. Comparison with the expected form of $\partial \Sigma_1$ from eqn. (\ref{firstone})  gives 
\begin{equation}
\Sigma_1 = e^{\alpha+ \bar \alpha} \delta(\gamma)\delta(\bar \gamma).
\end{equation}
The exponential prefactor provides the $\partial \alpha$ part of $J^0_{-1} \circ \Sigma_1$.

An $\SO(3,\C)$ rotation of this expression gives 
\begin{equation}
\Sigma_1(z;\mu) = e^{\alpha+ \bar \alpha} \delta(\gamma -\mu)\delta(\bar \gamma- \bar \mu).
\end{equation}
where $\mu$ is an inhomogeneous coordinate on $\bCP^1$.  
The integral over $\mu$ is easily done, resulting in
\begin{equation}
\h\Sigma_1(z) = e^{\alpha+ \bar \alpha} .
\end{equation}
This is a multiple of the identity and thus annihilated by all non-negative modes of the currents. Furthermore, the oper differential equation manifestly holds: $\partial_z^2 \h\Sigma_1(z) = S_2(z)\h\Sigma_1(z)$.

We expect this pattern to persist for all $G$ and $\m$. The critical Wakimoto realization gives central 
elements which take the form of a Miura oper built from the Cartan-valued $\alpha$. The spectral flow operators will take the form of spectral flow operators for the $\beta \gamma$ system combined with some function of $\alpha$. The averaged spectral flow operators will give multiples of the identity for the $\beta \gamma$ system, multiplied by certain functions of $\alpha$
which give the Miura expression for solutions of the oper differential equation. 

\section{Wavefunctions from Symplectic Bosons} \label{sec:symp}

\subsection{Basics of Symplectic Bosons}\label{sb}
As we discussed in the  Section \ref{Enriched}, Neumann boundary conditions enriched by 3d hypermultiplets create states described by a path integral 
\begin{equation}
\Psi[a] = \int DZ D\bar Z\, e^{\int_C \left[\langle Z, \bar \partial_a Z \rangle -  \langle 
\bar Z, \partial_a \bar Z \rangle\right]}
\end{equation} 
This is a non-chiral version of the path integral for {\it symplectic bosons}. Chiral symplectic bosons are the Grassmann-even analogue of chiral fermions. They are a special case of $\beta \gamma$ systems where the conformal dimension of both $\beta$ and $\gamma$ is set to $1/2$.

Concretely, chiral symplectic bosons are a collection of $2 n$ two-dimensional spin $1/2$ chiral bosonic fields $Z^a$ with action 
\begin{equation}
\int_{C}\left( \omega_{ab} Z^a \bar\partial Z^b + {\cal A}_{ab} Z^a Z^b\right)
\end{equation}
where $\omega_{ab}$ is a constant symplectic form and we included a coupling to a background  
 connection ${\cal A}_{ab}$ of type $(0,1)$ defining an $\Sp(2n)$ bundle on the Riemann surface $C$.\footnote{As the symplectic bosons are spinors, we do not strictly need to separately define a spin structure and  an $\Sp(2n)$ bundle. Instead, we can specify a $\Spin\cdot \Sp(2n)$ bundle, a notion that is precisely analogous
 to the $\Spin\cdot \SU(2)$ bundles of Section \ref{toposubt}.} We will employ Einstein summation convention in this section unless otherwise noted. 
 
 The analogy to chiral fermions is somewhat imperfect. Chiral fermions are a well-defined two-dimensional (spin)CFT. 
Chiral symplectic bosons are mildly anomalous. The anomaly manifests itself as a sign ambiguity of the chiral partition function
\begin{equation}
\int DZ \,e^{\int_{C}\left( \omega_{ab} Z^a \bar\partial Z^b + {\cal A}_{ab} Z^a Z^b\right)} = \frac{1}{\sqrt{\det \bar \partial_{\cal A} }}
\end{equation}
where we denote as $\bar \partial_{\cal A}$ the $\bar \partial$ operator acting on sections of $K_C^{1/2} \otimes E$ and $E$ is the rank $2n$ bundle associated to the $\Sp(2n)$ bundle.\footnote{Notice that generically $\bar \partial_{\cal A}$ has no zero modes and the functional determinant is well-defined. The partition function diverges for special choices of $\Sp(2n)$ bundle where zero modes appear. } 
The non-chiral partition function, though, 
\begin{equation}
\int DZ D\bar Z\, e^{\int_C \left[\langle Z, \bar \partial_A Z \rangle -  \langle 
\bar Z, \partial_A \bar Z \rangle\right]} =  \frac{1}{|\det \bar \partial_{\cal A} |}
\end{equation}
is unambiguous: it can be defined by an actual the path integral along the cycle $\bar Z{}^a = (Z^a)^*$ (that is, the integration cycle is defined by saying $\bar Z{}^a$ is the complex conjugate of $Z^a$).

We record here the OPE
\begin{equation}
Z^a(z) Z^b(w) \sim \frac{\omega^{ab}}{z-w},
\end{equation}
where $\omega^{ab}$ is the inverse symplectic form. The corresponding algebra of modes is 
\begin{equation}
[Z^a_m, Z^b_n] = \omega^{ab} \delta_{n+m,0}.
\end{equation}
Because of the half-integral spin, the mode indices $n$, $m$ are half-integral in the Neveu-Schwarz sector of the chiral algebra and in particular in the vacuum module. The vacuum satisfies 
\begin{equation}
Z^a_n |0\rangle =0 \qquad \qquad n>0.
\end{equation}
The mode indices are integral in Ramond sector modules, which we will discuss in Section \ref{sec:Ramond}.

The variation of the action with respect to the $\Sp(2n)$ connection ${\cal A}$ gives Kac-Moody currents
\begin{equation}
J^{ab} = \frac12 :Z^a Z^b:
\end{equation}
of level $- \frac12$. The fractional level is another manifestation of the global anomaly of the chiral theory.\footnote{The $J^{ab}$ currents actually generate a quotient of the $\widehat{\mathfrak{sp}}(2n)_{-\frac12}$ chiral algebra: some linear combinations of current bilinears and derivatives of the currents vanish. The number of level $2$ descendants in the symplectic boson vacuum module is smaller than the number of level $2$ descendants in the Kac-Moody vacuum module. } The partition function of the non-chiral theory defines a well-defined section of a bundle $|{\cal L}|$ on the space of $\Sp(2n,\C)$ bundles, where ${\cal L}$ is the line bundle corresponding to Kac-Moody level $-1$. 

Once we specialize to the gauge group $G \subset \Sp(2n)$ of  the 4d theory,  
we will obtain $G$ currents of a level which may not be critical. This signals a gauge anomaly obstructing the 
existence of a 2d junction between the deformed Neumann boundary condition and the enriched Neumann boundary condition. If the level is more negative than the critical level, we may attempt to cancel the anomaly by 
some auxiliary 2d system, such as a collection of free fermions. In the example we discuss momentarily, the anomaly will be absent from the outset. 

\subsection{The Trifundamental Example}\label{tri}

We now specialize to $n=4$ and focus on a $G\equiv \SL(2) \times \SL(2)\times \SL(2)$ subgroup of $\Sp(8)$.
In other words, we identify $\mathbb{C}^8$ with $\mathbb{C}^2 \otimes\mathbb{C}^2 \otimes\mathbb{C}^2$
and we couple the theory to a connection $a$ for $\SL(2) \times \SL(2)\times \SL(2)$.

This gives enriched Neumann boundary conditions which are conjecturally $S$-dual to a tri-diagonal interface, a BBB brane supported on the 
diagonal of $\M_H \times \M_H \times \M_H$ with trivial $\CP$ bundle \cite{Gaiotto:2016hvd,Benini:2010uu}.\footnote{The $S$-duality statement can be generalized to other $G$ or diagonal interfaces between more than three copies. The corresponding enriched Neumann boundary conditions employ the theories defined in \cite{Benini:2010uu}. The boundary chiral algebras for these theories are known from work by Arakawa \cite{Arakawa:2018egx} and have $G$ currents of critical level. The corresponding states would be the partition function of a non-chiral 2d CFT built from Arakawa's chiral algebras, which is currently unknown.} The basic consequence of this $S$-duality identification is that the state $\Psi[a]$ produced by the partition function should intertwine the action of the three copies of $\SL(2)$ Hitchin Hamiltonians and quantum Hecke operators. Our goal in the rest of this section is to confirm this.
 
We denote the symplectic boson fields as $Z^{\alpha \beta \gamma}(z)$ and the symplectic form as
$\epsilon_{\alpha \alpha'}\epsilon_{\beta \beta'}\epsilon_{\gamma \gamma'}$. We get three copies of 
$\widehat{\mathfrak{sl}}(2)_{-2}$ Kac-Moody currents such as 
\begin{equation}
J^{\alpha \alpha'} =\frac12 \epsilon_{\beta \beta'}\epsilon_{\gamma \gamma'} :Z^{\alpha \beta \gamma}Z^{\alpha' \beta' \gamma'}:
\end{equation}
We will denote the three sets of currents simply as $J$, $J'$, $J''$, avoiding indices when possible. 

Crucially, these currents have critical level. Accordingly, the Sugawara vectors are central. A remarkable observation is that the three Sugawara vectors actually coincide here:
\begin{equation} \label{eq:sugathree}
:JJ: = :J'J': = :J'' J'':
\end{equation}
 As the three Sugawara vectors coincide, the correlation functions of Sugawara vectors also coincide. These are obtained from the action of the 
$\mathfrak{sl}(2)$ quantum Hitchin Hamiltonians on the partition function $\Psi(a)$, seen as a half-density on $\M\times \M \times \M$. 

Concretely, the the kinetic operator of the symplectic bosons $Z$ is the $\bar \partial_{a}$ operator acting on the bundle $E \otimes E' \otimes E'' \otimes K_C^{\frac12}$.
The partition function is: 
\begin{equation}\label{sqr}
\Psi(a) = \frac{1}{|\det \bar \partial_a|}.
\end{equation}  
We have thus given a chiral algebra derivation of an intertwining property
\begin{equation}
\boxed{H_i \Psi = H'_i \Psi = H''_i \Psi}
\end{equation}
where $H_i$ run over the $\mathfrak{sl}(2)$ quantum Hitchin Hamiltonians acting on the three spaces of $\SL(2)$ bundles.
Similarly
\begin{equation}
\boxed{\bar H_i \Psi = \bar H'_i \Psi = \bar H''_i \Psi}
\end{equation}
for the conjugate quantum Hitchin Hamiltonians, acting as antiholomorphic differential operators on the space of $\SL(2)$ bundles.
These relations match the relations expected on the B-model side for the tri-diagonal interface. 
Our next objective is to demonstrate the analogous intertwining relations for Hecke operators. 

For simplicity, we will work with Hecke operators of minimal charge for $\SO(3)$, even though the symplectic bosons are coupled to $\SL(2)$ bundles rather than $\PSL(2)$. Minimal Hecke modifications map $\SL(2)$ bundles to $\SL(2)$ bundles twisted by a gerbe 
and viceversa, so we can consistently describe the action of pairs of Hecke operators on half-densities on $\M(C,\SU(2))$. 

A minimal Hecke operator will create the endpoint of a $Z \to -Z$ cut for the symplectic bosons. This leads us to consider Ramond vertex operators. 

\subsection{The Ramond Sector} \label{sec:Ramond}
The symplectic boson chiral algebra admits Ramond modules which are associated to a circle with 
non-bounding spin structure. In such a module, the mode expansion of the $Z^a$ fields involves modes $Z^a_n$ with integral $n$. 
The corresponding vertex operators introduce a cut across which the $Z^a$ flip sign. 

The zero modes $Z^a_0$ form a Weyl algebra. There is a rich collection of {\it highest weight} Ramond modules for the chiral symplectic bosons which is induced from a module for the zero mode Weyl algebra. Every element of the Weyl module is promoted to a highest weight vector/vertex operator which is annihilated by the positive Fourier modes $Z^a_n$. The negative Fourier modes act freely and the zero modes act as in the Weyl module.

The Kac-Moody current zero modes $J_0^{ab}$ act on a highest weight vector as $\Sp(2n)$ generators $Z^{(a}_0 Z^{b)}_0$. 
All Weyl modules break to some degree the $\Sp(2n)$ symmetry of the VOA. In other words, there is no Weyl module 
equipped with an $\Sp(2n)$-invariant vector. Thus the insertion of any such vertex operator into a correlation function will always reduce $\Sp(2n)$ gauge invariance at that point. 

The simplest way to produce a Ramond vertex operator is to consider a spectral flow operator of minimal charge in $\PSp(2n)=\Sp(2n)/\Z_2$. Select a Lagrangian splitting $\mathbb{C}^{2n} = V \oplus V^\vee$. Pick a singular gauge transformation which acts as $z^{\frac12}$ on $V$ and $z^{-\frac12}$ on $V^\vee$. The resulting spectral flow operator $S_V(0)$ is a Ramond module. Linear combinations of $Z^a$ in $V$ vanish as $z^{\frac12}$ as they approach $S_V(0)$, while linear combinations in $V^\vee$ diverge as $z^{-\frac12}$. This means that $S_V(0)$ is a highest-weight Ramond module annihilated by linear combinations of $Z_0^a$ in $V$. In particular, it only depends on $V$. 

$S_V(0)$, for any $V$, can be obtained by 
 an $\Sp(2n)$ rotation from some particular $S_{V_0}(0)$, which we choose as a reference. Without loss of generality, pick a basis where 
\begin{equation}
\omega^{ab} = \delta^{a-b-n}- \delta^{b-a-n},~~a,b=1,\cdots, 2n,\end{equation}
and choose the reference vertex operator $S_{V_0}(z)$ to be annihilated by $Z^{n+1}_0, \cdots, Z^{2n}_0$. 
Denote the remaining zero modes, which act as creation operators, as $u^a=Z_0^a$,  $a\leq n$ and denote the corresponding descendants of $S_{V_0}(z)$ as $S_{V_0}[u^a](z)$, $S_{V_0}[u^a u^b](z)$, etcetera. Annihilation zero modes act as $Z_0^{a+n}=\partial_{u^a}$.

Consider a coherent state in the Weyl module:
\begin{equation}
S_{V_0}[e^{\frac12 B_{ab} u^a u^b}](z)
\end{equation}
This is annihilated by linear combinations $Z^{a+n}_0 - B_{ac} Z^c_0$. That condition  defines a rotated 
Lagrangian subspace $V = B \circ V_0$. We thus identify
\begin{equation}
S_{B \circ V_0}(z) = S_{V_0}[e^{\frac12 B_{ab} u^a u^b}](z)
\end{equation}
Indeed, we have $\partial_{B_{ab}}S_{B \circ V_0}(z) = J_0^{ab}S_{B \circ V_0}(z)$. 

\subsection{Non-chiral Ramond Modules}
Next, we can consider the combined theory of chiral and antichiral symplectic bosons. In the Ramond sector, we now have 
an action of the chiral zero modes $Z_0^a$ and the antichiral zero modes $\bar Z_0^a$. As the path integration contour 
relates $\bar Z$ to the conjugate of $Z$, it is natural to consider a space of Ramond states such that the 
zero modes are adjoint to each other. 

Furthermore, the 2d theory in the zero momentum sector is essentially a quantum mechanics with target $\mathbb{C}^{2n}$. 
It is thus natural to pick a polarization and set the zero mode Hilbert space to be $\lmark^2(\mathbb{C}^{n})$. This answer is actually independent of the choice of polarization, as we can use generalized Fourier transform operations to relate different polarizations.\footnote{The metaplectic anomaly cancels out because we act simultaneously on holomorphic and anti-holomorphic variables.} 
The full Ramond Hilbert space ${\mathcal R}$ is induced from $\lmark^2(\mathbb{C}^{n})$ in the usual way, by having $Z^a_k$ and $\bar Z^a_k$ annihilate vectors in $\lmark^2(\mathbb{C}^{n})$ when $k>0$ and act freely with $k<0$. 

A variety of different highest weight Ramond vertex operators are realized as distributions on $\mathbb{C}^{n}$. 
Pick the same polarization as in the definition of $S_{V_0}(0)$. Then the reference spectral flow operator $S_{V_0}(0)$ is represented by the distribution ``$1$''. The rotated $S_{B \circ V_0}(z)$ is represented by the Gaussian
\begin{equation}
e^{\frac12 B_{ab} u^a u^b-\frac12 \bar B_{ab} \bar u^a \bar u^b }
\end{equation}
The distribution $\delta^{(2n)}(u)$, instead, represents a Ramond vertex operator annihilated by $Z^{1}_0, \cdots Z^{n}_0$,
which is a spectral flow operator with charge opposite to $S_{V_0}(0)$.

We can now specialize to $n=4$ and to the spectral flow operators associated to the $\SL(2) \times \SL(2) \times \SL(2)$ subgroup of $\Sp(8)$. 
We pick our polarization of $\mathbb{C}^{8}$ to be invariant under the second and third $\SL(2)$ groups in the product. Concretely, we identify 
$Z_0^{2 \beta \gamma} = u^{\beta \gamma}$ and $Z_0^{1 \beta \gamma}=\epsilon^{\beta\beta'}\epsilon^{\gamma\gamma'}\frac{\partial}{\partial u^{\beta'\gamma'}} .$

In this framework, the spectral flow operators $\Sigma_{1;\mu, \bar \mu}$, $\Sigma'_{1;\mu', \bar \mu}$ and $\Sigma''_{1;\mu'', \bar \mu}$ for the three $\SL(2)$'s are all special cases of $\Sp(8)$ spectral flow operators of minimal charge, i.e. of highest weight Ramond vertex operators. They are represented by distributions, which we can average over $\mu$ to obtain representations of $\hat \Sigma_{1}$, $\hat \Sigma'_{1}$, $\hat \Sigma''_{1}$. The calculations are straightforward:
\begin{itemize}
\item $\Sigma_1$ is represented by the distribution ``1''. $\Sigma_{1;\mu, \bar \mu}$ is represented by the Gaussian
\begin{equation}
e^{\mu  \epsilon_{\beta \beta'}\epsilon_{\gamma \gamma'} u^{2\beta \gamma} u^{2 \beta' \gamma'}- \mathrm{c.c.}}.
\end{equation} Averaging over $\mu$, $\hat \Sigma_{1}$ is represented by the distribution
\begin{equation}
\delta^{(2)}(\epsilon_{\beta \beta'}\epsilon_{\gamma \gamma'} u^{2\beta \gamma} u^{2 \beta' \gamma'}).
\end{equation}
\item $\Sigma'_1$ is represented by the distribution 
\begin{equation}
\delta^{(2)}(u^{211})\delta^{(2)}(u^{212}).
\end{equation}
 $\Sigma'_{1;\mu', \bar \mu'}$ is represented by the distribution 
\begin{equation}
\delta^{(2)}(u^{211}- \mu' u^{221})\delta^{(2)}(u^{212}- \mu' u^{222}).
\end{equation}
 Averaging over $\mu'$, we find that $\hat \Sigma'_{1}$ is represented by the same distribution as $\hat \Sigma_{1}$.
\item $\Sigma''_1$ is represented by the distribution 
\begin{equation}
\delta^{(2)}(u^{211})\delta^{(2)}(u^{221}).
\end{equation}
Similarly,  $\Sigma''_{1;\mu'', \bar \mu''}$ is represented by the distribution 
\begin{equation}
\delta^{(2)}(u^{211}- \mu'' u^{212})\delta^{(2)}(u^{221}- \mu'' u^{222}).
\end{equation}
 Averaging over $\mu''$, we find that $\hat \Sigma''_{1}$ is represented by the same distribution as $\hat \Sigma_{1}$
\end{itemize}
We conclude that  
\begin{equation}
\hat \Sigma_{1} = \hat \Sigma'_{1} =\hat \Sigma''_{1}
\end{equation}
in the theory of trifundamental symplectic bosons. These relations play an analogous role to (\ref{eq:sugathree}): inserted in correlation functions they prove that the partition function $\Psi(a)$ intertwines the action of minimal Hecke operators for the three $\SL(2)$ groups:
\begin{equation}
\boxed{H_z \Psi =  H'_z \Psi = H''_z \Psi}
\end{equation}

\subsection{A Marvelous Module}\label{marvel}
It is worth elaborating on the properties of the Weyl module generated by the distribution 
\begin{equation}
M=\delta^{(2)}(\epsilon_{\beta \beta'}\epsilon_{\gamma \gamma'} u^{2\beta \gamma} u^{2 \beta' \gamma'}).
\end{equation}
We learned some surprising properties which follow from $M$ representing averaged spectral flow operators: 
$M$ is invariant under $\SL(2) \times \SL(2) \times \SL(2)$ and the module treats the three $\SL(2)$ groups in a completely symmetric manner. The latter property is somewhat hidden in the analysis, so we can spell it out in detail here: we can change polarization 
by a Fourier transform and go to representations of the Weyl module in terms of functions of $u^{\alpha 2 \gamma}$ or $u^{\alpha \beta 2}$. The Fourier transform of $M$ produces distributions $\delta^{(2)}(\epsilon_{\beta \beta'}\epsilon_{\gamma \gamma'} u^{\beta 2 \gamma} u^{\beta' 2\gamma'})$ and $\delta^{(2)}(\epsilon_{\beta \beta'}\epsilon_{\gamma \gamma'} u^{\beta \gamma 2} u^{\beta' \gamma' 2})$ respectively. 

Acting with the Weyl algebra on $M$ we generate a remarkable module with an explicit action of $\SL(2) \times \SL(2) \times \SL(2)$.
As a vector space, the module decomposes as 
\begin{equation}\label{marmod}
\oplus_{d=1}^\infty V_d\otimes V_d \otimes V_d
\end{equation}
where $V_d$ is the $d$-dimensional irreducible representation  of $\SL(2)$. The Weyl generators act as a sum of two terms: one term raises $d$ by 1 and the other lowers it by $1$. 

\subsection{Some Generalizations}
The $\SU(2) \times \SU(2) \times \SU(2)$ gauge group can be seen as a special case of $\Sp(2n) \times \mathrm{Spin}(2n+2)$. 
Trifundamental hypermultiplets are a special case of bifundamental hypermultiplets for $\Sp(2n) \times \mathrm{Spin}(2n+2)$.
Bifundamental hypermultiplets engineer an ``NS5'' interface between $\Sp(2n)$ and $\mathrm{Spin}(2n+2)$ 4d gauge theories. 

The levels of the corresponding Kac-Moody currents in the theory of bifundamental symplectic bosons are critical, so the NS5 interface 
has non-anomalous corners with deformed Neumann boundaries and our construction applies. 

The NS5 interface is $S$-dual to a ``D5'' interface between $\mathrm{\Spin(2n+1)}$ and $\mathrm{Spin}(2n+2)$ 4d gauge theories,
at which the $\mathrm{Spin}(2n+2)$ gauge group is reduced to $\mathrm{\Spin(2n+1)}$. The D5 interface will descend in the B-model to a BBB interface supported on the space of $\mathrm{Spin}(2n+1)$ flat connections embedded in the space of $\mathrm{Spin}(2n+2)$ flat connections. This gives simple predictions for the action of quantum Hitchin Hamiltonians and 't Hooft operators on the NS5 interface. 
Our calculations in this section should be generalized to verify these predictions. 

Bifundamental hypermultiplets can also be used to engineer an ``NS5'' interface between $\U(n)$ 4d gauge theories which has a relatively simple dual and should be amenable to a 2d chiral algebra analysis. The Kac-Moody currents for the $\SU(n)$ subgroups are critical, and the $\U(1)$ anomalies can be cancelled by a single complex chiral fermion of charge $(1,-1)$ under the diagonal $\U(1)$ gauge symmetries. 
Again, our calculations in this section should be generalized to verify these predictions. For example, the minimal spectral flow operators 
are represented by the distribution $\delta^{(2)}(\det u)$. 

\noindent{\it Acknowledgment}   
Research at Perimeter Institute is supported by the Government of Canada through Industry Canada and by the Province of Ontario 
through the Ministry of Research $\&$ Innovation.
Research  of EW supported in part by  NSF Grant PHY-1911298.   We thank D. Baraglia, P. Etingof, E. Frenkel, D. Kazhdan, and L. Schaposnik for discussions.
We also thank P. Etingof for a careful reading of the manuscript and suggesting a number of clarifications.

\appendix

\section{Local Operators Along The Canonical Coisotropic Brane and Quantum Corrections in $\varphi {\cal F}$ Theory}\label{coho}
In this appendix, we will discuss the full space of boundary local operators for the deformed Neumann boundary condition,
which corresponds to the canonical coisotropic brane ${\mathcal B}_{\mathrm{cc}}$. This problem was originally addressed in Section 12.4 of \cite{KW}. 
Here we will extend and in some ways correct the analysis given there.

At the classical level, the BRST cohomology of the relevant space of boundary local operators is as follows.   One type of generator is a local operator ${\mathcal P}(\phi_z)$,
where ${\mathcal P}$ is an invariant polynomial on the Lie algebra $\mathfrak g$ of the gauge group $G$.    A second type of generator is an operator
${\mathcal P}'(\phi_z,\psi_z)$ defined by  ${\mathcal P}'(\phi_z+\epsilon\psi_z)={\mathcal P}(\phi_z)+\epsilon {\mathcal P}'(\phi_z,\psi_z)$, where $\epsilon$ is
an odd parameter.   It turns out that classically, the BRST cohomology of boundary local operators for ${\mathcal B}_{\mathrm{cc}}$ is generated by operators ${\mathcal P}$,
${\mathcal P}'$, and their derivatives with respect to $z$.    This fact will be explained presently.  
One expects that the BRST cohomology in the full quantum theory is the cohomology of some further differential $Q$ acting on the classical cohomology.   Our goal here
is to show that this differential vanishes and hence the cohomology in the quantum theory is the same as it is classically.  As explained in section \ref{hh}, this implies
in particular that the commuting Hitchin Hamiltonians can be quantized to commuting holomorphic differential operators.

Since the cohomology is generated classically by operators of the form $\mathcal P$ and $\mathcal P'$ and their derivatives with respect to $z$,
and since $\partial_z$ commutes with $Q$, it suffices to show that $[Q,\mathcal P]=\{Q,\mathcal P'\}=0$ for any invariant polynomial $\mathcal P$ on $\mathfrak g$.
For this, one can look for a symmetry that will enforce this result.

In Section 12.4 of \cite{KW}, it was observed that $Q$ commutes with a time-reversal symmetry that acts by
\begin{equation}\label{tadd}
{\sf T}: t \to - t \qquad \qquad \varphi_z \to - \varphi_z \qquad \qquad \psi_z \to \psi_z.
\end{equation}
For $G=\mathrm{SU}(2)$, this symmetry suffices to imply the result that we want.   For this group, the ring of invariant polynomials is generated by $\mathcal P(\phi_z)={\mathrm{Tr}}\,
\phi_z^2$, so it suffices to prove that $[Q,\mathcal P]=\{Q,\mathcal P'\}=0$ for this particular $\mathcal P$.    To prove that $[Q,\mathcal P(\phi_z)]=0$, we observe that the only operator
in the cohomology with the same spin and fermion number\footnote{We use the fact that $\phi_z,\psi_z$ and $\partial_z$ all have spin 1, while
$\phi_z$ has fermion number 0 and $\psi_z$ has fermion number 1. $Q$ has spin zero and fermion number 1.} as  $[Q,\mathcal P(\phi_z)]$ is $\mathcal P'(\phi_z,\psi_z)$, 
but $\mathcal P(\phi_z)$ is even under $\sf T$, while  ${\mathcal P}'(\phi_z,\psi
_z)$ is odd.
  It is also true that $\{Q,\mathcal P'\}=0$, since the classical
cohomology does not contain any nonzero operator with the spin and fermion number of $\{Q,\mathcal P'\}$.

This argument does not suffice, in general, for groups of higher rank.   Once one takes into account that the classical cohomology in this problem is constructed
with $z$ derivatives $\partial_z$ as well as operators $\phi_z$ and $\psi_z$, one finds that there are indeed candidate operators that could appear on the right hand
side of formulas $[Q,\mathcal P]=?$ and $\{Q,\mathcal P'\}=?$ and are consistent with the $\sf T$ symmetry.    However, there is another relevant symmetry.  This is the symmetry $\sf C$
 that physically is called charge conjugation; mathematically it is called the Chevalley involution.   The root diagram of any simple Lie algebra $\mathfrak g$ is invariant under
 the operation $\sf C$ that acts as $-1$ on every root vector. Since $\mathfrak g$ 
 can be reconstructed from the lengths of root vectors and the angles between them,
 $\sf C$ extends to an involution of $\mathfrak g$ and therefore of a corresponding Lie group $G$.   This symmetry is charge conjugation or the Chevalley involution.
 (It is an inner automorphism for some groups, such as ${\mathrm{SO}}(2n+1)$, and an outer automorphism for other groups, notably ${\mathrm{SU}}(n),\,n>2$, 
 ${\mathrm{SO}}(2n)$, and $E_6$. For $G$ such that $\sf C$ is inner, one could complete the following argument just using $\sf T$.)    
 
 Because an invariant polynomial $\mathcal P$ on $\mathfrak g$ has a nontrivial restriction to the Lie algebra of a maximal torus, 
one can determine whether $\mathcal P(\phi_z)$ is even or odd under $\sf C$ by restricting to a maximal torus.   As a result, one 
 one finds that
 in all cases $\sf C \mathcal P(\phi(z)) \sf C^{-1}=\mathcal P(-\phi_z)$  (even though it is not true that $\sf C \phi_z\sf C^{-1}=-\phi_z$).    Likewise
 $\sf C\mathcal P'(\phi_z,\psi_z)\sf C^{-1} =\mathcal P'(-\phi_z,-\psi_z)=-\mathcal P'(-\phi_z,\psi_z)$.    When this is combined with eqn. (\ref{tadd}),
 we learn that in all cases 
 \begin{equation}\label{wadd} \sf C\sf T \mathcal P(\phi_z)(\sf C\sf T)^{-1}=\mathcal P(\phi_z),~~~
  \sf C\sf T \mathcal P'(\phi_z,\psi_z)(\sf C\sf T)^{-1}=-\mathcal P'(\phi_z,\psi_z). \end{equation}
  Therefore, building the classical cohomology out of polynomials in operators $\mathcal P$, $\mathcal P'$, and their $z$ derivatives, we learn that $\sf C\sf T$
  acts as $(-1)^k$ on the part of the classical cohomology that is of fermion number $k$ (that is, the part that is of degree $k$ in operators $\mathcal P'$ and their
  derivatives).   Since the BRST differential $Q$ commutes with $\sf C\sf T$, it follows that this operator must act trivially on the classical
  cohomology in this problem. Indeed the left and right hand sides 
 of a formula $[Q,\mathcal P]=?$ or $\{Q,\mathcal P'\}=?$ would transform oppositely under $\sf C\sf T$.
 
 In the rest of this appendix, we will explain the assertion that at the classical level, the BRST  cohomology of the relevant space of local operators
 is generated by operators $\mathcal P$, $\mathcal P'$, and their derivatives.
We study this problem in the BV formalism for (analytically continued) holomorphic-topological three-dimensional $\varphi {\cal F}$ theory. 
It is convenient to employ a superspace with odd coordinates ${\mathrm d} t$ and ${\mathrm d} \bar z$, so that superfields become sums of forms of various degree. 
The physical fields, ghosts and antifields can be collected in two superfields which we can denote as ${\cal B}$ and ${\cal A}$, with action
\begin{equation}
\int \mathrm{Tr} \,{\cal B} \,{\cal F}_{\cal A}
\end{equation}
This is completely analogous to the BV formalism for 3d Chern-Simons theory, which employs the standard Chern-Simons action for a superfield 
which is a sum of forms of all degrees. 

The 0-form component of ${\cal B}$ is $\varphi_z$.  The 0-form component of ${\cal A}$ is the ghost field $c$ and the 1-form (which only has ${\mathrm d}t$ and ${\mathrm d} \bar z$ components) is the gauge connection. The ghost $c$ itself is not allowed in the construction of local operators: only its derivatives are. This is a standard feature of the BRST formalism for a gauge-theory with a compact gauge group. 
The holomorphic derivative $\partial_z c$ in the holomorphic $\varphi {\mathcal {F}}$ theory plays the role of $\psi_z$ in the full four-dimensional gauge theory.

Classically, in the free theory, the BRST differential comes from the kinetic term and involves the partial differential
\begin{equation}
\mathrm{d} \equiv {\mathrm d}t \frac{\partial}{\partial t}+ {\mathrm d} \bar z \frac{\partial}{\partial \bar z}.
\end{equation}
Both superfields are annihilated by $Q_{\mathrm{free}} + \mathrm{d}$. The cohomology of the free theory consists of 
$G$-invariant polynomials in $\varphi_z$, $\psi_z$ and their holomorphic or $\partial_z$ derivatives.\footnote{The kinetic term pairs the $0$-form components 
of the superfields with the $1$-form components. This means that such polynomials in the $0$-form components are well-defined, with no renormalization ambiguities. It also means that the BV Laplacian of the quantum free theory does not affect the final answer. }

The cohomology in the classical $\varphi \mathcal F$ theory is not the same as the cohomology of the free theory, since 
interactions change the BRST differential already at tree level. The tree-level BRST variation of the elementary superfields is
\begin{equation}
\left[Q_{\mathrm{tree}} + \mathrm{d}_{\cal A} \right] {\cal B} = 0 \qquad \qquad Q_{\mathrm{tree}} {\cal A} + {\cal F}_{\cal A} = 0
\end{equation}

The classical or tree-level cohomology can be computed by first passing to the cohomology of the free theory, and then restricting the nonlinear part of the differential 
to the free cohomology. This is a standard step in the analysis of twisted supersymmetric field theories.
Taking the cohomology of the free theory amounts to  
dropping $D_t$ and $D_{\bar z}$ derivatives and higher form components of the fields (leaving only $\partial_z$ derivatives).   
The resulting differential takes the form 
\begin{equation}
Q \varphi_z = [c,\varphi_z] \quad \qquad Q c = \frac12 [c,c]
\end{equation}

This cohomology problem can be stated mathematically as a relative Lie algebra cohomology problem $H^*(\mathfrak{g}[z],\mathfrak{g};S \mathfrak{g}^*[z])$. 
The underlying Lie algebra $\mathfrak{g}[z]$ is associated to the ghost $c$ and its holomorphic derivatives. The problem is relative to $\mathfrak{g}$ because 
we omit the $c$ field itself. The cohomology has coefficients in $S \mathfrak{g}^*[z]$, representing polynomials in $\varphi_z$ and its derivatives. 

This cohomology problem was solved by Theorem B in \cite{FGT}, with the result that was claimed earlier: the classical cohomology is generated by operators
$\mathcal P$, $\mathcal P'$, and their $z$ derivatives.  As we have already explained, there is a symmetry $\sf {CT}$ which ensures that there are no quantum
corrections to this result.

The  space we have described of boundary local operators for the $A$-brane $\mathcal B_{\mathrm{cc}}$ 
agrees with the space of local operators at the $S$-dual oper boundary condition. 
Indeed, the $B$-model endomorphisms of the oper brane are holomorphic forms on the oper manifold, which is an affine vector space parameterized by
operators associated to invariant polynomials $\mathcal P$ on $\mathfrak g$.  For every affine coordinate $w$ on the oper manifold, associated to some
$\mathcal P$, there is a corresponding 1-form ${\mathrm d} w$, associated in the same way to $\mathcal P'$.   Thus the holomorphic forms on the oper manifold correspond in
a natural way to what one finds in the $A$-model.

\section{The $B$-Model Hilbert Space}\label{bmodel}

Here we will describe some properties of the physical state space of the $B$-model.   The goal is to describe the conditions needed for the hermitian metric on this
space to be positive.  
First we describe what we will call $B$-model quantum mechanics.   This is, most simply, the quantum mechanics that arises in compactification of the $B$-model
on a circle, though our application involves a different occurrence of the same model.

The input is a Calabi-Yau manifold $X$ of complex dimension $n$ with a holomorphic function $W$ (the superpotential)  and a holomorphic volume form $\varOmega$. 
We generally assume that $W$ has only isolated, though possibly degenerate, critical points.   The full $B$-model quantum mechanics is
defined  with a Kahler metric
$g$ on $X$, but the $B$-model (which describes only a topological sector of the theory) does not depend on $g$.      Actually, because of $B$-model
localization, one does not lose much by specializing to $X=\C^n$, but there is no need to do so.
We locally parametrize $X$ by holomorphic coordinates $x^i$, $i=1,\cdots, n$.
The model also has two sets of canonically conjugate fermion variables\footnote{To compare to more complete descriptions of
this model, note that in  \cite{Vafa}, for example, our
 $\chi^{\bar i}$ is denoted $\psi^{\bar i}+\bar\psi^{\bar i}$, and our $\theta_i$ is $-\g_{i\bar i}\bar\psi^{\bar i}$.   A more
 complete description can also be found in \cite{WittenE}.}
\be\label{nimbo} \{\chi^{\bar i},\t \chi_{\bar j}\}=\delta^i_j,~~~~ \{\theta_i,\t\theta^j\}=\delta^j_i. \ee    $\chi$ and $\t\theta$ have fermion number 1 and
the conjugate variables $\t\chi,\,\theta$ have fermion number $-1$.   
We represent the anticommutators by  $\t\chi_{\bar j}=\partial/\partial \chi^{\bar j}$ and $\t\theta_i=\partial/\partial\theta^i$.  
Then a quantum wavefunction is a function $\Psi(x,\bar x, \chi, \theta)$.    A wavefunction  that is homogeneous in $\chi,\theta$ of degree $(p,q)$ takes the form
$\chi^{\bar i_1}\chi^{\bar i_2}\cdots \chi^{\bar i_p}
\theta_{i_i}\theta_{i_2}\cdots \theta_{i_q} \Phi_{\bar i_1\bar i_2\cdots \bar i_p} ^{i_1 i_2 \cdots i_q}(x,\bar x)$, and can be interpreted geometrically
as a $(0,p)$-form valued in $\wedge^q TX$, $TX$ being the holomorphic tangent bundle of $X$.  
  Acting on such a wavefunction, the differential of the $B$-model is the operator
\be\label{bdiff} Q=\sum_j \chi^{\bar j}\frac{\partial}{\partial \bar x^{ \bar j}} +\sum_i  \frac{\partial W}{\partial x^i}\frac{\partial}{\partial \theta_i} \ee  of fermion number 1.  Evidently $Q^2=0$.
The space of physical states of the $B$-model is defined as the cohomology of $Q$.     This cohomology is typically easy to describe explicitly.
In the basic case of an affine variety, the cohomology can be represented by holomorphic
 functions of the $x^i$ modulo the ideal generated by the partial derivatives
of $W$.   If $X=\C^n$ and $W$ is a polynomial, it suffices to consider polynomial functions:
\be\label{hilbdef} \H=\C[x_1,\cdots,x_n]/(\partial W/\partial x^1,\cdots,\partial W/\partial x^n). \ee
  The description of the state space in terms
of  holomorphic functions modulo an ideal makes it clear that the state space is a ring $\RR$, which in the application to the closed string sector
of a Landau-Ginzburg model is called the chiral ring of the model.    Assuming that the critical points of $W$ are isolated, $\RR$ has a decomposition
as a direct sum of subrings associated to critical points $p_s$:  
\be\label{bco} \RR=\oplus \RR_s.\ee
  In the context of the closed string $B$-model, $\RR_s$ is the
chiral ring of  the theory in the vacuum associated to the critical point $p_s$.

As a first step, we want to define the natural $B$-model
pairing on this space of states.   First, we need a way to integrate a wavefunction.   This will depend on the Calabi-Yau 
volume form $\varOmega$.  The antiholomorphic variables $\bar x^{\bar i}, \chi^{\bar i}$ have opposite statistics and transform similarly in
a change of coordinates, so they possess a natural integration measure $\dD(\bar x,\chi)=\d \bar x^{\bar 1}\cdots \ \bar x^{\bar n}\d\chi^{\bar 1}\cdots\d\chi^{\bar n}$.
   A choice of $\varOmega$ gives an integration measure $\dD x=\frac{1}{n!} \varOmega_{i_1\cdots i_n}\d x^{i_1}\cdots \d x^{i_n}$ for
 $x$.   Because the $\theta_i$ have opposite statistics to $x^i$ but also transform under coordinate changes in a dual fashion, a natural measure $\dD\theta$
 is likewise proportional to $\varOmega$.   The combined system therefore has a measure $\dD(x,\bar x, \chi,\theta)$.  This measure is proportional to $\varOmega^2$, a standard fact about the $B$-model.
We use this measure to define integration.   If $\alpha$ is any wavefunction that vanishes sufficiently rapidly at infinity, 
 one can integrate by parts, in the sense that
 \be\label{intparts} \int \dD(x,\bar x,\chi,\theta)\, Q\alpha=0.\ee

One cannot immediately apply this definition to physical states characterized as in eqn. (\ref{hilbdef}) as holomorphic functions of the $x^i$, because
such functions do not vanish sufficiently rapidly at infinity.   However, consider the $Q$-exact quantity
\be\label{zoro} V=\left\{Q, \sum_i \theta_i g^{i\bar j}\frac{\partial \bar W}{\partial \bar x^{\bar j}}\right\} =\left|\d W\right|^2 +\sum_{i,\bar j,\bar k} \chi^{\bar k}\theta_i g^{i\bar j}\frac{\partial^2\bar W}{\partial \bar x^{\bar j}\partial\bar x^{\bar k} } .  \ee  
The quantity $\exp\left(-\frac{1}{2\epsilon}V\right)$, for any positive $\epsilon$, is of the form $1+\{Q,\cdots\}$, and vanishes rapidly at infinity assuming that
$|\d W|^2$ grows at infinity.   So instead of representing a physical state  as in eqn. (\ref{hilbdef}) by a holomorphic function $f_0(x)$ (usually taken to be a polynomial)
we can use the representative 
\be\label{todo}f =f_0(x)\cdot \exp\left(-\frac{1}{2\epsilon}V\right)\ee
 of the same physical state.   Assuming that $|\d W|^2$ grows at infinity, this  means that we can assume that physical
states are represented by wavefunctions that vanish rapidly at infinity.   Hence one can define the natural $B$-model pairing on $\H$,
\be\label{zinparts} (f,g)=\int \dD(x,\bar x,\chi,\theta)\, f g . \ee
This is the usual nondegenerate, bilinear pairing of the $B$-model, compactified on a circle and reduced to quantum mechanics.   
 Because $f =f_0\cdot \exp\left(-\frac{1}{2\epsilon}V\right)$, for small $\epsilon$,  is strongly localized near critical points of $W$, that is points with $\d W=0$,
the evaluation of $(f,g)$ can be expressed as a sum over contributions of critical points.   With respect to the decomposition of eqn. (\ref{bco}),
$(~,~)$ is block diagonal: it restricts to a nondegenerate pairing
\be\label{restro}(~,~):\RR_p\otimes \RR_p\to \C\ee
for each $p$, and annihilates $\RR_{p'}\otimes \RR_p$ for $p'\not=p$.

However, we want a hermitian inner product in the $B$-model, not a bilinear one.   For this, we assume that $X$ has an antiholomorphic
involution $\tau$ under which $W$ is real in the sense that $\tau^*(W)=\bar W$.    An equivalent statement is that $W$ is real when restricted to the fixed points
of $\tau$, or that for $p\in W$, $\bar W(\bar p)=W(p)$, where  $\bar p=\tau(p)$. We assume that $\varOmega$ is likewise real, meaning that
$\tau^*\varOmega=\bar\varOmega$, and we 
 pick the local holomorphic coordinates $x^i$ to similarly satisfy $\tau^*x =\bar x$.
    Let $\Theta$ be complex conjugation (which descends from the $\sf{CPT}$ symmetry of an underlying two-dimensional $\sigma$-model)
and set $\Theta_\tau=\Theta\tau$.   Then $\Theta_\tau$ commutes with $Q$, so we can define a natural hermitian pairing on the physical
state space of the $B$-model by
\be\label{inparts}\la f,g\ra =(\Theta_\tau f, g). \ee
This is a hermitian pairing because it is linear in the second variable and antilinear in the first.

In this generality, the inner product $\la~,~\ra$ is nondegenerate, but it is not positive-definite.
First of all, the inner product is never positive-definite if $W$ has critical points that are not real. 
   Indeed, since the original bilinear pairing  annihilates $\RR_{p'}\otimes \RR_p$ for $p'\not=p$,
    the hermitian pairing $\la~,~\ra$ annihilates $\RR_{p'}\otimes \RR_p$ for $p'\not=\bar p$.   In particular, if $\bar p\not=p$, then the pairing
$\la~,~\ra$ annihilates $\RR_p \otimes \RR_p$, so elements of $\RR_p$ are null vectors and the pairing is not positive-definite.

Even if all critical points are real, the inner product is never positive-definite if there are degenerate critical points.
To understand why,  consider a simple example with $X=\C$, $\varOmega=\d x$,  and $W(x)=\lambda x^N/N$, for some 
integer $N\geq 2$ and real $\lambda$. The critical point at $x=0$ is degenerate for $N>2$ and nondegenerate for $N=2$.
 For representatives of the space of physical states, we can take $f_k=x^k \exp(-V/2\epsilon)$, $k=0,\cdots, N-2$.  One then has
\begin{align}\label{ninparts}\la f_r,f_s\ra=&\frac{\lambda(N-1)}{\epsilon}\int _\C\frac{|\d x\d\bar x|\d\chi\d\theta}{2\pi}\, \,x^{r+s}\bar x^{N-2}\theta\chi\, \exp(-\lambda^2|x|^{2N-2}/\epsilon) \cr
=&\frac{\lambda(N-1)}{\epsilon}\int _\C\frac{|\d x\d\bar x|}{2\pi} \,x^{r+s}\bar x^{N-2}\exp(-\lambda^2|x|^{2N-2}/\epsilon),\end{align}
where we included a convenient normalization factor $1/2\pi$.
This expression is independent of $\epsilon$, as expected because the cohomology class of $f_k$ does not depend on $\epsilon$.
The integral vanishes unless $r+s=N-2$.   Therefore, $f_k$ is a null vector for this inner product unless $2k=N-2$.   In particular, the inner product
$\la~,~\ra$ is not positive-definite if $N>2$, that is, if $W$ has a degenerate critical point.  The general case of a degenerate critical point is similar.   
It is a standard result in the $B$-model that, if $p$ is a degenerate critical point, the element\footnote{The element $1_p$ can be represented
by any polynomial that is 1 to sufficiently high order near $p$ and vanishes to sufficiently high order near other critical points.}   $1_p\in \RR$ which is the identity in $\RR_p$ and zero in the other
subrings is a null vector for $(~,~)$.    If $p$ is real, then $1_p$ is $\Theta_\tau$-invariant and so is also a null vector for $\la~,~\ra$; $1_p$ is also a null vector if
$p$ is not real, since then $\RR_p$ consists entirely of null vectors.

Even if all critical points of $W$ are real and nondegenerate, we do not get positivity for free.   Indeed, for $N=2$,  eq. (\ref{ninparts}) gives $\la f,f\ra=1/\lambda$, so it is only positive for one sign of $\lambda$.    As long as there is only one critical point, the sign of $\lambda$ is not really important,
because there was an arbitrary overall sign in the definition of the inner product.    However, if there are multiple nondegenerate critical points $p_1,p_2,\cdots , p_s$,
then for $r=1,\cdots, s$, we can pick a function $f_{r}$ that vanishes at all critical points except $p_r$ and is nonvanishing at $p_r$. 
Still with $X=\C$ and assuming the $p_r$ are real, 
the evaluation of $\la f_r,f_r\ra$ will reduce to the $N=2$ version of the above computation, with $\lambda$ replaced by $W''(p_r)$.
Thus to achieve positivity, we would want $W''(p_r)$ to have the same sign for all $r$.  For real critical points in one dimension, 
 it is impossible  to satisfy
this condition, since inevitably 
the sign of $W''$ alternates from one critical point of $W$ to the next.

However, with more variables, positivity is possible for any number of critical points.   Suppose that $W(x_1,x_2,\cdots, x_n)$ is a holomorphic function
on $\C^n$ that is real when restricted to $\R^n$, and has real, nondegenerate critical points.   At any given critical point $p$, we can pick real 
local coordinates $x_{i,p}$ such
that
\be\label{morse} W=\frac{1}{2}\sum_{i=1}^n\lambda_{i,p} x_{i,p}^2 \ee
with real, nonzero $\lambda_{i,p}$.   
We restrict the function $W$ (or equivalently $\Re\,W$) to $\R^n$ and view it as a Morse function on $\R^n$.
The number of $\lambda_{i,p}$ that are negative is defined to be the Morse index $d_p$ of $W$ at $p$.  The generalization
of the above one-variable computation shows that, for a state $f_p$ localized near $p$, and normalized so that $f_p(p)=1$, 
\be\label{hessdet} \la f_p,f_p\ra= \frac{1}{\lambda_{1,p}\lambda_{2,p}\cdots
\lambda_{n,p}}=\frac{1}{\det H_p}, \ee
where $H_p$ is the Hessian (or matrix of second derivatives) of the function $W$ at its critical point $p$.
Thus, $\la f_p,f_p\ra$ is positive or negative depending on whether $d_p$ is even or odd.  If there are multiple critical points, then $\la~,~\ra$ can be defined to be
positive if and only if the Morse indices at critical points are all even or all odd.  For a simple example in two dimensions showing that such behavior is
possible with any number of critical points, consider the case that
\be\label{nvu} W(x,y)=y U(x). \ee
Assuming that there are no solutions of $U=\d U=0$, the critical points are at $y=U(x)=0$.  They are nondegenerate if and only if $U(x)$ has only simple zeroes, and if so all critical points have Morse index 1.

The $B$-model quantum mechanics that we have described has four supercharges, though we have emphasized just one that plays the role
of a differential in  a topological version of the model.   This quantum mechanics is the low energy limit of the closed string sector $B$-model, with target $X$ and
superpotential $W$.   In the present paper, we are interested in open strings, not closed strings; that is, we are interested in the $B$-model formulated
on an interval $I$, with boundary conditions set by branes.   This breaks at least half of the supersymmetry, and accordingly $B$-model open strings,
with a generic Calabi-Yau target, are not governed by the sort of quantum mechanics that we have described.   However, if $X$ is hyper-Kahler,
this doubles the amount of supersymmetry.   For hyper-Kahler $X$, open strings that end on complex Lagrangian branes are governed by
the same $B$-model quantum mechanics that we have described.

We consider the $B$-model  of $X$ in one of its complex structures.
In our application, $X$ is the Higgs bundle moduli space $\M_H(G,C)$, and we choose
the complex structure $J$, with holomorphic symplectic form $\Omega_J=\omega_K+\i\omega_I$.   (The holomorphic volume form of the above
discussion is then $\varOmega=\frac{1}{d!}\Omega_J^d$, with $d=n/2$.)   
Locally, we can pick coordinates on $X$ with $\Omega=\sum_i \d p_i \d q^i$.  Because of $B$-model localization, this local description is enough
for many purposes; we make a more global statement presently.

We start by considering closed strings in a way that uses the complex symplectic structure of $X$.
Let $\L X$ be the free loop space of $X$, parametrizing maps $\Phi:S^1\to X$.
We can describe $\Phi$ by functions $p_i(\sigma)$, $q^j(\sigma)$, and we define a holomorphic function on $\L X$ by
\be\label{knon} W(\Phi)=\oint_{S^1} p_i \d q^i. \ee  
It turns out that the supersymmetric $\sigma$-model with target $X$ can be understood as $B$-model quantum mechanics on $\L X$ with this
superpotential.   For example, with this choice of $W$, the usual potential energy $|\d W|^2$ of $B$-model 
quantum mechanics becomes $\oint\d\sigma \sum_i\left(\left|\frac{\d p_i}{\d \sigma}\right|^2
+\left|\frac{\d q^i}{\d \sigma}\right|^2\right)$, which is the usual kinetic energy of a $\sigma$-model with target space coordinates $p_i$, $ q^j$.   The condition
for a critical point of $W(\Phi)$ gives
\be\label{nonj}\frac{\d p_i}{\d \sigma}=\frac{\d q^i}{\d \sigma}=0.\ee  Since  $B$-model quantum mechanics localizes on critical points, as described above, this
means that the $B$-model localizes on constant maps to $X$.  That is a standard result.

Now let us consider open strings.   We consider an open-string model in which the superpotential is the same integral as in eqn. (\ref{knon})
plus some boundary terms.  To see what boundary terms should be considered, first consider without boundary terms the functional
\be\label{jnon}W_0(\Phi)=\int_I p_i \d q^i =\int_0^1 p_i \frac{\d q^i}{\d t} \d t\ee
on the interval $I:0\leq t\leq 1$.      Looking for a critical point, we have the same condition $\d p_i/\d t= \d q^i/\d t =0$ as before, but now there
are also endpoint conditions $p_i(0)=p_i(1)=0$.   Since $p$ vanishes at the endpoints in the unperturbed theory, in adding boundary terms to $W_0(\Phi)$, we can assume
that these boundary terms are functions of $q$ only.   Thus we choose holomorphic functions $P_\ell(q^i)$, $P_r(q^i)$, and define the superpotential
\be\label{koln}W(\Phi)=\int_0^1 p_i \frac{\d q^i}{\d t}\d t - P_r(q^i(1)) +P_\ell(q^i(0)). \ee 
   The endpoint conditions for a critical point of $W$ are now
\begin{align}\label{endpt} p_i(1) &=\frac{\partial  P_r(q(1))}{\d q^i}\cr 
                                         p_i(0) &=\frac{\partial  P_\ell(q(0))}{\d q^i}.\end{align}
 In general, if $P(q^i)$ is a holomorphic function, the condition $p_i=\partial P/\partial q^i$ defines a complex Lagrangian submanifold of $X$.   Conversely
 a generic complex Lagrangian submanifold can be described locally in this form; $P(q)$ is called the generating function of the canonical
 transformation that maps the complex Lagrangian submanifold $p_i=0$ to the complex Lagrangian submanifold $p_i=\partial_i P$.   So eqn. (\ref{endpt})
 describes two complex Lagrangian submanifolds $L_\ell$ and $L_r$ that are defining the boundary condition at $t=0$ and $t=1$.
 
 What we have found then is a version of $B$-model quantum mechanics for open strings propapagating on $X$ with left and right boundary conditions
 set by complex Lagrangian submanifolds $L_\ell$ and $L_r$, respectively.     The construction can also be slightly modified to incorporate nongeneric
 complex Lagrangians, such as the one defined by $q^i=0$, which cannot be put in the form $p_i=\partial_i P$ for some $P$. 
 
   In this construction,
critical points of $W$ correspond to constant maps from $I$ to $X$ that map $I$ to a point in $L_\ell$ (because of the boundary condition at $t=0$)
and in $L_r$ (because of the boundary condition at $t=1$).   So in short, the localization is on constant maps of $I$ to $L_\ell\cap L_r$. 

 Now we make a few remarks of a more global nature.
 In general, the superpotential $W$ of eqn. (\ref{koln}) is well-defined
if $\Omega$ is cohomologically trivial, meaning that $\Omega=\d \lambda$ for a holomorphic 1-form $\lambda$, and the Lagrangians $L_\ell$ and $L_r$
are exact, meaning that $\lambda$ can be chosen to be exact (and not just closed) when restricted to $L_\ell$ or to $L_r$.   When these conditions    are not satisfied,
$W$ is in general multi-valued -- it is only well-defined modulo periods of $\Omega$ in $X$ and of $\lambda$ in $L_\ell$ and $L_r$.   That actually does not prevent the
$B$-model from being well-defined, as the Lagrangian of the model only involves derivatives of $W$ and is not sensitive to a constant shift in $W$.

 In our application, we take $X$ to be the Higgs bundle moduli space, and $L_\ell$, $L_r$ to be the varieties $L_\op$ and $L_\bop$ of holomorphic
 and antiholomorphic opers.   Based on our finite-dimensional analysis, we can immediately state necessary conditions for the $B$-model hermitian 
 form on $\H=\Hom(\B_\bop,\B_\op)$  to be positive-definite.   The intersection $\Upsilon=L_\op\cap L_\bop$ must consist only of real points.  And those intersections
 must be transverse, since that is the condition that makes the critical point of $W$ corresponding to an intersection nondegenerate.   This accounts
for   the claim  in section   \ref{centertriv} that  for positivity of the hermitian inner product of the physical state
space of  the $B$-model, the intersections must be transverse.

These are necessary conditions, but the finite-dimensional discussion showed that in general they are not sufficient.   In the finite-dimensional case,
the remaining necessary condition is that, if $\Re\,W$ is viewed as a Morse function on the real locus, then the differences in Morse indices of critical
points should be even.   In our application, $\Re\,W$ (but not $\Im\,W$) is actually single-valued.   That is because, in the case of the Higgs bundle
moduli space, $\Re\,\Omega_J=\omega_K$ is exact; moreover, as $L_\op$ and $L_\bop$ (being topologically equivalent to $\C^{n/2}$)
are contractible, they are automatically exact.   

$\Re\,W$ is a Morse function on an infinite-dimensional space, but nevertheless it is possible to make sense of the Morse index of a critical point,
in a regularized sense.   In the context of $\N=4$ super Yang-Mills theory compactified on a Riemann surface $C$, $\Re\,W$ is naturally written
as a Chern-Simons integral on the three-manifold $I\times C$:
\be\label{jnu} \Re\,W=\int_{I\times C}\Tr\,\left(\phi\wedge F-\frac{1}{3}\phi\wedge\phi\wedge\phi\right).\ee
This is the imaginary part of the Chern-Simons function  $\frac{1}{2}\int \Tr\,\left(\cA\d\cA +\frac{2}{3}\cA^3\right)$ 
of the complex connection $\cA=A+\i\phi$; $F=\d A+A\wedge A$ is the Yang-Mills curvature.
At a critical point of $\Re\,W$, its Hessian  is an elliptic differential operator modulo the gauge
group, and the Atiyah-Patodi-Singer (APS)  $\eta$-invariant of this operator is a regularized version of the Morse index.    It is possible to use index theory
to show that the differences between the values of $\eta$ at critical points are integers.  For this, one uses the fact that the gradient flow equation
for the function $\Re\,W$ is actually, and rather exceptionally, an elliptic differential equation (the KW equation \cite{KW})  in four dimensions \cite{WittenAnalytic}.
The APS index theorem implies that the difference between the values of $\eta$ at critical points is the index of the linearized KW equation for
a four-dimensional field (not necessarily a solution of the KW equation) that interpolates between the two critical points.   The index is an integer,
so the differences between $\eta$-invariants  are integers.   However, for positivity of the hermitian form on $\H$, we need the differences to be
even integers.   A proof of this is not immediately apparent.   

If it is true that the differences of $\eta$-invariants are even integers, then $\Re\,W$ can be
viewed, in the sense of \cite{AB}, as an equivariantly perfect Morse function on the real locus, which here consists of complex gauge fields on $I\times C$ that satisfy $\A(t,p)=\bar\A(1-t,p)$, up to a gauge transformation, for $p\in C$, along with the oper boundary condition at $t=0$.   Alternatively, letting $I'$ be
the smaller interval $0\leq t\leq 1/2$, the real locus consists of complex gauge fields on $I'\times C$ that satisfy the oper boundary condition 
at $t=0$ and which are real at $t=1/2$, in the sense that the structure group of the complex connection $\A=A+\i\phi$ reduces at $t=1/2$ to a real
form of $G_\C$.    The claim is that for gauge fields on $I'\times C$ with complex gauge group $G_\C$ satisfying the indicated boundary conditions at the two ends,
the function $\Re\,W$ is an equivariantly perfect Morse function. 

Apart from the question of whether the inner product is positive, the formula (\ref{hessdet}) for the normalization of the state $f_p$ is noteworthy.  The formula
says that if  $f_p$ is
a $B$-model state localized at $p$ and normalized by $f_p(p)=1$, then its norm is $\la f_p,f_p\ra=1/\det\,H_p$, where $H_p$ is the Hessian of the
Morse function at $p$.   For the case of gauge theory
on $I'\times C$ with the boundary conditions that we have imposed, the critical points are flat $G_\C$ bundles over $C$ that are real opers. 
Consider a real oper that corresponds to a
transverse intersection point $p\in L_\op\cap L_\bop$.   Associated to such a point is a 1-dimensional space of states in the $B$-model; this
is simply the space of complex-valued functions at the point $p$.   There is a distinguished vector $\Psi_p$ in this space, corresponding to the function 1.  
The point $p$ determines a real oper bundle $E\to I'\times C$; let $E_\ad\to I'\times C$ be the flat bundle associated to $E$ in the adjoint representation.   The analog of $1/\det H_p$
in this situation is the Ray-Singer analytic torsion\footnote{Analytic torsion was originally defined for flat bundles with compact structure group
but the definition can be generalized to flat bundles with noncompact structure group, for example structure group $G_\C$.}
 of the flat bundle $E_\ad\to I'\times C$, with the relevant boundary conditions at the ends.   Thus
this is a conjectural formula for the Hilbert space norm of the state $\Psi_p$, with its natural $B$-model normalization.   It would be desirable to find a combinatorial
analog of the analytic torsion in this situation, but we do not immediately know how to do this with oper boundary conditions.

 \begin{figure}
 \begin{center}
   \includegraphics[width=1.2in]{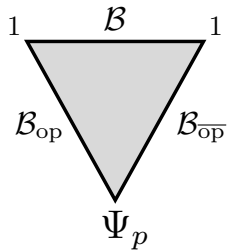}
 \end{center}
\caption{\small  A triangle with the bottom corner labeled by $\Psi_p\in \Hom(\B_\bop,\B_\op)$ and the other corners labeled by the canonical
corners between the branes $\B_\op$ and $\B_\bop$ and the dual $\B$ of the regular Nahm pole.  The path integral on this triangle
computes $(\chi,\Psi_p)$.  \label{example10}}
\end{figure} 

The normalization condition that we have imposed on $\Psi_p$
has another interpretation.  We explained in Section \ref{nbc} that it makes sense to define the inner product of 
$\Psi_p$ with a state created in the $A$-model
by the brane $\B_N$ associated to the regular Nahm pole (with suitable corners), even though the support of this  brane is entirely outside $T^*\M(G,C)$.  This
computation is actually quite straightforward in the $B$-model.   The $B$-model dual of $\B_N$ is simply the structure sheaf of $\M_H$; that is, it is a rank
1 brane $\B$ whose support is all of $\M_H$, with trivial $\CP$ bundle.    $\B$ has a canonical  corner  with $\B_\cc$, associated to the constant function 1 on the intersection of the support
of the two branes, namely on $\M_H\cap L_\op=L_\op$.   Likewise, $\B$ has a  canonical corner with $\bar\B_\cc$, represented by the function 1 on
$\M_H\cap L_{\bop}=L_{\bop}$.   Let $\chi\in \Hom(\bar\B_\cc,\B_\cc)$ be
the state created by the brane $\B$ with these canonical corners.    The $B$-model pairing $(\chi,\Psi_p)$ is computed by a  path integral
on a triangle (fig. \ref{example10}) whose sides are labeled by the branes $\B$, $\B_\cc$, and $\bar\B_{\cc}$ and whose corners are labeled by the canonical corners of $\B$
and the state $\Psi_p$.    The $B$-model localizes on constant maps, and in this case the only constant map that satisfies the necessary conditions is
a constant map of the triangle to the point $p$.   We have chosen all the wavefunctions to equal 1.   So $(\chi,\Psi_p)=1$.   In other words, the normalization
condition on the eigenstate $\Psi_p$ of the Hitchin Hamiltonians that leads to the prediction stated previously for its Hilbert space norm is $(\chi,\Psi_p)=1$.    
If it is possible to compute $(\chi,\Psi_p)$ on the $A$-model side, this would give a prediction for the Hilbert space norm of $\Psi_p$ purely
in $A$-model terms.  The pairing $(\chi,\Psi_p)$ is somewhat analogous to the Whittaker coefficient in the categorical approach to geometric Langlands.

\section{Some Examples of Relations Between Canonical Oper Sections}\label{somex}
\subsection{The Case of $\SL(2,\C)$}
The fundamental representation has a basis of canonical sections $s_2 \equiv s$, $Ds$, with $D^2 s = t s$. 
Irreducible representations of $\SL(2,\C)$ can be presented as symmetric products of the fundamental representation. 
The symmetric powers $s^{\otimes n}$ give canonical sections $s_{n+1}$ in the $(n+1)$ dimensional representation.  What differential equation is
satisfied by $s_{n+1}$?
After replacing every $D^2 s$ with $t s$, the object
$D^r s_{n+1}$, for $r\leq n$, can be expressed as a linear combination of symmetrized forms of $s^{\otimes (n-k)} Ds ^{\otimes k}$, for $k\leq r$.
In order  $n+1$, no new object appears and one gets a differential equation.  For example,
\begin{equation}
D^3 s_3= D^2(s \otimes Ds + Ds \otimes s) =2 D(Ds \otimes Ds + 2 t s \otimes s) = 6 t Ds_3 + 4 Dt s_3.
\end{equation}

As we multiply representations, antisymmetrizations can be simplified by $s \wedge Ds=1$. For example, we have 
\begin{equation}
s_2 \otimes Ds_2 = \frac12 D s_3 + \frac12 s_1
\end{equation}
where $s_1=1$ and we left implicit the embedding maps of $2 \otimes 2 = 3\oplus 1$. We could even write an OPE:
\begin{equation}
s_2(z) \otimes s_2(w) = s_3(w) + \frac12(z-w)Ds_3(w) + \cdots - \frac12 (z-w) s_1(w)+ \cdots
\end{equation}
 
\subsection{The Case of $\SL(3,\C)$}
The fundamental representation has a basis of sections $s_3 \equiv s$, $Ds$, $D^2s$, with $D^3 s = t_2 Ds +t_3 s$. 
The exterior product $s_{\bar 3} = s_3 \wedge Ds_3$ is a canonical section in the anti-fundamental representation. We indeed have
\begin{equation}
D^3s_{\bar 3} = D^2(s_3 \wedge D^2s_3) = D(D s_3 \wedge D^2s_3+t_2 s_3 \wedge D s_3) = t_2 Ds_{\bar 3} + (D t_2-t_3) s_{\bar 3}
\end{equation}

The square $s_{6,5} = s_3 \otimes s_3$ is a canonical section of the symmetric square representation. 
The first five derivatives $D^n s_{6,5}$ are independent. The sixth derivative can be expressed in terms of the 
lower derivatives, but with coefficients which are rational functions in $t_2$, $t_3$ and their derivatives. 
It is better to define a second basis element 
\begin{equation}
s_{6,1} \equiv s_3 \otimes D^2 s_3 + D^2 s_3 \otimes s_3 - D s_3 \otimes Ds_3- t_2 s_3 \otimes s_3
\end{equation} 
We then get relations of the form 
\begin{align}
&D^5 s_{6,5}- 5 t_2 D^3 s_{6,5}+ (7 t_3 + 4 D t_2)D^2 s_{6,5} + \cdots = (4 t_3 - 2 D t_2) s_{6,1} \cr
&D s_{6,1} = (2 t_3- D t_2) s_{6,5} 
\end{align} 
We can capture these relations and more in an OPE 
\begin{equation}
s_3(z) \otimes s_3(w) \sim s_{6,5} + \cdots + \frac12 (z-w) s_{\bar 3} +\cdots + \frac14 (z-w)^2 s_{6,1} + \cdots
\end{equation}

Notice that $s_3$ and $s_{\bar 3}$ are orthogonal to each other, so $s_{8,5} = s_3 \otimes s_{\bar 3}$ is a canonical adjoint section and
 $s_{8,3} = s_3 \otimes D s_{\bar 3}-D s_3 \otimes s_{\bar 3}$ is a second one. These satisfy the expected differential equations 
 of the form $D^5 s_{8,5} = \cdots$ and $D^3 s_{8,3}= \cdots$, with derivatives of both sections on both right hand sides.

 \section{A Local Model Of Singular Behavior Of Wavefunctions }\label{localmodel}
 
 \subsection{The Wobbly Divisor}\label{wobbly}
 
An eigenfunction $\Psi$ of the Hecke operators or the Hitchin Hamiltonians is naturally represented, as discussed in Section \ref{wkb},  by a rank 1 Lagrangian
 brane $\B_F$ supported on a fiber $F$ of the Hitchin fibration.   $F$ must satisfy a quantum-corrected WKB condition.
 
 To represent $\Psi$ as an explicit half-density on $\M(G,C)$, one can take its inner product with a delta function state $\delta(x,x_0)$ 
 to get $\Psi(x_0)=\la \delta(x,x_0),\Psi\ra$.   The delta function state $\delta(x,x_0)$ is represented by a Lagrangian brane $\B_{x_0}$ supported on the fiber $L_{x_0}$
 of the cotangent
 bundle $T^*\M(G,C)\to \M(G,C)$ over the point  $x_0\in \M(G,C)$.   Though the evaluation of the inner product $\la \delta(x,x_0),\Psi\ra$ is a difficult problem that is not
 likely to admit any fairly explicit answer, it is well-defined and will produce a result that varies with $x_0$ in a real-analytic fashion as long as
 the brane $\B_{x_0}$ in the $A$-model of $\M_H(G,C)$ is well-defined and varies with $x_0$ in a real-analytic fashion.     This in turn is true as long as 
$L_{x_0}$
 is closed in $\M_H(G,C)$.   
 
 It is known, however, that this fails along a divisor $\sD\subset \M(G,C)$ that is sometimes called the wobbly divisor (see for example \cite{DP,PP,PPN,HZ}).   A stable bundle $E\to C$ is said to be very
 stable if any Higgs field $\varphi:E\to E\otimes K$ that is nilpotent is actually 0.   Conversely, if there is some nilpotent and nonzero $\varphi:E\to E\otimes K$,
 then $E$ is said to be wobbly.   The divisor $\sD$ parametrizes wobbly bundles.  
 
  The support of a brane must be closed, so when $L_{x_0}$ fails
 to be closed, it is not the support of a brane.   Hence the wavefunction $\Psi(x_0)$ may become singular along $\sD$.
 
 In this appendix, we will analyze the behavior along $\sD$  for the case that $G=\SU(2)$; with minor changes, the discussion also applies for
 $G=\SO(3)$.     Explicit computations for the behavior of eigenfunctions of the Hecke operators and the Hitchin Hamiltonians are available in genus
 0 for bundles with parabolic structure \cite{EFK3}.   Though we have not explicitly incorporated parabolic structure in the present article, and will not
 do so in the present appendix,  all of our considerations extend naturally for bundles with parabolic structure.   (In gauge
 theory, parabolic structure is described by incorporating a certain class of surface operators, associated to codimension 2 singularities \cite{GuW2}.)
 Therefore, we will compare our results to what has been found in \cite{EFK3}.   
 
 In what follows, $C$ is a Riemann surface of genus $g$ with canonical bundle $K$ and tangent bundle $T$, and $E\to C$ is a holomorphic vector bundle of rank 2 with
 $\det\,E$ trivial.  
If $E$ is not very stable, then there is a nonzero Higgs field $\varphi:E\to E\otimes K$ with $\varphi^2=0$.   Then $E$ has a line subbundle $\L=\mathrm{ker}\,\varphi$ and
so $E$ is an extension
\be\label{extform} 0\to \L \to  E\to \L^{-1}\to 0.\ee
We are interested in a nontrivial extension, but we will want to compare to the direct sum $E_0=\L\oplus \L^{-1}$.   The main case that we are interested in will be that $E_0$
is unstable but $E$ is stable.   This is true generically if\footnote{A rank 2 bundle $E\to C$ of degree 0 is stable if and only if
any rank 1 subbundle $\M\subset E$ has negative degree.  
 If $E$ has a line subbundle of degree zero but none of positive degree,
it is said to be semistable.
 By definition, $E$ in the exact sequence (\ref{extform}) has the subbundle $\L$; if $\L$ has
negative degree, this does not destabilize $E$ (and an $E$ appearing in such an exact sequence is generically stable).   However, for $\L$ of negative degree,
$\L^{-1}$ has positive degree and the direct sum $E_0=\L\oplus \L^{-1}$ is unstable.}
$$c_1(\L)<0. $$
For generic $\L$, the automorphism group of $E_0$, as an $\SL(2,\C)$ bundle, is the group of diagonal matrices, with we will call $P$.

The nilpotent Higgs field in this situation is $\varphi_0\in H^0(C,\Hom(\L^{-1},\L\otimes K))=H^0(C, \L^2\otimes K)$.   We may expect that at a generic point of the wobbly divisor $\sD$ the space of nilpotent Higgs fields is  1-dimensional, so we anticipate that we will want\footnote{The dimension of $H^q(C,\M)$ is denoted as $h^q(C,\M)$ or just $h^q(\M)$.} $h^0(\L^2\otimes K)= 1$.  
For simplicity, we will study a component of $\sD$ with the property that $\L$ can be generically deformed (keeping fixed only $c_1(\L)$) within $\sD$.
(There are components of $\sD$ such that this is not possible; they can be studied similarly, with similar results.)  
 Under this assumption, asking for  1  to be the generic value of $h^0(\L^2\otimes K)$, 
 the generic value of $h^1(\L^2\otimes K)$ must be 0 and the degree of $\L$ is determined by the Riemann-Roch formula:
\be\label{degl} 1=h^0(\L^2\otimes K)-h^1(\L^2\otimes K) = 1-g+ c_1(\L^2\otimes K)=g-1 +2c_1(\L)\ee
so
\be\label{egf}c_1(\L)=1-\frac{g}{2},\ee
Therefore, this construction as stated is only possible for even $g$.   However, for odd $g$, we can do something very similar: the half-integral value of $c_1(\L)$
means that we should take the gauge group to be $\SO(3)$ instead of $\SU(2)$ and replace $E$ with $\ad(E)$ everywhere.
  As it is notationally simpler, we will continue with the language of
$\SU(2)$ gauge theory, but the following remarks also apply to the other case.
   For $g>2$, $c_1(\L)<0$, so an extension $E$ as in eqn. (\ref{extform}) can be stable but the direct sum $E_0$ is unstable.  For $g=2$, $E$ can be stable
   while $E_0$ is semistable.

The condition $h^0(\L^2\otimes K)=1$ is Serre dual to $h^1(\L^{-2})=1$.   Since $H^1(C,\L^{-2})$ parametrizes deformations of $E_0$ that do {\it not}
preserve the triangular structure (\ref{extform})    (and the counting of deformations of $E$ that do not preserve this structure is the same), the condition
$h^1(\L^{-2})=1$ is indeed what we want so that the structure we are discussing occurs in codimension 1 in $\M(G,C)$, and thus can represent the generic behavior along $\sD$.
We can also now count the deformations of $E_0$ that {\it do} have the triangular structure of eqn. (\ref{extform}).  
 The dimension of this space of deformations is
$h^1(\L^2)=g-1-2c_1(\L) =2g-3$.    As a check, the dimension of the family of stable bundles that we make by deforming $E_0$ is 
$h^1(\ad(E_0)) - h^0(\ad(E_0))-\mathrm{dim}\,P = g + (2g-3)+1 -1 =3g-3$, the expected value.  (We included $g$ diagonal deformations of $E_0$.
In counting the parameter space of deformations of $E_0$,
we subtract ${\mathrm{dim}}\,P=1$ to account for the fact that generic deformations break the $P$-invariance of $E_0$.)
For what follows, define
\be\label{weddo} N = h^1(\L^2)= 2g-3,\ee
the number of upper triangular deformations of the unstable bundle $E_0$ along the divisor $\sD$.    Thus the construction as stated gives only positive odd values of $N$,
though positive even values can be obtained by a similar construction with parabolic structure or by dropping the assumption that $\L$ can be deformed
arbitrarily while remaining within $\sD$.

Now we are going to look at the local behavior of the Higgs bundle moduli space  in this situation.   
In the basis with $E_0=\begin{pmatrix}\L\cr \L^{-1}\end{pmatrix}, $ the unstable bundle $E_0$ can be described explicitly by a $\bar\partial$ operator
\be\label{barpar}\bar D=\bar\partial +\begin{pmatrix} a_0 & 0 \cr 0 & -a_0\end{pmatrix}. \ee
In what follows, the ``diagonal'' deformations that change $\L$ are not interesting because we have assumed that any generic $\L$ (of the appropriate degree)
can lead to the situation we are considering.   In other words, our assumptions are such that deformations of $\L$ just move us along $\sD$.
For the same reason, we will not be interested in diagonal Higgs fields.  
By constrast, we have to look at the ``above diagonal'' and ``below diagonal'' deformations, which move us away from $\sD$.
  By our assumptions, the space of above diagonal bundle deformations
has dimension $N$, and the space of below diagonal bundle deformations has dimension 1.  We pick a basis $b_1,\cdots, b_n$ of $H^1(C,\L^2)$ and write
a generic element of this group as $\sum t_i b_i$ with complex parameters $t_i$.   Similarly, we pick a nonzero $c\in H^1(C,\L^{-2})$ and write a generic
element as $uc$ with a complex parameter $u$.  
So a generic off-diagonal perturbation of $\bar D$ will give us
\be\label{arpar}\bar D'=\bar D +\begin{pmatrix} 0 & \sum_{i=1}^n t_i b_i \cr uc & 0  \end{pmatrix}. \ee
Similarly, for the Higgs field, we introduce a nonzero element $e$ of $H^0(C, K\otimes \L^2)$  and a basis $f_1,f_2,\cdots,f_N$ of $H^0(C,K\otimes \L^{-2})$, 
and then, with additional complex parameters $s$ and $r_1,r_2,\cdots, r_N$,   the off-diagonal part of the Higgs field is
\be\label{offed}\varphi_\perp =\begin{pmatrix} 0 & se \cr \sum_{i=1}^N r_i f_i&0 \end{pmatrix}. \ee
On this data, we need to impose the Higgs bundle equation $[\bar D',\varphi_\perp]=0$.  Since we already have $[\bar D,\varphi_\perp]=0$ by virtue of the definition
of $\varphi_\perp$, the condition reduces to $\mu=0$ with
\be\label{kinny} \mu=\sum_{i=1}^N r_i t_i -su. \ee 
(Some constants were set to 1 here by suitably normalizing the wavefunctions.)
To get a local description of the Higgs bundle moduli space, or more exactly a slice of it transverse to the uninteresting diagonal deformations, we also
have to divide by $P\cong \C^*$, the automorphism group of the unstable bundle $E_0$.   $P$ acts on these variables with weights 1 for above diagonal
parameters $t_i$ and $s$, and $-1$ for below diagonal parameters $u$ and $r_i$.    The combined operation of dividing by $P$ and imposing the condition
(\ref{kinny}) has a simple interpretation.      Introducing the symplectic form
\be\label{inny}\Omega = \sum_{i=1}^N \d r_i\, \d t_i +\d s \,\d u, \ee 
we see that $\mu$ is the moment map for the action of $P$ on these parameters.   The combined operation of setting $\mu=0$ and dividing by $P$
 is a complex symplectic quotient and the result will be a complex symplectic manifold, as expected for the Higgs bundle moduli space.

However, we have to be careful about what we mean by the quotient.   A Higgs bundle in which the weight 1 parameters $t_i$ and $s$ all vanish
is unstable.\footnote{Let $E$ be a rank 2 bundle with trivial determinant.  A Higgs bundle $(E,\varphi)$ is unstable if $E$ has a $\varphi$-invariant
subbundle  of positive degree. If $E$ has a $\varphi$-invariant subbundle of zero degree but none of positive degree, then $(E,\varphi)$ is said to be semistable.
  If $E$ has a subbundle of positive degree, but any  $\varphi$-invariant subbundle of $E$ has negative degree, then the bundle $E$ is
unstable but the Higgs bundle $(E,\varphi)$ is  stable.   In the present example, if $t_i=s=0$, then $E$ has the $\varphi$-invariant subbundle 
$\L^{-1}$ of positive degree.}  So we want to take the quotient
with a Mumford stability condition such that $t_i$ and $s$ are not allowed to all vanish.   Modulo the action of $P$, the $N+1$ variables 
$t_i$ and $s$ therefore define a point in 
$\bCP^N$ and   the local model for the Higgs bundle moduli space that we get from this construction  is therefore
\be\label{hb} M_H= T^*\bCP^N. \ee
To be more exact, this is a transverse slice that captures the relevant aspects of the Higgs bundle moduli space near $\sD$.

What about a corresponding local model for the moduli space $M$ of semistable bundles?   For this we forget the Higgs parameters $s$ and $r_i$
and just remember the bundle parameters $t_i$ and $u$.   If we set all $t_i$ to vanish, we get an unstable bundle.   So the $t_i$ are taken to not
all vanish and, modulo the action of $P$, they define an element of $\bCP^{N-1}$.   
Taking the weight $-1$ variable $u$ into account, we see
that the local model of $M$ is the total space of the line bundle $\O(-1)\to \bCP^{N-1}$.    This is the same as the blowup of $\C^N$ at a point.
The divisor $\sD$ of not very stable bundles is defined in this model by the equation $u=0$; in other words, it is the exceptional divisor 
$\bCP^{N-1}\subset M$.    It is convenient to define $x_i= u t_i$.   So $M$ is parametrized by the $x_i$ with the point $\vec x=0$ blown up.
For variables canonically conjugate to the $x_i$, we can take $p_i=r_i/u$, since
\be\label{wobo}\Omega =\sum_{i=1}^N \d p_i \d x_i. \ee
Of course, these coordinates are not good near the exceptional divisor.

  The case $N=1$ is exceptional in many ways.  
For $N=1$, the bundle $E_0$ is semistable rather than unstable, since $\L$ has degree 0.  Moreover, for $N=1$, there is actually no blowup
in the construction just described.  Only for  $N>1$ is  the divsor
$\sD\cong \bCP^{N-1}\subset M$  wobbly.    For $N=1$, it is semistable; that is, it parametrizes a family of semistable bundles.

 We can now explain the assertion made at the beginning of this appendix that along a wobbly
divisor, the fiber of the cotangent bundle to the moduli space of bundles fails to be closed in the Higgs bundle moduli space.  In this analysis, we
replace the moduli space of bundles $\M(G,C)$ and $\M_H(G,C)$ by their local models $M$ and $M_H$.  

First we consider a point in $M$ that is not in $\sD$.   The complement to $\sD$ in $M$ consists of nonzero $N$-plets $\vec x=(x_1,x_2,\cdots, x_N)$.
For example, take the point $p$ defined by $x_1\not=0$, $x_2=\cdots= x_N=0$.    Since $x_1=ut_1$, $x_1\not=0$ 
 implies that $u$ and $t_1$ are both nonzero.  We can fix
the action of $P$ to set $t_1=1$.   The Higgs field is constrained by 
\be\label{higgscon} r_1-su=0, \ee
with no constraint on $r_2,\cdots, r_N$.   After using eqn. (\ref{higgscon}) to solve for $r_1$ in terms of $s$, we see that the space of Higgs fields
at $p$ is parametrized by $s,r_2, \cdots, r_N$.   This is the expected copy of $\C^N$; it  is the fiber at $p$ of the cotangent 
bundle of $M$, and it is closed in $\M_H$. We will denote it as $L_p$.   Now instead let us look at the
fiber of the cotangent bundle at a point $p'\in \sD$.     For this, we can simply take $u=0$, still with $t_1=1$ and $t_i=0,$ $i>1$.   Eqn. (\ref{higgscon})
still holds and tells us to set $r_1=0$, with no constraint on $u$.    Thus the cotangent fiber $L_{p'}$ is in this case still parametrized
by $s$ and $r_2,\cdots, r_N$.    But $L_{p'}$ is not closed in $M_H$.   To see this, consider the copy of  $\C\subset L_{p'} $ defined by  $(t_1,s)=(1,\lambda),\,\lambda\in\C$
and $r_2=\cdots = r_N=0$.   By the action of $P$, the condition $(t_1,s)=(1,\lambda)$ is equivalent to $(t_1,s)=(1/\lambda,1)$.    
Evidently,  the limit $\lambda\to\infty$ exists.
Therefore this copy of $\C$ is not closed
in $M_H$; its closure contains a point with $(t_1,s)=(0,1)$, compactifying $\C$ to  $\bCP^1$.   More generally, the closure of $L_{p'}$ in $M_H$
is parametrized by $(t_1,s,r_2,\cdots, r_N)$ modulo the action of $P$, with the constraint that $t_1$ and $s$ are not allowed to be both zero.  It is a $\C^{n-1}$
bundle over $\bCP^1$.    The points
with $t_1=0$, $s\not=0$ correspond to Higgs bundles $(E_0,\varphi)$ such that the underlying bundle $E_0$ is unstable but the pair $(E_0,\varphi)$ is
stable.   These points are contained in $M_H$ but not in $T^*M$.   If we set $t_1=s=0$, we get an unstable Higgs bundle.

Now let us return to a generic point $p\in M$, not contained in $\sD$, and ask what happens to $L_p$ as $p$ approaches $\sD$.  We consider the same
point $p$ as before, but since we now know that it is going to be important to allow the possibility $t_1=0$, we write the constraint (\ref{higgscon}) without
setting $t_1=1$:
\be\label{iggscon} t_1 r_1-su=0. \ee
We also have the definition of the point $p$:
\be\label{defp} ut_1= x_1.\ee
$L_p $ is parametrized by  $t_1,u,s$ and $r_1,r_2\cdots, r_N$, satisfying these equations, and modulo the action of $P$. 
To take the limit of $L_p$ as $p\to p'$, we just set $x_1=0$ in eqn. (\ref{defp}).   The resulting variety has two components.  On one component, 
$u=r_1=0$ and $t_1,s$ are generically nonzero.   This component, which we will call $L_{p',1}$, is the closure of $L_{p'}$ in $M_H$, as described earlier; it is a $\C^{N-1}$ bundle over
$\bCP^1$.   On the second component, $t_1=0$, so that $s$ must be nonzero (for stability of $(E,\varphi)$) and hence eqn.
(\ref{iggscon}) implies that $u=0$.    The second component is a copy of $\C^N$, parametrized by $r_1,\cdots, r_N$.   Since $t_1=u=0$, this second component $L_{p',2}$
is entirely contained in the complement of  $T^*M$ in $M_H$.

For generic $p$, there is a rank 1 $A$-brane $\B_p$ supported on $L_p$; it is unique, as $L_p \cong \C^N$ is simply-connected.   For $p\to p'$,
$L_p$ splits up into two components, either of which can suppport an $A$-brane.   Since $\B_p$ only varies smoothly away from the wobbly divisor,
an eigenfunction of the Hitchin Hamiltonians can potentially become singular along this divisor.   
We investigate this question in Section \ref{singular}.

For $N=1$, there is no wobbly divisor and no blowup in this construction. (The genus 2 moduli space does have a wobbly divisor, but not with the
properties that we have assumed.)  The $N$-plet $(t_1,\cdots, t_N)$ collapses to a single variable
$t$, and the moduli space $M$, in this local model, is a copy of $\C$ parametrized by $x=ut$.   The behavior at $x=0$ is exceptional, but for a different
reason.  At $x=0$, the bundle $E$ reduces to the direct sum $E_0=\L\oplus \L^{-1}$, which for $N=1$ is semistable. There is an unbroken $\U(1)$ gauge symmetry along the divisor $\sD$, which must be taken into account in understanding the behavior along
this divisor. 

\subsection{Singular Behavior of Wavefunctions}\label{singular}

A generic Hitchin Hamiltonian is $\int_C \alpha \Tr \varphi^2$,
for $\alpha\in H^1(C,T)$.    Because of the form of the Higgs field $\varphi$ in eqn. (\ref{offed}), any such Hamiltonian is
homogeneous in $s$ of degree 1 and likewise homogeneous in the $r_i$ of degree 1.   Since the number of linearly independent Hitchin Hamiltonians is the same
as the number of  Higgs bundle parameters, we can generically pick a set of Hitchin Hamiltonians such that
\be\label{genbas} H_i = s r_i,~~~i=1,\cdots,N.\ee  (Because there are also diagonal perturbations that we are ignoring, this is not
the full set of Hitchin Hamiltonians. It is the set of Hitchin Hamiltonians that depend on the off-diagonal perturbations in this approximation.)

In terms of the canonical variables $p_i, x_i$ that are good away from the origin, this is $H_i=p_i\sum_{k=1}^N x_k p_k$.   Upon quantization,
this becomes
\be\label{qut} H_i=\frac{1}{2}\sum_{k=1}^n \left(\frac{\partial}{\partial x_i} x_k \frac{\partial}{\partial x_k}+\frac{\partial}{\partial x_k} x_k\frac{\partial}{\partial x_i}\right). \ee
Operator ordering was chosen to make these operators formally symmetric (the Hitchin Hamiltonians are known to have this property).   
The $H_i$ act on half-densities
\be\label{murky} F(x_1,x_2,\cdots, x_N)(\d x_1\d x_2\cdots \d x_N)^{1/2}. \ee
 Of course, what we have in eqn. (\ref{qut}) is not an exact formula
for the Hitchin Hamiltonians.   It is only an asymptotic formula valid for small $\vec x$.   Consider a scaling in which $x, p$ have respectively weights $1,-1$.
The operators $H_i$ in eqn. (\ref{qut}) scale with weight $-1$.   The claim is that the Hitchin Hamiltonians are given by these expressions modulo 
terms of nonnegative weight.

To use this model to predict the behavior of eigenfunctions of the Hitchin Hamiltonians near the exceptional divisor $\sD$, we need to use
coordinates that are good near $\sD$.    A convenient choice is to use $x_1$ along with $y_k=x_k/x_1$ for $k=2,3,\cdots,N$.   These are good coordinates
near a large open set in $\sD$.   

As well as changing variables from $x_1,x_2,x_3,\cdots,x_N$ to $x_1,y_2,y_3,\cdots, y_N$, we also want to express a half-density in a way that is natural
in these coordinates.   An appropriate form is
\be\label{urky} G(x_1,y_2,\cdots, y_N) (\d x_1 \d y_2 \d y_3\cdots \d y_N)^{1/2}. \ee
Thus the relation between $F$ and $G$ is to be
\be\label{rky} F(x_1,x_2,\cdots,x_N)=x_1^{-(N-1)/2} G(x_1,y_2,\cdots, y_N), \ee
so that $F(x_1,\cdots, x_N)(\d x_1\cdots \d x_N)^{1/2}=G(x_1,y_2,\cdots, y_N) (\d x_1 \d y_2\cdots \d y_N)^{1/2}$.
The Hitchin Hamiltonians acting on $G$ turn out to be in the new coordinates
\begin{align} H_1& = \frac{\partial}{\partial x_1}x_1\frac{\partial}{\partial x_1} -\frac{1}{2}\sum_{k=2}^N\left( y_k \frac{\partial}{\partial y_k}+\frac{\partial}{\partial y_k} y_k\right)
\frac{\partial}{\partial x_1} \cr
H_k& =\frac{\partial^2}{\partial x_1 \partial y_k},~~~~~~~~ k>1. \end{align}
Looking for a joint eigenfunction that behaves as $x_1^\alpha$ for $x_1\to 0$, one sees that for even $N$, the possibilities are $\alpha=0$ and $\alpha=(N-1)/2$,
in accord with  computations \cite{EFK3} for  parabolic bundles in genus 0. In the case of a singular solution $x_1^{(N-1)/2}g(x_1,\vec y)$ with $g(x_1,\vec y)$
holomorphic at $x_1=0$, $g(x_1,\vec y)$ is actually independent of $\vec y$ at $x_1=0$.  For odd $N$, the corresponding statement is that, in addition to holomorphic
solutions, 
there are solutions of the form 
  $f(x_1,\vec y)+g(x_1,\vec y)x_1^{(N-1)/2}\log x_1$, where $f(x_1,\vec y)$ and $g(x_1,\vec y)$ are holomorphic at $x_1=0$ and again $g(x_1,\vec y)$ is
  independent of $\vec y$ at $x_1=0$.

 These formulas only capture the most singular terms near the divisor $D$ and have a rather nongeneric behavior.
In particular, the eigenvalue equations $H_s\Psi=\lambda_s\Psi$, $s=1,\cdots, N$, do not have the expected $2^N$ linearly independent joint solutions.
The expected behavior is restored if one adds generic  higher order terms to the Hitchin Hamiltonians, preserving their commutativity.
For example (abbreviating $\partial/\partial x_1$ as $\partial_1$ and $\partial/\partial y_k$ as $\partial'_k$), a simple perturbation of the Hitchin
Hamiltonians that preserves their commutativity to first order in $\alpha$ is to replace the above formulas by
\begin{align}\label{perturbed} \t H_1& = \partial_1 x_1\partial_1 -\frac{1}{2}\sum_{k=2}^N(y_k \partial'_k+\partial'_k y_k)\partial_1
+2 \sum_{l,m=2}^N \alpha_{klm}\partial'_l y^k \partial'_m \cr
\t H_k&= \partial_1\partial'_k +\sum_{l,m=2}^N \alpha_{klm}\partial'_l\partial'_m,~~~~~k>1, \end{align}  
with generic complex coefficients $\alpha_{klm}=\alpha_{kml}$.   The singular behavior is similar to before.

\bibliographystyle{unsrt}

\end{document}